%% file: interfPaper.tex
\documentclass[a4paper,11pt]{article}

\pdfoutput=1

\usepackage{graphicx,amsmath,amssymb,amsthm,xcolor,soul}
\usepackage[T1]{fontenc}

\usepackage[lmargin=3cm,rmargin=3cm,tmargin=3.5cm,bmargin=3.5cm]{geometry}
\usepackage{authblk} 
\usepackage[square,numbers]{natbib}
\usepackage[linkbordercolor=red]{hyperref}

\makeatletter

\numberwithin{equation}{section}

\renewcommand\section{\@startsection{section}{1}{\z@}%
                                   {-3.5ex \@plus -1.3ex \@minus -.7ex}%
                                   {2.3ex \@plus.4ex \@minus .4ex}%
                                   {\normalfont\large\bfseries}}
\renewcommand\subsection{\@startsection{subsection}{2}{\z@}%
                                   {-2.3ex\@plus -1ex \@minus -.5ex}%
                                   {1.2ex \@plus .3ex \@minus .3ex}%
                                   {\normalfont\normalsize\bfseries}}

\def\fnum@figure{\textbf{\figurename\nobreakspace\thefigure}}
\def\fnum@table{\textbf{\tablename\nobreakspace\thetable}}

\long\def\@makecaption#1#2{%
  \vskip\abovecaptionskip
  \sbox\@tempboxa{\small #1. #2}%
  \ifdim \wd\@tempboxa >\hsize
    \small #1. #2\par
  \else
    \global \@minipagefalse
    \hb@xt@\hsize{\hfil\box\@tempboxa\hfil}%
  \fi
  \vskip\belowcaptionskip}

\makeatother

\def\beq{\begin{equation}}
\def\beqn{\begin{eqnarray}}
\def\eeq{\end{equation}}
\def\eeqn{\end{eqnarray}}
\input{newCommands}

\begin{document}

\hfill ADP-20-17/T1127, CERN-TH-2020-124 \hspace{0.5cm}

\title{Model-independent approach for incorporating interference effects in collider searches for new resonances}
\author[1]{Stefano Frixione}
\author[2]{Lydia Roos}
\author[3]{Edmund Ting}
\author[4]{Eleni Vryonidou}
\author[3]{Martin White}
\author[3]{Anthony G. Williams}

\affil[1]{INFN, Sezione di Genova, Via Dodecaneso 33, I-16146, Genoa, Italy}
\affil[2]{Sorbonne Universit\'e, CNRS, Laboratoire de physique nucl\'eaire et de hautes \'energies (LPNHE-IN2P3), Paris, France}
\affil[3]{ARC Centre of Excellence for Particle Physics at the Terascale \& CSSM, School of Chemistry and Physics, University of Adelaide, Adelaide, Australia}
\affil[4]{Theoretical Physics Department, CERN, Geneva, Switzerland}

\date{} 

{\let\newpage\relax\maketitle}
\flushbottom
\thispagestyle{empty}

\begin{abstract}
\noindent The presence of large-mass resonances in the data collected at the Large Hadron Collider would provide direct evidence of physics beyond the Standard Model.
A key challenge in current resonance searches at the LHC is the modelling of signal--background interference effects,
which can severely distort the shape of the reconstructed invariant mass distribution relative to the case where there is no interference.
Such effects are strongly dependent on the beyond the Standard Model theory that must be considered as unknown if one aims to minimise any theoretical bias on the search results.
In this paper, we describe a procedure which employs a physically-motivated, model-independent template functional form that can be used to model interference effects,
both for the characterisation of positive discoveries, and in the presentation of null results.
We illustrate the approach with the example of a scalar resonance decaying into a pair of photons.
\end{abstract}

\cleardoublepage
\setcounter{page}{1}

\tableofcontents
\noindent\rule{\linewidth}{0.3pt}
\vspace{0.5cm}

\section{Introduction}
\label{sec:intro}

With the discovery of the Higgs boson at the Large Hadron Collider (LHC) in 2012~\cite{Aad:2012tfa,Chatrchyan:2012ufa},
the experimental evidence for the particle spectrum of the Standard Model (SM) is seemingly complete.
In addition to this, there is a fast-growing body of successful comparisons between accurate SM predictions and the corresponding LHC data,
for a very large number of observables measured in a significantly diverse set of production processes.
Nevertheless, and regardless of this satisfactory phenomenological framework,
theoretically the SM is almost universally understood to be the low-energy manifestation of a theory whose validity extends up to the Planck scale.
It is hoped that glimpses of such a theory could be found at the TeV scale,
which is what motivates the large variety of searches for physics beyond the Standard Model (BSM) at the LHC and at future colliders.

A common feature of BSM physics is the existence of new resonances,
whose discovery and characterisation can, for example, be achieved by studying the invariant mass distributions of their decay products.
A statistically meaningful quantification of a discovery, or indeed a null result, can be inferred from experimental data after the selection of a benchmark model.
In the absence of strong interference effects between the resonant signal and the relevant SM backgrounds,
it is straightforward to take a lineshape for the signal (which depends on both the mass and width of the resonance),
convolve it with the known detector resolution, and perform a fit to the observed data with both the signal and background invariant mass distributions.
In the case of positive results in the data, this allows for the extraction of the discovery significance;
for null results, one can extract exclusion limits on the product of the cross section for the resonance production and the branching ratio for decay into the final state of interest.
Needless to say, both of these results will depend crucially on the underlying theoretical assumptions;
in other words, the same data might lead to different results had different benchmark models been adopted.

Unfortunately, neglecting interference effects is a pragmatic compromise rather than a well-motivated assumption.
The details of signal--background interference are highly dependent on the new physics that gives rise to the resonance, which is unknown \emph{a priori}.
A scalar being produced via gluon-fusion through a fermion loop, for example, interferes with existing gluon-induced SM processes that produce the same final state as the scalar,
with the precise effects depending on both the mass of the scalar and the masses of any fermions that can run in the loop.
Different models will generate different patterns of interference,
the ultimate effect of which is to change the production rate of the resonance,
whilst distorting the invariant mass distribution of the decay products in such a way as to change the apparent mass of the scalar resonance~\cite{Djouadi:2016ack,Hespel:2016qaf,BuarqueFranzosi:2017jrj,Djouadi:2019cbm}.
The potential presence of these effects complicates both the interpretation of a new discovery in a resonance search,
and the presentation of null results in the form of cross-section times branching ratio limits, which are not well-defined in the case of interference.

In this work, we adopt an alternative point of view and present a practical approach for incorporating interference effects in resonance searches in a model-independent way.
The key idea is that although the actual lineshape (of the resonance invariant mass) depends on the unknown parameters of an unknown physics model,
the space of its possible functional forms is largely dictated by general Quantum Field Theory arguments.
Thus, we employ a physically-motivated functional form that is capable of describing the distortions of the lineshape encountered in the presence of signal--background interference,
and illustrate how LHC-experiment fitting procedures can be modified to use this functional form in the presentation of both positive and null results.
We demonstrate the technique using an assumed model of a scalar resonance produced via gluon fusion and decaying to pairs of photons,
but the approach easily generalises to other models and final states.
A model-independent approach to resonance searches
has also recently been presented in ref.~\cite{Xia:2019opu},
but we note that the method we propose is vastly different;
in particular, ref.~\cite{Xia:2019opu} relies heavily on a Fourier representation of non-periodic functions,
which need not be introduced in this paper.

This paper is structured as follows.
Preliminary considerations are first presented in section~\ref{sec:prem}.
We then introduce a general, model-independent functional form in section~\ref{sec:template}.
A benchmark signal model is presented in section~\ref{sec:bm}, which will serve as our assumed choice of a scenario that exists in Nature.
In section~\ref{sec:toytruth},
we demonstrate that the general functional form of section~\ref{sec:template} is able to describe the physics of the benchmark signal model of section~\ref{sec:bm},
and assume a generalisability of its description to other signal models due to the wide range of behaviour covered.
In section~\ref{sec:detector},
we make our tests more realistic by using fully-generated Monte Carlo (MC) samples of the signal, interference and background diphoton invariant mass distributions for the benchmark model,
along with simulated detector effects.
Finally, we conclude in section~\ref{sec:conclusions}.

\section{Preliminary considerations}
\label{sec:prem}

The most straightforward way to present both positive and null search results is that of working in the context of a given BSM theory;
an approach of this type is, by construction, a top-down one.
While statistically clean, top-down procedures have two main drawbacks.
Firstly, they often have to be repeated, even if the datasets are unchanged, whenever a different theoretical model is chosen.
Secondly, the details of how BSM theories are treated in such procedures are under the control of the experimental collaborations, which,
among other things, renders it difficult for theorists to assess how tweaking different aspects of the models might improve, worsen, or otherwise affect the search results.

For these reasons, it is interesting to consider the opposite viewpoint,
namely that of a bottom-up approach in which data are manipulated,
and the search results presented using the fewest possible number of theoretical assumptions.
This is the goal of the present paper.
More specifically, a model-independent functional form for describing the lineshape of a resonance and its interference with the background is employed,
and the search results presented as allowed or forbidden regions in the space of parameters relevant to such a form.
The idea is that if experimental results are given in this way,
any theoretical model can be quickly checked to be compatible or incompatible with the data by means of a simple computation whose results are expressed in terms of the same parameters.

In order to simplify the approach we are proposing, a number of assumptions need to be made.
In particular:
\begin{enumerate}
	\item We consider one resonance at a time;
	if several resonances are present, they must be sufficiently well separated for the procedure to work independently for each of them.
	
	\item We work with the invariant mass of the resonance,
	which can be reconstructed by means of the four-momenta of the decay products.
	
	\item A single partonic process is responsible for the signal--background interference pattern.
\end{enumerate}
There is at least one implication of item \#3 that requires an immediate explanation.
The overarching understanding is that we presently have a solid confidence in the SM,
as well as in the correctness of the theoretical tools that are used to simulate both SM and BSM physics processes with a good control on the systematics.
Thus, in the entirely realistic possibility that the background to the search proceeds through more than one partonic channel ($H\to\gamma\gamma$ being a chief example of this situation), the channels that are not interfering
must be subtracted from the data prior to the fitting procedure that we shall describe below.
This operation will contribute to the overall systematics of the procedure we are proposing.

\section{Model-independent template functional form\label{sec:template}}
Given the assumptions listed in section~\ref{sec:prem}, let us denote by $m$
and $\Gamma$ the mass and width, respectively, of the resonance whose
characteristics we seek to determine. We write the amplitude for the partonic
process that features the signal--background interference as follows:
\beq
\bA_h(\qt)=\frac{S_h(\qt)}{\qt-m^2+im\Gamma}+\frac{\hB_h(\qt)}{m^2}\,,
\label{amph}
\eeq
where we have denoted by $\qt$ the resonance virtuality\footnote{For the
sake of the present paper, $\qt$ (i.e.~a quantity defined at the level
of Feynman diagrams) is assumed to coincide with the squared invariant 
mass of the decay products of the resonance (i.e.~with an observable).},
and by $h$ (with \mbox{$1\le h\le N$}) the label of the helicity 
configurations. Loosely speaking, one can identify the first and the
second term on the r.h.s.~of eq.~\eqref{amph} with the ``signal'' 
and ``background'' contributions, respectively. Indeed, the amplitude
\beq
\bB_h(\qt)=\frac{\hB_h(\qt)}{m^2}
\label{hBdef}
\eeq
is by construction the one relevant to the production process of interest 
when all BSM effects are neglected; we remark that we find it convenient to
work with $\hB_h$, rather than directly with $\bB_h$, owing to the fact that 
its canonical dimensions are equal to those of $S_h$. We write the complex 
numbers $S_h$ and $\hB_h$ by making their dependences on complex phases 
explicit, as follows:
\beqn
S_h(\qt)&=&\abs{S_h(\qt)}\exp\left[i\,\xi_h(\qt)\right]\,,
\\
\hB_h(\qt)&=&\abs{\hB_h(\qt)}\exp\left[i\,\chi_h(\qt)\right]\,.
\eeqn
Thus, the square of the amplitude in eq.~(\ref{amph}) is:
\beqn
\abs{\bA_h(\qt)}^2&=&\frac{\abs{S_h(\qt)}^2}{(\qt-m^2)^2+m^2\Gamma^2}+
\frac{\abs{\hB_h(\qt)}^2}{m^4}
\label{amph0}
\\*&+&
\frac{2}{m^2}\,
\frac{\abs{S_h(\qt)}\abs{\hB_h(\qt)}}{(\qt-m^2)^2+m^2\Gamma^2}\,
\Big[(\qt-m^2)\cos\phi_h(\qt)+m\Gamma\sin\phi_h(\qt)\Big]\,,
\nonumber
\eeqn
with:
\beq \label{relative-phase}
\phi_h(\qt)=\xi_h(\qt)-\chi_h(\qt)\,.
\eeq
In order to make the forthcoming discussion as transparent as possible,
we assume that only one helicity configuration exists, i.e.~$N=1$; later,
we shall consider the case $N>1$. We simplify our notation accordingly, by
dropping the index $h$ wherever it appears. With this assumption, the
amplitude squared of eq.~(\ref{amph0}), when multiplied by the flux 
and phase-space factors, is the differential cross section for the signal 
plus background plus signal--background interference; henceforth, we shall
refer to this quantity as to the ``full'' cross section.
By computing its ratio over its analogue stemming from eq.~(\ref{hBdef}) (which is thus the background-only cross section),
flux and phase-space factors mutually cancel,
and we obtain what follows:
\beq
\frac{\abs{\bA(\qt)}^2}{\abs{\bB(\qt)}^2}=
\frac{m^4E(\qt)}{(\qt-m^2)^2+m^2\Gamma^2}
+\frac{m^2(\qt-m^2)\,O(\qt)}{(\qt-m^2)^2+m^2\Gamma^2}
+1\,,
\label{xsecrat0}
\eeq
where the ``even'' and ``odd'' dimensionless functions $E(\qt)$ and $O(\qt)$, respectively, are:
\beqn
E(\qt)&=&\hR(\qt)^2+2\,\frac{\Gamma}{m}\,\hR(\qt)\,\sin(\phi(\qt))\,,
\label{Efdef}
\\
O(\qt)&=&2\,\hR(\qt)\cos(\phi(\qt))\,,
\label{Ofdef}
\eeqn
with:
\beq
\hR(\qt)=\frac{\abs{S(\qt)}}{\abs{\hB(\qt)}}\,.
\label{hRdef}
\eeq
In the vicinity of the resonance mass, $\qt\simeq m^2$, by neglecting all
dynamical effects, i.e.~by replacing $E(\qt)$ with $E(m^2)$ and
$O(\qt)$ with $O(m^2)$, eq.~(\ref{xsecrat0}) exhibits the well-known
interference pattern of pure kinematical origin. Namely, the 
functional form in $\qt$ is a linear combination of a Breit-Wigner 
(BW henceforth), which is even under the \mbox{$(\qt-m^2)\to (m^2-\qt)$}
transformation, and of a BW times a \mbox{$(\qt-m^2)$} factor, which
is odd under the said transformation. This does {\em not} imply
that the functions $E(\qt)$ and $O(\qt)$ are even and odd, respectively.
However, we do expect that in a neighbourhood of $m^2$ the kinematical
effects be dominant over the dynamical ones. We can formalise this
statement by re-writing eq.~(\ref{xsecrat0}) by Taylor-expanding $E(\qt)$ 
and $O(\qt)$ around $\qt=m^2$:
\beq
\frac{\abs{\bA(\qt)}^2}{\abs{\bB(\qt)}^2}=
\frac{m^4}{(\qt-m^2)^2+m^2\Gamma^2}\,
\sum_{k=0}^\infty \frac{\ha_k}{k!}\,\left(\frac{\qt}{m^2}-1\right)^k
+1\,,
\label{xsecrat1}
\eeq
where\footnote{The quantity $O^{(-1)}(m^2)$ that appears in
eq.~(\ref{havsEO}) when $k=0$ need not be defined, since it
is multiplied by a null coefficient.}
\beq
\ha_k=E^{(k)}(m^2)+k\,O^{(k-1)}(m^2)\,,
\label{havsEO}
\eeq
having denoted:
\beq
E^{(k)}(\qt)=\frac{d^kE(\qt)}{d(\qt/m^2)^k}\,,\;\;\;\;\;\;\;\;
O^{(k)}(\qt)=\frac{d^kO(\qt)}{d(\qt/m^2)^k}\,.
\label{dEOdq}
\eeq
Note that for $k=0$, eq.~(\ref{dEOdq}) implies \mbox{$E^{(0)}(m^2)=E(m^2)$}
and \mbox{$O^{(0)}(m^2)=O(m^2)$}.

Equation~(\ref{xsecrat1}) gives us the first opportunity to discuss
the bottom-up approach introduced in section~\ref{sec:prem}. One first
truncates the Taylor expansion to some order $K$, i.e.~one writes:
\beq
\frac{\abs{\bA(\qt)}^2}{\abs{\bB(\qt)}^2}=
\frac{m^4}{(\qt-m^2)^2+m^2\Gamma^2}\,
\sum_{k=0}^K \frac{\ha_k}{k!}\,\left(\frac{\qt}{m^2}-1\right)^k
+1+{\cal O}\Big((\qt-m^2)^{K+1}\Big)\,.
\label{xsecrat1T}
\eeq
The terms of \mbox{${\cal O}((\qt-m^2)^{K+1})$} and higher are then discarded,
and the parameters that appear in eq.~(\ref{xsecrat1T}), namely the
following ones:
\beq
\Big\{m\,,\Gamma\,,\ha_0\,,\ldots\ha_K\Big\}\,,
\label{fitapar}
\eeq
have to be regarded as parameters to be determined by a fit to the data.
The results of such a fit will be compared with the theoretical predictions
for the same set of parameters (bar for $m$ and $\Gamma$, which must be 
considered as inputs to theoretical simulations).

We point out that, for any given choice of $K$, the values of the 
parameters of eq.~(\ref{fitapar}) emerging from fitting eq.~(\ref{xsecrat1T})
to the data will differ from those one would obtain if one had retained all
orders in the Taylor expansion as is done in eq.~(\ref{xsecrat1}), even in 
the ideal case of infinite statistics. This is because the fit based on eq.~\eqref{xsecrat1T} will
tend to compensate for the lack of the missing higher-order terms by
suitably adjusting the fit parameters, which can happen rather effectively
(i.e.~without changing significantly the quality of the fit) if the
fitting range in $\qt$ is chosen in an appropriate manner. It is obvious
that, by progressively enlarging such a range, the fit quality will
degrade, and eventually lead to an unstable procedure. We shall comment
at length on this point in the following, and show that the flexibility
in choosing the fitting range is an effective self-diagnostic tool.

The set in eq.~(\ref{fitapar}) constitutes a convenient choice since, 
for any given $K$, it allows one to include all of the information resulting
from the fit in a minimal number of parameters. On the other hand,
a possible drawback associated with it is the fact that
the parameters $\ha_k$ do not have one-to-one relationships with quantities
that emerge directly from matrix-element computations, such as the 
Taylor coefficients of $\hR(\qt)$ and $\phi(\qt)$. However, one can
express the former parameters in terms of the latter ones. By exploiting
eqs.~(\ref{Efdef}), (\ref{Ofdef}), and~(\ref{havsEO}) we obtain, for $K=2$ (which will be our default choice henceforth):
\beqn
\ha_0&=&\hRzt+2\,\frac{\Gamma}{m}\,\hRz\spz\,,
\label{ha0}
\\
\ha_1&=&2\left[\hRz\hRo+\hRz\cpz+
\frac{\Gamma}{m}\left(\hRz\spo+\hRo\spz\right)\right]\,,
\label{ha1}
\\
\ha_2&=&2\Bigg[\hRot+\hRz\hRt+2\hRz\cpo+2\hRo\cpz
\nonumber\\*&&\phantom{\Bigg[}
+\frac{\Gamma}{m}\left(\hRz\spt+2\hRo\spo+\hRt\spz\right)
\Bigg]\,,
\label{ha2}
\eeqn
where, analogously to eq.~(\ref{havsEO}), we have defined:
\beq
\hR^{(k)}=\left.\frac{d^k\hR(q^2)}{d(\qt/m^2)^k}\right|_{q^2=m^2}\,,\;\;\;\;\;\;\;\;
c_\phi^{(k)}=\left.\frac{d^k\cos\phi(q^2)}{d(\qt/m^2)^k}\right|_{q^2=m^2}\,,\;\;\;\;\;\;\;\;
s_\phi^{(k)}=\left.\frac{d^k\sin\phi(q^2)}{d(\qt/m^2)^k}\right|_{q^2=m^2}\,.
\label{dRcsdq}
\eeq
Thus, after having determined the values of the $\ha_k$ parameters,
one solves eqs.~(\ref{ha0})--(\ref{ha2}) for the Taylor coefficients of 
the $\hR(\qt)$ and $\phi(\qt)$ functions. There are two issues with
the procedure. Firstly, the system of eqs.~(\ref{ha0})--(\ref{ha2}) is 
underconstrained: there are more unknowns than equations, the more so the 
larger $K$. This implies that the solutions can not be given as central values 
plus uncertainties for each parameter, but rather as allowed hyperplanes
in the space of parameters. For example, the set of possible solutions
for $\hRz$ and $\hRo$ will sketch out a band in the
\mbox{$\langle\hRz,\hRo\rangle$} plane,
with finite width due to uncertainties and the effect of projecting out the remaining parameters.
Secondly, the system of eqs.~(\ref{ha0})--(\ref{ha2}) is in any case
not easy to solve, particularly owing to the presence of trigonometric
functions whose argument is $\phi(\qt)$. This problem can be alleviated
by solving directly for the sine and cosine of $\phi(\qt)$, which is
what the notation of eqs.~(\ref{ha0})--(\ref{ha2}) already implicitly
suggests. While this implies that the system of equations is even
more underconstrained, it is mostly an academic issue: in fact, 
we shall see that it is inevitable in the realistic case of multiple
helicity configurations.

An alternative, and much more practical, procedure is that of 
regarding the parameters on the r.h.s.~of eqs.~(\ref{ha0})--(\ref{ha2}) 
directly as fit parameters. This implies employing the r.h.s.~of
those equations in the fitting template of eq.~(\ref{xsecrat1T}),
thereby replacing eq.~(\ref{fitapar}) with:
\beq
\Big\{m\,,\Gamma\,,\hRz\,,\ldots\hR^{(K)}\,,
\cpz\,,\ldots c_\phi^{(K)}\,,
\spz\,,\ldots s_\phi^{(K)}\Big\}\,.
\label{fitRcspar}
\eeq
The comparison of eq.~(\ref{fitapar}) with eq.~(\ref{fitRcspar})
renders it manifest the first issue discussed above: there are more
parameters in the latter set than in the former one. Conversely,
since the parameters of eq.~(\ref{fitRcspar}) are more directly
related to physical quantities (or rather, to quantities that naturally
emerge in theoretical computations), it is possible to reduce their
number by means of physics considerations.
For example, we expect the complex phases to be more 
slowly-varying than the absolute values of the amplitudes,
and thus we may neglect the $\qt$ dependence of the former.
This implies trimming eq.~(\ref{fitRcspar}) down to:
\beq
\Big\{m\,,\Gamma\,,\hRz\,,\ldots\hR^{(K)}\,,\cpz\,,\spz\Big\}\,.
\label{fitRcspar2}
\eeq
Furthermore, rather than regarding eq.~(\ref{xsecrat1T}) as emerging
from the Taylor expansion of eq.~(\ref{xsecrat0}), one can start 
from Taylor-expanding the functions that appear on the r.h.s.~of
eqs.~(\ref{Efdef}) and~(\ref{Ofdef}), and then replacing the resulting
expressions into eq.~(\ref{xsecrat0}), discarding consistently
the terms of orders higher than those stemming from the original
expansions. As an explicit example, we consider again our default
case $K=2$, where eq.~(\ref{fitRcspar2}) reads as follows:
\beq
\Big\{m\,,\Gamma\,,\hRz\,,\hRo\,,\hRt\,,\cpz\,,\spz\Big\}\,.
\label{fitRcspar2K2}
\eeq
Conversely, by using a first-order Taylor expansion for $\hR$
(and by still neglecting the $\qt$ dependence of the complex phases)
the fit parameters are:
\beq
\Big\{m\,,\Gamma\,,\hRz\,,\hRo\,,\cpz\,,\spz\Big\}\,.
\label{fitRcspar3K2}
\eeq
The corresponding template functional form is the same as in
eq.~(\ref{xsecrat1T}), with the $\ha_0$, $\ha_1$, and $\ha_2$ coefficients
given in eqs.~(\ref{ha0})--(\ref{ha2}), where all the parameters
on the r.h.s.~of those equations which are not explicitly present in 
the sets of eqs.\eqref{fitRcspar2K2} and~(\ref{fitRcspar3K2}) must be set equal to zero.

We can repeat here the comment made after eq.~(\ref{fitapar}). Namely,
the simplifying assumptions that lead from eq.~(\ref{fitRcspar}) 
to eq.~(\ref{fitRcspar2K2}) and thence to eq.~(\ref{fitRcspar3K2})
imply that the values of the parameters that are common to these
three sets will in general be different in the three cases. However, at 
variance with the case of the $\ha_k$ parameters, such differences will not
necessarily be small even in fits of comparable good quality, since 
the system is underconstrained: thus, the individual parameter has 
more latitude to accommodate for the neglected terms than any of
the $\ha_k$ ones. The trigonometric parameters $\cpz$ and $\spz$
will give the clearest example of this behaviour.

Furthermore, at variance with the case of eq.~(\ref{fitapar}) which
is unambiguously determined once $K$ is chosen, the sets in
eqs.~(\ref{fitRcspar})--(\ref{fitRcspar3K2}) differ from each other
owing to considerations stemming from their underlying physics
meaning, which is more direct than for eq.~(\ref{fitapar}).
While this is an appealing characteristic, it must be kept in mind 
that the parameters in eqs.~(\ref{fitRcspar})--(\ref{fitRcspar3K2}) 
are still not measurable quantities. Thus, the considerations
mentioned above must be subject to a level of scrutiny that is
deeper than that relevant to the parameters of eq.~(\ref{fitapar});
we shall further this point in section~\ref{sec:toytruth}.

Before closing this section, we return to considering the case of
multiple helicity amplitudes, i.e.~we work with $N>1$ and start
from eq.~(\ref{amph0}). The amplitudes squared relevant to the full
and background-only cross sections are:
\beqn
\abs{\bA(\qt)}^2&=&\sum_{h=1}^N\abs{\bA_h(\qt)}^2\,,
\\
\abs{\bB(\qt)}^2&=&\sum_{h=1}^N\abs{\bB_h(\qt)}^2\,.
\eeqn
It is then a matter of simple algebra to show that 
eqs.~(\ref{xsecrat1})--(\ref{fitRcspar3K2}) are unchanged, provided
that the function $\hR(\qt)$ is defined as follows:
\beq
\hR(\qt)=\sqrt{\frac{\sum_{h=1}^N\abs{S_h(\qt)}^2}
{\sum_{h=1}^N\abs{\hB_h(\qt)}^2}}\,,
\eeq
and that the functions $\cos\phi(\qt)$ and $\sin\phi(\qt)$ are replaced
by $c_\phi(\qt)$ and $s_\phi(\qt)$, respectively, where:
\beqn
c_\phi(\qt)&=&\frac{\sum_{h=1}^N\abs{S_h(\qt)}\abs{\hB_h(\qt)}
\cos\phi_h(\qt)}{\sqrt{\sum_{h=1}^N\abs{S_h(\qt)}^2
\sum_{h=1}^N\abs{\hB_h(\qt)}^2}}\,,
\label{cphidef}
\\
s_\phi(\qt)&=&\frac{\sum_{h=1}^N\abs{S_h(\qt)}\abs{\hB_h(\qt)}
\sin\phi_h(\qt)}{\sqrt{\sum_{h=1}^N\abs{S_h(\qt)}^2
\sum_{h=1}^N\abs{\hB_h(\qt)}^2}}\,.
\label{sphidef}
\eeqn
While eqs.~(\ref{cphidef}) and~(\ref{sphidef}) imply that:
\beq
-1\le c_\phi(\qt)\,,s_\phi(\qt)\le 1\,,
\eeq
and therefore that both $c_\phi(\qt)$ and $s_\phi(\qt)$ can indeed
be seen as the cosine and the sine of an angle, in general this is
not the same angle. This fact, which has been anticipated before, 
is what forces one to treat the Taylor coefficients of $c_\phi(\qt)$ 
and $s_\phi(\qt)$ as independent fit parameters, as we have done in
eqs.~(\ref{fitRcspar})--(\ref{fitRcspar3K2}).

We conclude this section by summarising our fit setup. In what follows
we will show results based on two template fits: \textbf{T}$_R$ and \textbf{T}$_a$ corresponding to the fit parameters listed in  eq.~(\ref{fitRcspar2K2}) and eq.~(\ref{fitapar}) respectively,
the latter with $K=2$.
For completeness, the two sets of parameters are:
\beqn
\textbf{T}_R:&& \Big\{m\,,\Gamma\,,\hRz\,,\hRo\,,\hRt\,,\cpz\,,\spz\Big\}\,,\label{tempRcs}\\
\textbf{T}_a:&& \Big\{m\,,\Gamma\,,\ha_0\,,\ha_1\,,\ha_2\,\Big\}\,,\label{tempA}
\eeqn
which, along with eq.~\eqref{xsecrat1T}
(and substitutions similar to those of eqs.~\eqref{ha0}, \eqref{ha1}, and~\eqref{ha2} in the case of the \textbf{T}$_R$ set),
define the two functional forms that we will use.
We point out again that the results obtained by employing either of the \textbf{T}$_R$ or \textbf{T}$_a$ sets constitute alternative descriptions of the same underlying physics;
their different characteristics can be exploited depending on the emphasis of the specific new-physics search or modelling.

\section{Benchmark physics model}
\label{sec:bm}

\begin{figure}
	\centering
	\includegraphics[width=\textwidth]{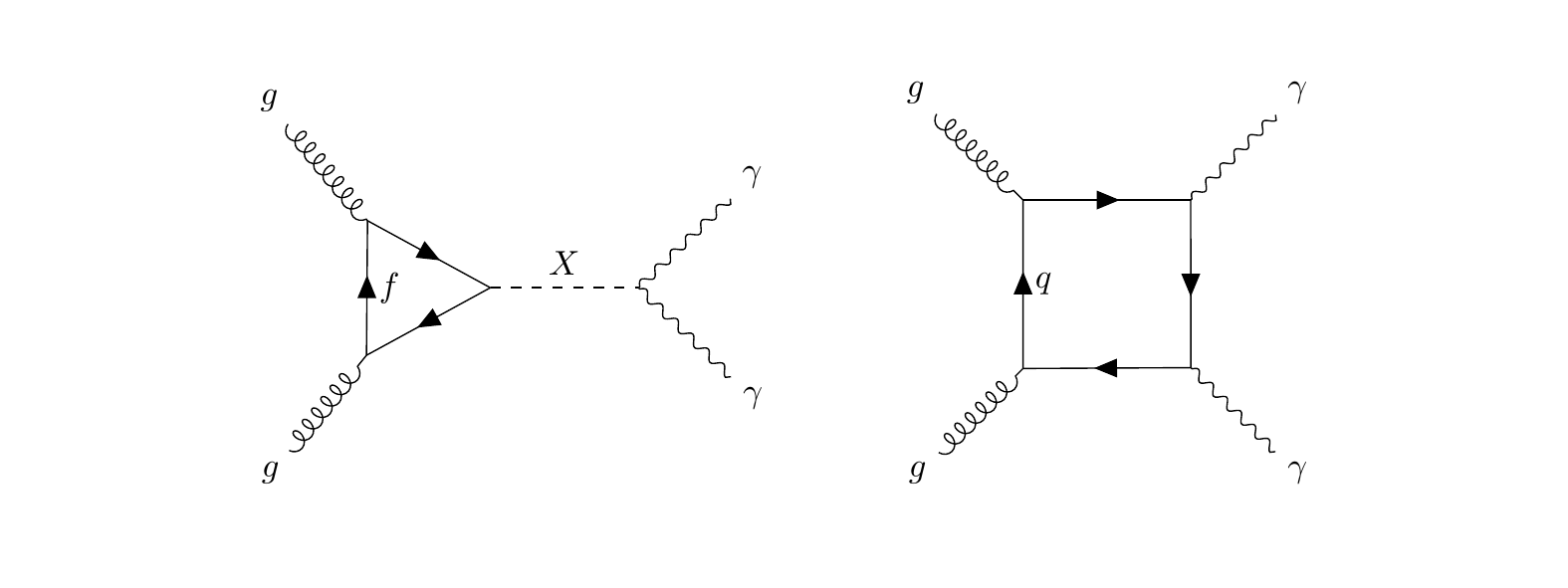}
	\caption{
		Left: the $gg\rightarrow X\rightarrow\gamma\gamma$ signal process.
		The resonance $X$ is a $CP$-even spin-0 particle.
		The $f$ denotes heavy virtual fermions.
		Right: the leading order $gg\rightarrow\gamma\gamma$ interfering SM background, with circulating quarks $q$.
	}
	\label{feyn-diag}
\end{figure}

The discussion of the previous section is general, model-independent, and can be used to interpret the results of any discovery (or null result),
even without making assumptions of an underlying resonant physics model.

In order to test that this model-independent functional form is indeed appropriate for a realistic physics analysis,
we now consider a benchmark model that contains a new Higgs-like $CP$-even scalar (spin-0) resonance produced via gluon fusion and decaying to two photons.
The leading-order (LO) Feynman diagram of the process that we consider is displayed on the left panel of fig.~\ref{feyn-diag}.
The LO diagram relevant to the SM process that interferes with this signal is shown on the right panel.
The two template forms of the previous section will be tested against the invariant mass distribution of the $\gamma\gamma$ pairs produced in these interactions,
which are assumed to represent the model chosen by Nature.

For testing purposes, it is useful to have semi-analytic descriptions of the signal process,
the dominant SM background, and the expected resonant--background interference to facilitate the generation of very high-resolution distributions of the diphoton invariant mass.
To distinguish these descriptions from the contents of the previous section, we will refer to them collectively as the \emph{physics model} (PM) functional form.
We now discuss the signal, the background, and their interference in turn.

\subsection{Signal model}
\label{sec:bm-subsec:signal}

In the assumed benchmark model, the interaction of the scalar resonance, $X$, with gluons is mediated by heavy fermion loops,
and can therefore be described by the effective interaction:
\begin{equation} \label{dim-5-op-prod}
	\lagr_0^G \propto G_{\mu\nu} G^{\mu\nu}X\,,
\end{equation}
where $G_{\mu\nu}$ is the gluon field strength tensor.
Its decay into photons is described by the dimension-5 operator:
\begin{equation} \label{dim-5-op-decay}
	\lagr_0^A \propto A_{\mu\nu} A^{\mu\nu}X\,,
\end{equation}
with $A_{\mu\nu}$ the electromagnetic field strength tensor.

These effective interactions can be used to compute the amplitude of the production and decay of the scalar resonance.
The differential cross section with respect to the diphoton invariant mass is given by:
\begin{equation} \label{dxs-gen-lumin-2}
	\frac{d\sigma_S}{dq} \propto \frac{\lagr_{gg}(q)}{q}\, |A_S(q^2)|^2\,,
\end{equation}
where $\lagr_{gg}(q)$ is the gluon-gluon luminosity function,
and $A_S(q^2)$ is the signal amplitude,
which can be written as:
\begin{equation} \label{signal-ampl}
	|A_S|^2 = f_\text{BW}|A_{ggX}A_{X\gamma\gamma}|^2\,,
\end{equation}
with $A_{ggX}$ and $A_{X\gamma\gamma}$ the amplitudes of the production loop and decay vertex respectively,
and $f_\text{BW}$ the BW function:
\begin{equation} \label{breit-wigner}
f_\text{BW}(q^2)=\frac{1}{(\qt-m_X^2)^2+m_X^2\Gamma_X^2}\,,
\end{equation}
where $m_X$ and $\Gamma_X$ are the mass and the width of the resonance, $X$, respectively.

Both the effective production and decay vertices contribute a factor to the amplitude with a simple $q$-dependence, $A_{X\gamma\gamma/XGG}\propto q^2$.
Thus, using eq.~\eqref{dxs-gen-lumin-2}, we posit that the diphoton invariant mass distribution of our chosen signal can be described by~\cite{yee}:
\begin{equation} \label{pm-signal}
	\frac{d\sigma_S}{dq} = f_s\, \lagr_{gg}(q)\, q^{7} f_\text{BW}(q^2)\,,
\end{equation}
where $f_s$ is a proportionality factor involving all other $q$-independent factors.
An approximation of the gluon luminosity lineshape was extracted using APFEL~\cite{Bertone:2013vaa} and the NNPDF2.3 set of parton luminosity functions (PDFs)~\cite{Ball:2013hta} to leading order:
\begin{equation} \label{glumin}
	\lagr_{gg}(q) = \left(1-\left( \frac{q}{E_\text{CM}}\right)^{1/3}\right)^{10.334} \left(\frac{q}{E_\text{CM}}\right)^{-2.8}\,,
\end{equation}
where $q$ is expressed in GeV,
and $E_\text{CM}=13$\,TeV is the centre-of-mass energy corresponding to LHC Run 2 specifications.

\begin{figure}
	\centering
	\includegraphics[width=0.5\textwidth]{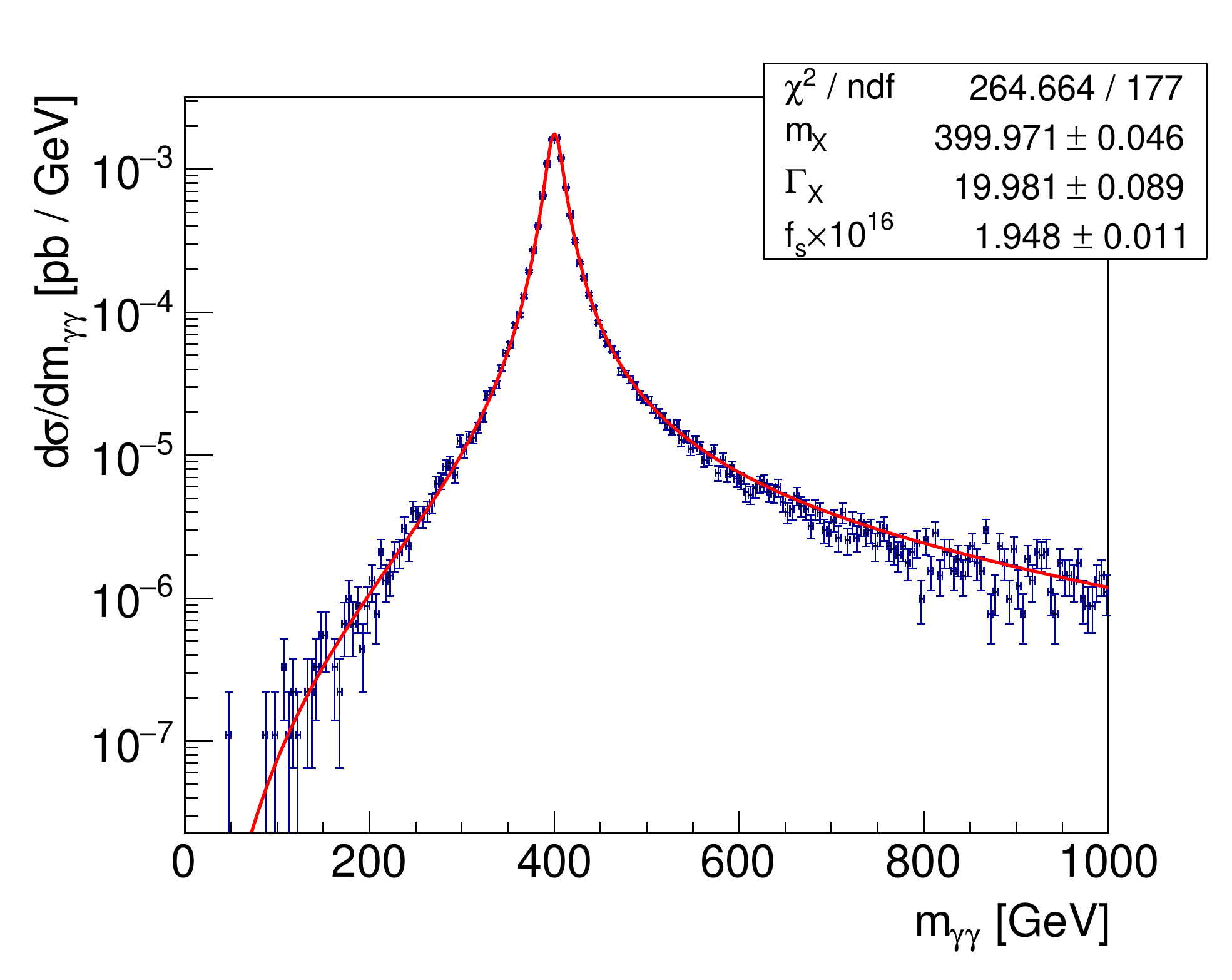}
	\caption{
		A fit of the signal differential cross section functional form to a histogram of the MC events
		generated at the $m_X=400$\,GeV, $\Gamma_X/m_X=5\%$ point.
	}
	\label{fit-pm-signal}
\end{figure}
A sample of MC signal-only events was generated using \madgraph{},
interfaced with \pythia{} for a parton shower simulation~\cite{Alwall:2014hca,Frixione:2002ik,Sjostrand:2014zea}.
The Higgs characterisation (HC) framework~\cite{Artoisenet:2013puc} was used for an implementation of our assumed resonant model.
Events were generated for a resonant mass $m_X=400$\,GeV, with a width $\Gamma_X/m_X=5\%$.
Using the ROOT data analysis framework~\cite{Antcheva:2009zz},
the analytic form of eqs.~\eqref{pm-signal} and~\eqref{glumin} was tested against a binned histogram of the diphoton invariant mass distribution, for events at the generator level (i.e. those obtained without performing any detector simulation).
The result is presented in fig.~\ref{fit-pm-signal}.
The values extracted for $m_X$ and $\Gamma_X$ agree well with the inputs selected,
and a good description of the data is found, with $\chi^2/\text{ndf} \approx 1.5$.
Thus, it is clear that the analytic form of eqs.~\eqref{pm-signal} and~\eqref{glumin} indeed provide a description equivalent to that of the HC model,
but which will be quicker to run, and is unaffected by statistical fluctuations.

\subsection{Background parametrisation}
\label{sec:bm-subsec:background}

The interfering SM $gg\rightarrow \gamma\gamma$ diagram at the LO is shown in fig.~\ref{feyn-diag} (right).
While it is possible to proceed in a similar manner as in the signal case in order to obtain a description of the background contribution to the diphoton invariant mass distribution,
the physics of the background is not the main interest of our study.
We adopt the methodology of experimental collaborations and employ a template functional form for an \emph{ad hoc} description of the background differential cross section~\cite{Aaboud:2016tru}:
\begin{equation} \label{bg-func}
	\frac{d\sigma_B}{dq} \equiv F_B(q) = \frac{f_b}{q}\left( 1-\left(\frac{q}{E_\text{CM}}\right)^{1/3}\right)^A \left(\frac{q}{E_\text{CM}} \right)^B\,,
\end{equation}
where $E_\text{CM}=13$\,TeV. The parameters of this template consist of a normalisation constant, $f_b$, and exponents, $A$ and $B$.

\begin{figure}
	\centering
	\includegraphics[width=0.5\textwidth]{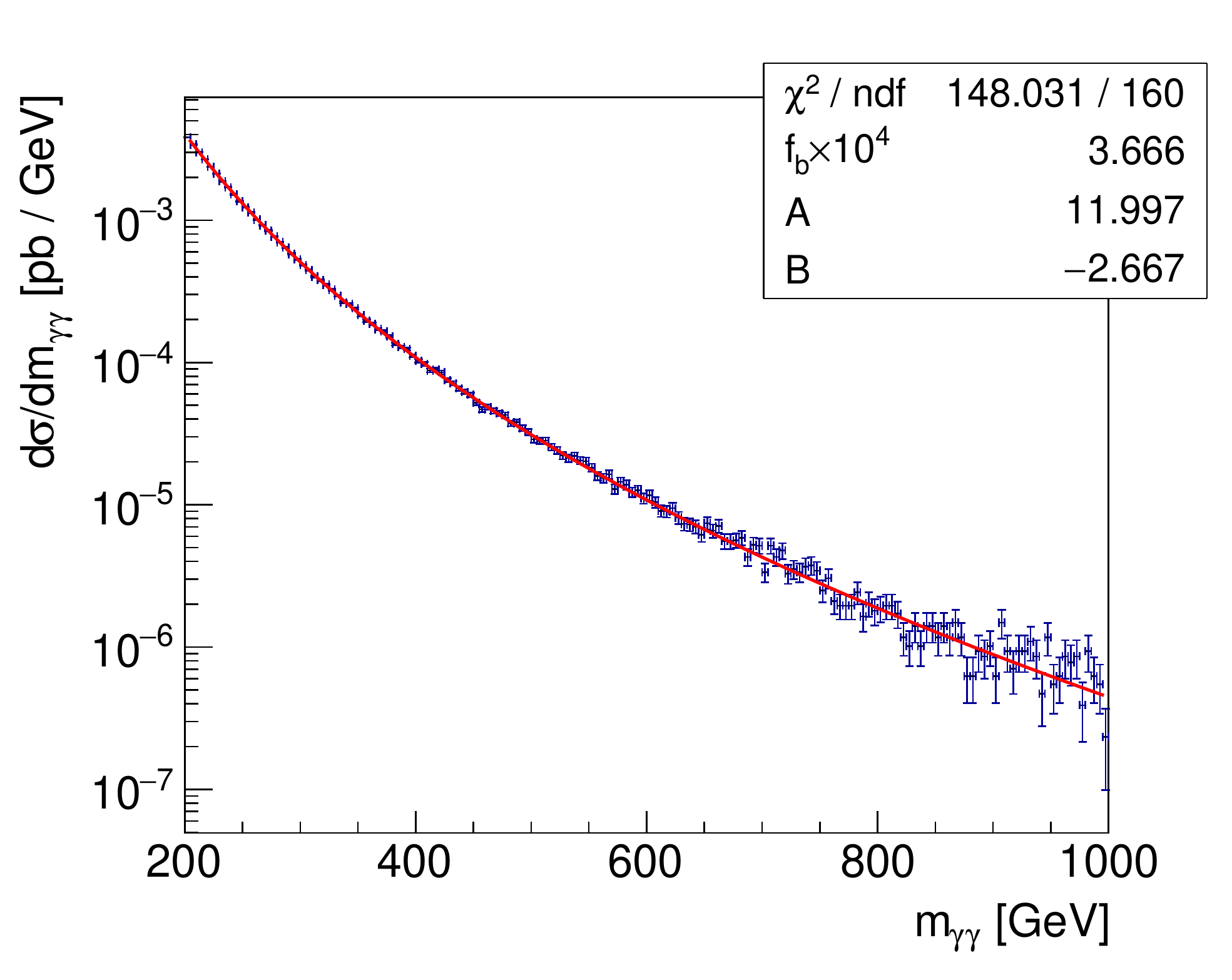}
	\caption{
		A fit of the background template functional form to the histogram of generated background $gg\to\gamma\gamma$ events.
	}
	\label{fit-background}
\end{figure}

The description of eq.~\eqref{bg-func} was tested against a sample of background $gg\rightarrow\gamma\gamma$ events generated using \madgraph{}.
The result of its fit to the events is presented in fig.~\ref{fit-background};
a good description of the simulated background diphoton invariant mass distribution is found, with $\chi^2/\text{ndf}\approx 0.9$.

\subsection{Interference between signal and background}
\label{sec:bm-subsec:interf}

Using eqs.~\eqref{pm-signal} and~\eqref{bg-func},
we can write the interference contribution to the full differential cross section as follows:
\begin{equation} \label{dxs-interf}
	\frac{d\sigma_I}{dq} = 2\sqrt{f_s}\, \sqrt{\frac{\lagr_{gg}(q)}{q}}\, \sqrt{F_B(q)}\, f_\text{BW}\, q^4 \left[ (q^2-m_X^2)c_{\phi_X} +m_X\Gamma_X s_{\phi_X} \right],
\end{equation}
where $c_{\phi_X}$ and $s_{\phi_X}$ are analogous to the quantities of
eqs.~(\ref{cphidef}) and~(\ref{sphidef}), but specifically defined under the
assumed benchmark model.

For a heavy resonance that decays via an effective contact interaction,
and in the limit of infinite fermion masses for the loop-induced resonant production and background interaction,
the phase difference $\phi_h(q^2)$ (eq.~\eqref{relative-phase}) vanishes for all interfering helicity amplitudes.
Thus, to generate interference-only event samples corresponding to non-trivial phase differences,
we have modified the signal amplitude appearing within the HC framework by means of the replacement
$S \to S \times e^{i\theta}$, such that the value chosen for the artificial phase $\theta$ then corresponds to the phase difference between the signal and background helicity amplitudes, $\theta\equiv\phi_h$.

To verify that we can extract the expected phase from interference-only event samples generated in this way,
we note that since the same phase difference is defined for each of the interfering helicity configurations,
eq.~\eqref{dxs-interf} can be written in the following form: 
\begin{equation} \label{dxs-interf-2}
	\frac{d\sigma_I}{dq} = 2 f_i \, \sqrt{\frac{\lagr_{gg}(q)}{q}}\, \sqrt{F_B(q)}\, f_\text{BW}\, q^4 \left[ (q^2-m_X^2)\cos\theta+m_X\Gamma_X \sin\theta \right],
\end{equation} 
where $f_i$ is defined as:
\begin{equation} \label{pm-interf-norm}
	f_i \equiv \sqrt{f_s}\,\times \frac{\sum_{h=1}^N |S_h||B_h|} {\sqrt{\sum_{h=1}^N |S_h|^2 \sum_{h=1}^N |B_h|^2}}\,,
\end{equation}
with $S_h$ and $B_h$ the helicity amplitudes of the signal and background respectively, as defined in eq.~\eqref{amph}.
The four free parameters of eq.~\eqref{dxs-interf-2} are $m_X$, $\Gamma_X$, $\theta$, and $f_i$,
where the invariant mass dependence of $f_i$ is neglected.

\begin{figure}
	\centering
	\includegraphics[width=0.5\textwidth]{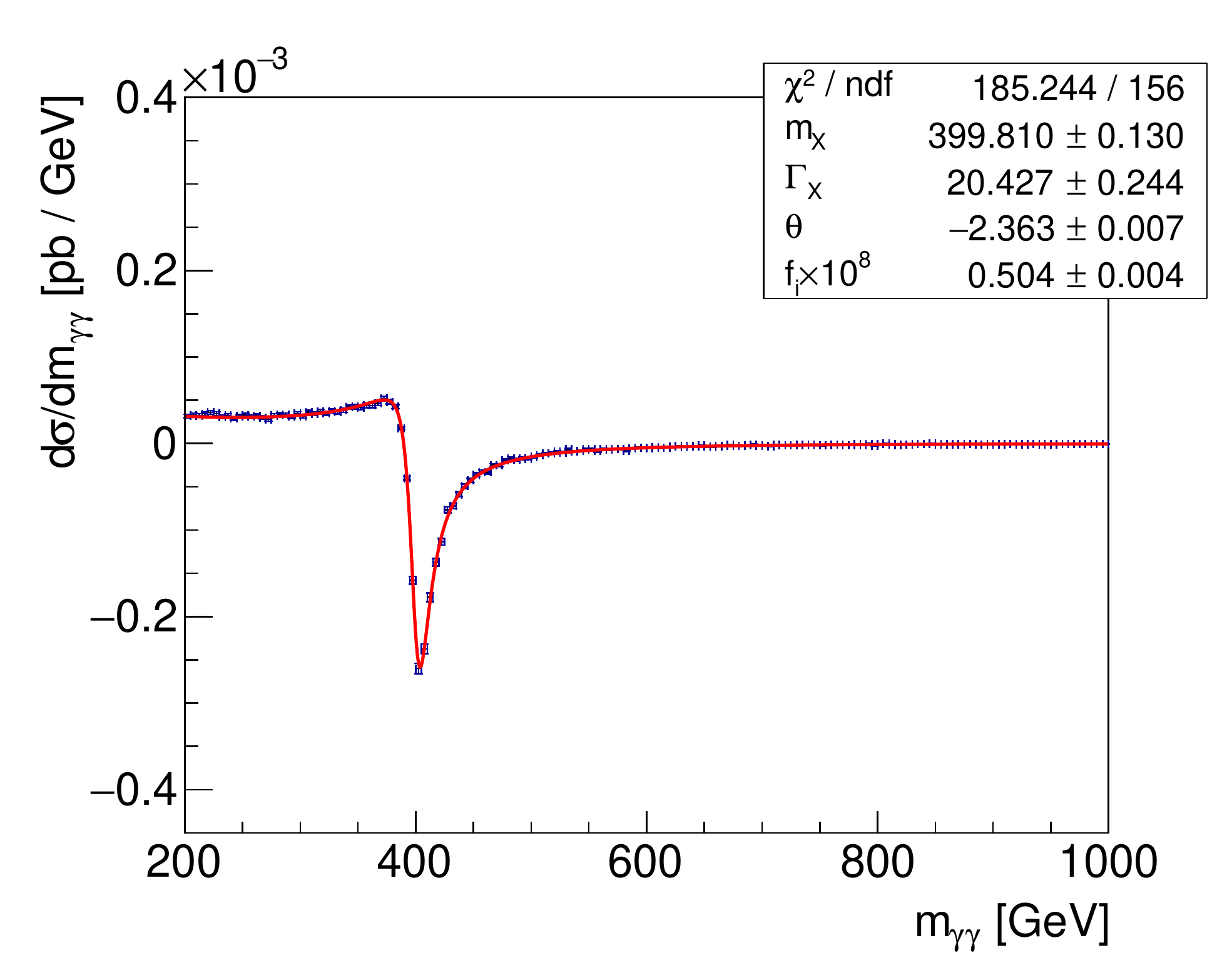}
	\caption{
		A fit of the analytic description of the interference between the benchmark signal model and the SM background to a MC sample of interference-only events.
	}
	\label{fit-interference}
\end{figure}

Eq.~\eqref{dxs-interf-2} was tested against an interference MC sample generated with $m_X=400$\,GeV, $\Gamma_X=20$\,GeV,
and a complex phase $\theta=-3\pi/4\approx-2.36$.
The result of a fit to the event sample is shown in fig.~\ref{fit-interference}.
The values extracted for $m_X$, $\Gamma_X$, and $\theta$ are in good agreement with those used to generate the events,
and a good description of the interference differential cross section lineshape is found, with $\chi^2/\text{ndf}\approx 1.2$.

\section{Template fits to physics model toys}
\label{sec:toytruth}

In this section, we shall test the model-independent functional form of
eq.~(\ref{xsecrat1T}) against the chosen benchmark physics model.
These tests will be performed using toy samples obtained from the analytic description of the benchmark model,
rather than from a Monte Carlo generator,
in order to achieve very fine resolutions of the invariant mass distributions.

We consider the template functional form as expressed in terms of the \textbf{T}$_R$ or \textbf{T}$_a$ parameter sets,
of eqs.~\eqref{tempRcs} and~\eqref{tempA} respectively,
and compare and contrast the results that we obtain for the two different template fits.
By design, the parameters of the functional form should be extracted
through a fit of the ratio of full to (interfering) background-only differential cross
sections, so as to reduce the bias of a result on the particular partonic
luminosity assumed for the data.
However, in the current study, we shall
assume that the background is known exactly (i.e. without any uncertainties associated with its description).
In this case, it is equivalent to perform the fits in terms
of absolute differential cross sections,
using a description that follows from eq.~\eqref{dxs-gen-lumin-2}:
\begin{equation} \label{rc-dxs}
    \frac{d\sigma_\text{full}}{dq} = F_B(q)\,\,
    \frac{\abs{\bA(\qt)}^2}{\abs{\bB(\qt)}^2}\,,
\end{equation}
where $F_B(q)$ is the background-only differential cross section (eq.~\eqref{bg-func}),
and the ratio of amplitudes squared is given by eq.~(\ref{xsecrat1T}).

Binned likelihood fits are performed using the \textsc{MultiNest} implementation of the nested sampling algorithm~\cite{Feroz:2008xx}.
We employ a log-likelihood function defined assuming independent Gaussian random variables for each bin:
\begin{equation} \label{log-like}
	\log\mathcal{L}(\bsym{\Theta}) = \log\prod_i^\text{bins} G_i(\boldsymbol{\Theta};q_i) = -\frac{1}{2}\, \sum_i^\text{bins} \frac{\left(y_i - Y(q_i;\boldsymbol{\Theta})\right)^2}{\sigma_i^2}\,,
\end{equation}
where $y_i$ and $\sigma_i$ respectively represent the content and uncertainty of the $i^\text{th}$ bin,
and $Y(q_i;\boldsymbol{\Theta})$ denotes the fit function evaluated at the central bin value $q_i$
(in the case of template fits, $Y$ is given by eq.~\eqref{rc-dxs},
with $\bsym{\Theta}$ corresponding to either the \textbf{T}$_a$ or \textbf{T}$_R$ parameter sets).

\subsection{Construction of Asimov toys}
\label{sec:toytruth-subsec:pmtoys}

The analytic description of the PM invariant mass distribution,
given by the sum of eqs.~\eqref{pm-signal}, \eqref{bg-func} and~\eqref{dxs-interf},
is used to construct the Asimov toy histograms that will be used in our tests.
Asimov datasets contain no statistical fluctuations~\cite{Cowan:2010js},
and are constructed by setting the content of the $k^\text{th}$ bin equal to the value of the input distribution,
evaluated at the central invariant mass value, $q_k$, of that bin.
Bin uncertainties are given by:
\begin{equation} \label{asimov-bin-uncertainty}
    \Delta_k = \sqrt{N_\text{total}\,\frac{\mathcal{A}(q_k)}{\sum_i^\text{bins} \mathcal{A}(q_i)}}\, \frac{\sum_i^\text{bins} \mathcal{A}(q_i)}{N_\text{total}}\, ,
\end{equation}
where $\mathcal{A}(q)$ denotes the generating distribution of the Asimov (in this case, the analytic PM description),
and $N_\text{total}$ is the total number of events assumed for the dataset.
The square-rooted contribution arises from regular counting statistics,
while the rightmost factor accounts for the fact that bin contents do not correspond to a number of events.
Thus, the toys perfectly represent the PM invariant mass distributions used to generate them, and provide a clean test-bed for the general template.

\begin{figure}
	\centering
	\includegraphics[width=0.49\textwidth]{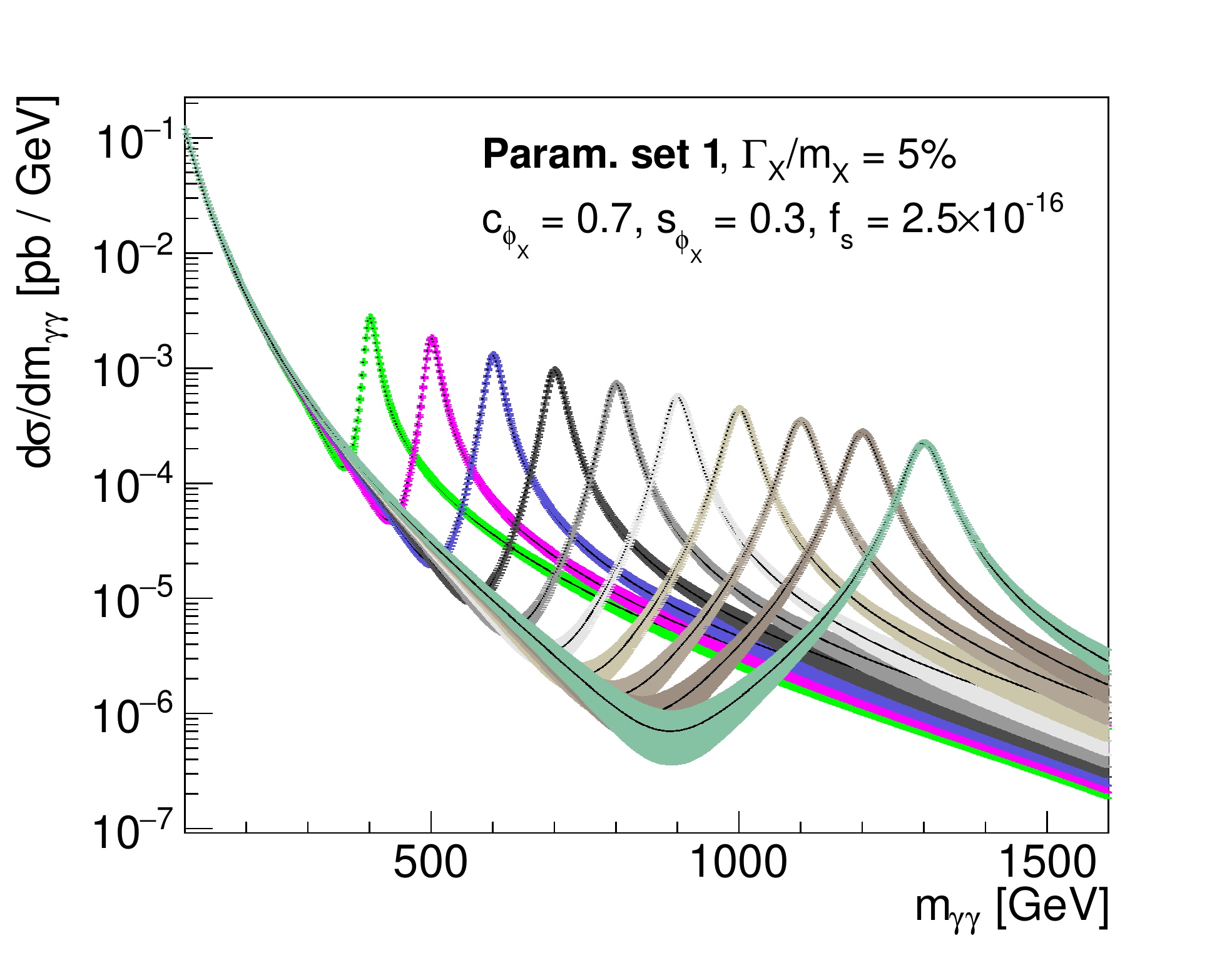}
	\includegraphics[width=0.49\textwidth]{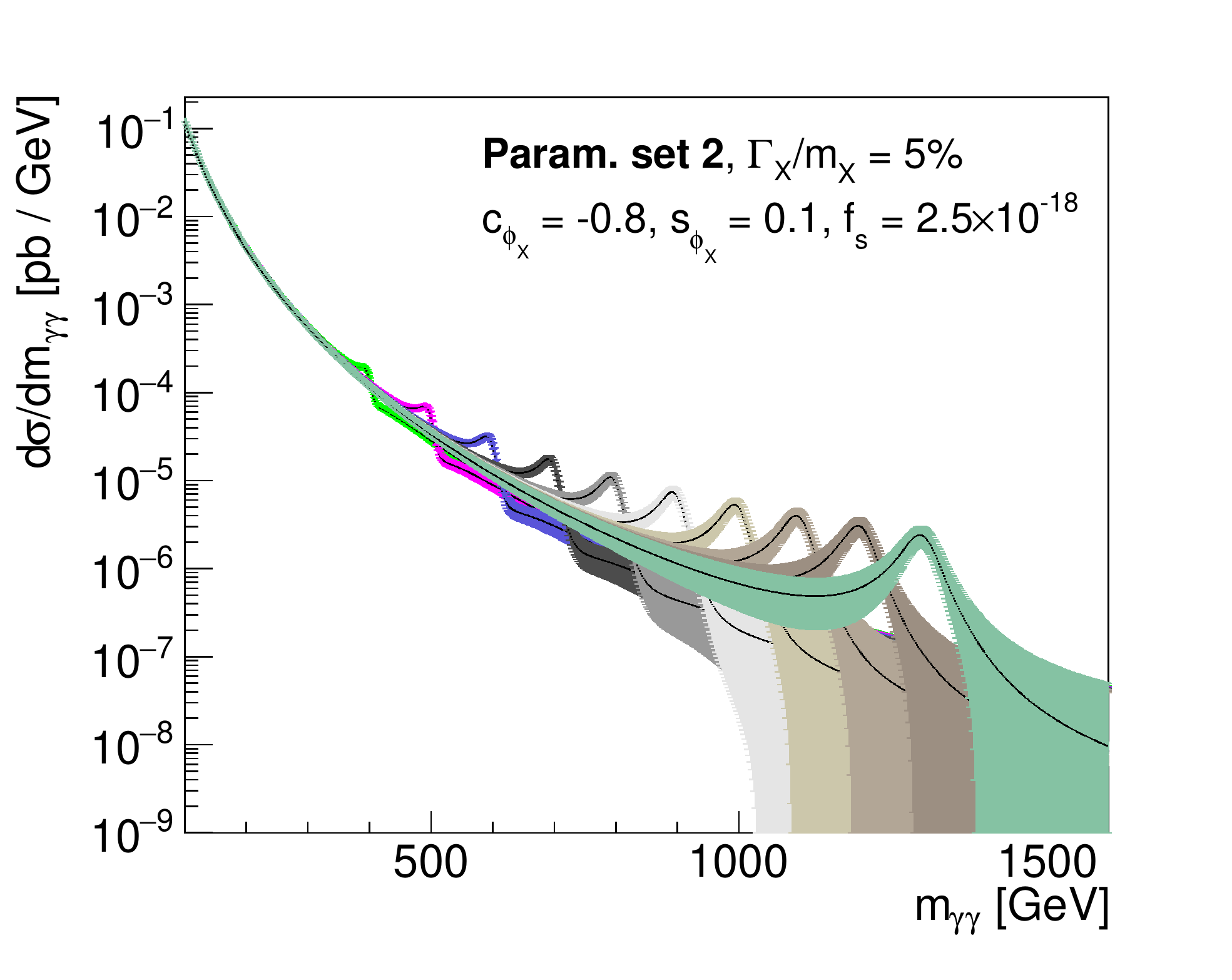}
	\caption{
		Visualisation of Asimov datasets, generated using the PM functional form at various mass points with $\Gamma_X/m_X=5\%$,
		for the two different sets of input $c_{\phi_X}$, $s_{\phi_X}$, and $f_s$.
	}
	\label{pm-input-dists}
\end{figure}

Ten input PM mass points are considered, with values ranging from $m_X=400$\,GeV to $m_X=1300$\,GeV in increments of $100$\,GeV.
Input widths are set equal to $\Gamma_X/m_X=5\%$,
and two different sets of values for the remaining PM parameters are chosen as follows:
\begin{equation} \label{pm-input-sets}
	\{c_{\phi_X},s_{\phi_X},f_s\}=
	\begin{cases}
		\{0.7,\, 0.3,\, 2.5{\times}10^{-16}\} \quad &\text{``set 1''}\,,\\
		\{-0.8,\, 0.1,\, 2.5{\times}10^{-18}\} &\text{``set 2''}\,,
	\end{cases}
\end{equation}
with a notable difference being the ``height'' of the signal, $f_s$, which is two orders of magnitude smaller in set~2.
The collection of Asimov histograms constructed for these inputs is shown in fig.~\ref{pm-input-dists},
for 2\,GeV bin widths and $N_\text{total}=10$ million events across the 100--1600\,GeV invariant mass range.
We note that this number of diphoton events is approximately of the same order as that collected in Run 2 of the LHC.

Finally, we point out that, as can be inferred from fig.~\ref{pm-input-dists},
the contribution of the signal to the full cross section in the high-mass
tail is larger than that of the background in the adopted PM.
While this might not be the case in actual physics scenarios,
we stress that it does not have any implications on the procedure we present in this paper,
owing to the limited fit window that is central to the latter;
we shall comment explicitly on this matter in section~\ref{sec:toytruth-subsec:fitwindows}.

\subsection{Test of fit windows}
\label{sec:toytruth-subsec:fitwindows}

The general template we consider retains only terms up to $\mathcal{O}((q^2-m^2)^2)$ in its form.
Thus, it is expected that its description of an invariant mass distribution will deteriorate if one considers masses too far away from the resonance peak.
In this section, we fit a range of invariant mass windows centred on the true mass for each toy.
Window widths from $2w=40$\,GeV to $2w=400$\,GeV are chosen,
in increments of $40$\,GeV, such that fit windows are equal to $m_X\pm w$.
From such a collection of fit results,
one can (approximately) determine an appropriate range of fit windows by noting that
$-2\log\mathcal{L}(\boldsymbol{\Theta})$ follows a $\chi^2_\nu$ distribution,
where for a given fit $\nu$ is the number of fitted bins minus one for each free parameter.
A 1$\sigma$ cut-off in the fit quality, for example,
can then be determined by evaluating the quantile function of the $\chi^2_\nu$ distribution, $Q_{\chi^2_\nu}(p)$, at $p\approx 0.68$.
If a fit returns a best-fit $\chi^2$ smaller than this value,
then we can conclude that the general functional form is able to provide a good description of the data,
to within $1\sigma$, over the corresponding mass range.

\begin{figure}
	\centering
	\includegraphics[width=0.49\textwidth]{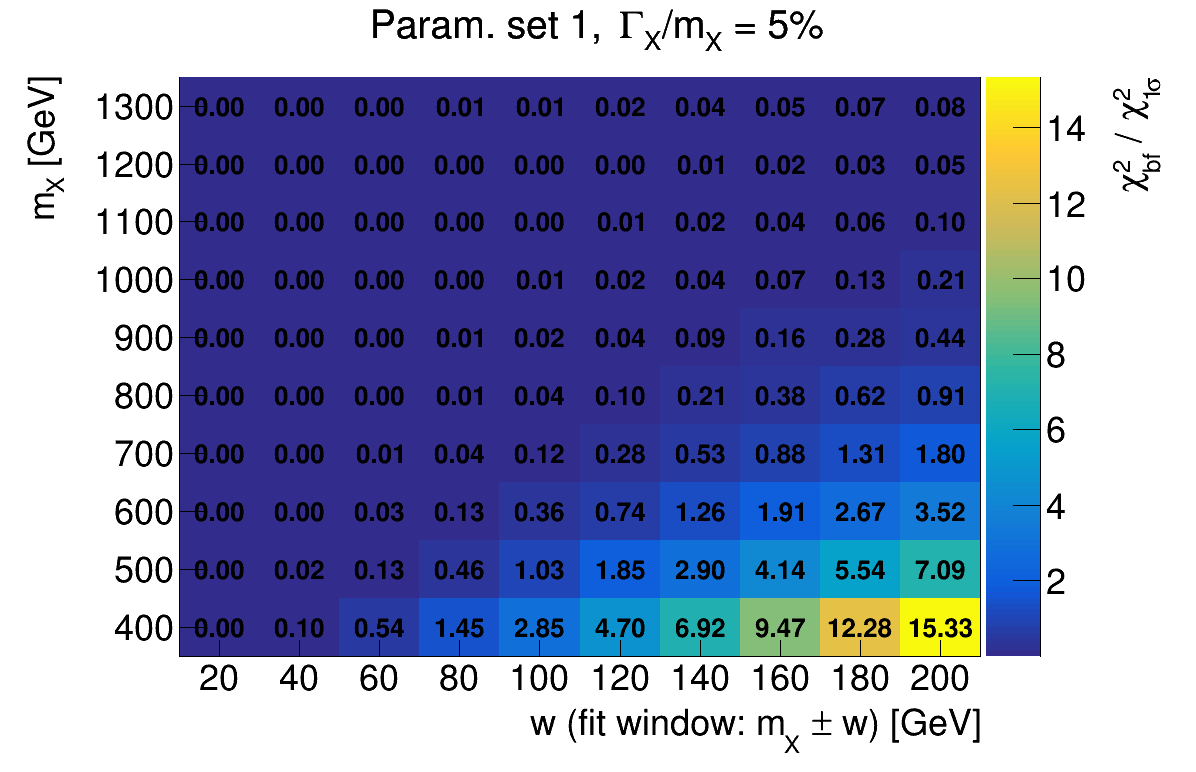}
	\includegraphics[width=0.49\textwidth]{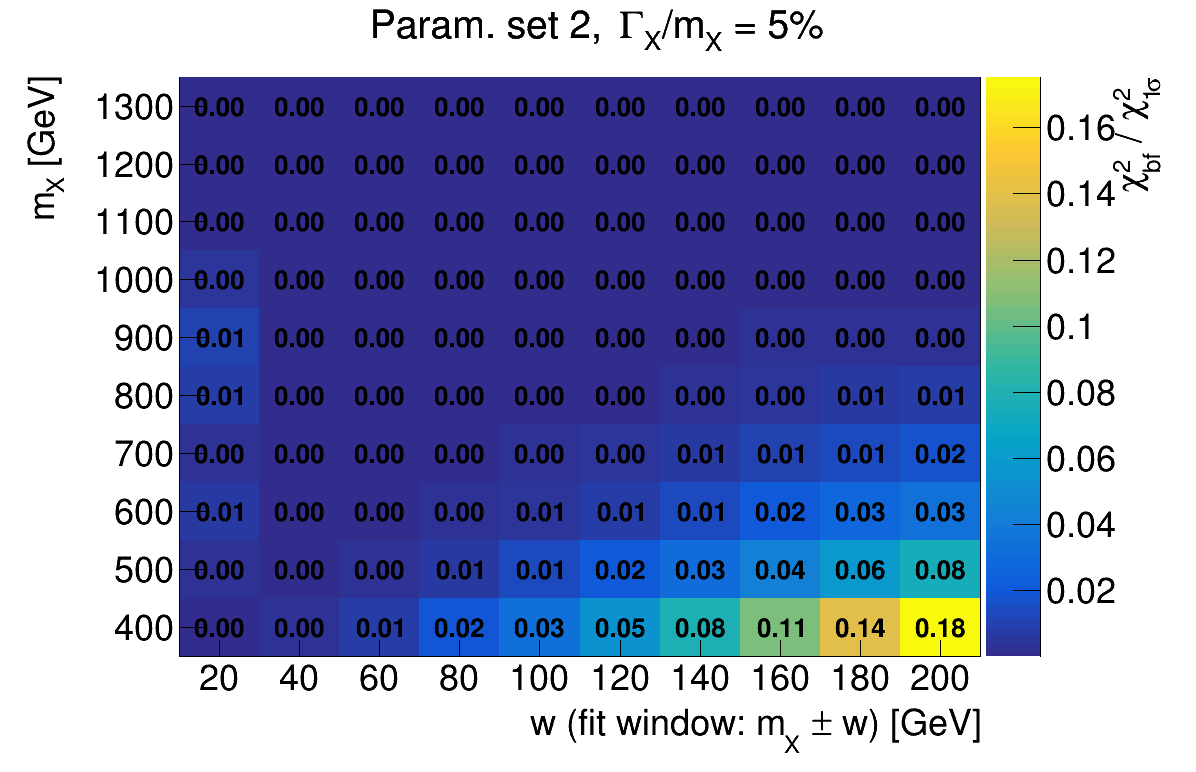}
	\caption{
		Plots depicting the quality of \textbf{T}$_a$ template fits relative to the quantile function of a $\chi^2_\nu$ distribution.
		The $z$-axis is represented by means of different colours, with a scale displayed on the right of each plot,
		and shows the ratio of best-fit $\chi^2$ to the $1\sigma$ cut-off.
		Lower values indicate better fits.
		Left (right) panel: the result using set 1 (2) of the input PM parameters.
	}
	\label{auto-5pc}
\end{figure}

The result of fitting the \textbf{T}$_a$ parameters to PM toys for all of the masses and windows chosen is presented in fig.~\ref{auto-5pc}.
The left and right panels are obtained using the set~1 and set~2 input PM parameter values of eq.~\eqref{pm-input-sets}, respectively. The half-width of the fit window, $w$, and the input PM mass, $m_X$, are reported on the $x$ and $y$ axes.
For each combination of these quantities,
the $z$-axis is represented as colour-coded values according to the scale depicted on the right of the two panels,
corresponding to the fit quality in terms of the ratio of best-fit to 1$\sigma$ cut-off chi-squares, $\chi^2_\text{bf}/\chi^2_{1\sigma}$,
with the latter obtained from a $\chi^2$ distribution of appropriate dimensions.
Thus, larger $z$-axis values indicate poorer fits, with a value of one marking the 1$\sigma$ boundary.
Note that the same colour code is used in the two results,
but corresponds to different $z$-axis scales.
We do not show the analogous results in terms of the \textbf{T}$_R$ parameters,
as both parametrisations yield identical lineshapes.

Considering each input mass point separately, we find the trend of decreasing fit quality with increasing fit window size, as is expected.
The results of set~1 show a much faster deterioration in contrast to those of set~2, with the latter indicating a good fit for every window tested.
Nevertheless, provided that a sufficiently small fit window is chosen,
these results show that the general parametrisation is capable of correctly characterising a wide range of physical lineshapes.

\subsection{Profile likelihood contours in template parameter space}
\label{sec:toytruth-subsec:proflike}

Once a suitable choice for the fit window has been made,
aided by results akin to those of fig.~\ref{auto-5pc} or otherwise,
the next step of an analysis is to extract a result in terms of the general parameters.
As an example, let us refer to the particular instance of fig.~\ref{auto-5pc} corresponding to the set~1 input PM parameters,
$m_X=700$\,GeV, $\Gamma_X/m_X=5\%$, and the $m_X\pm40$\,GeV fit window.
A ratio of chi-squares that is close to zero is found for this configuration,
indicating a very good fit of the toy data.

\begin{figure}
	\centering
	\includegraphics[width=0.49\textwidth]{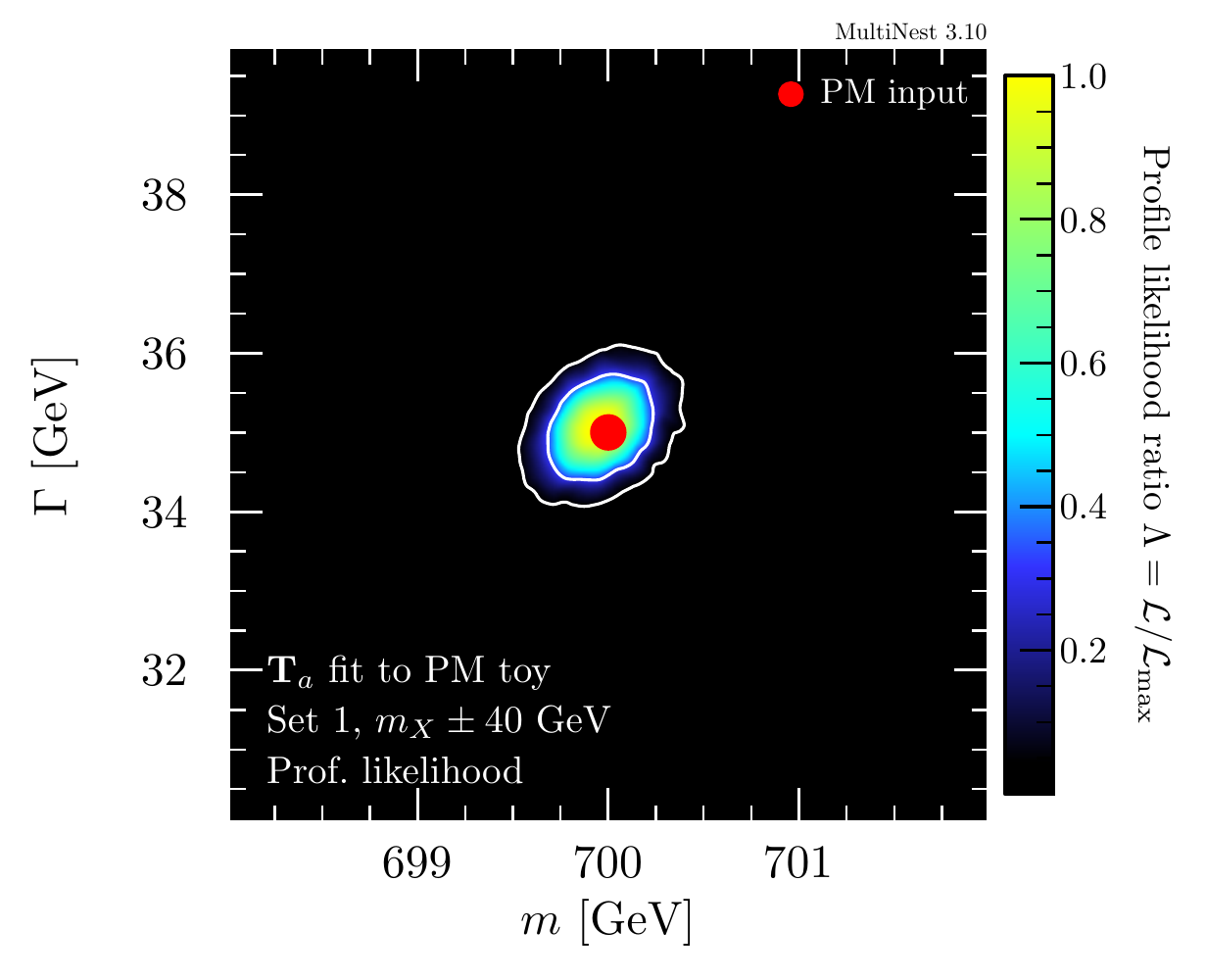}
	\includegraphics[width=0.49\textwidth]{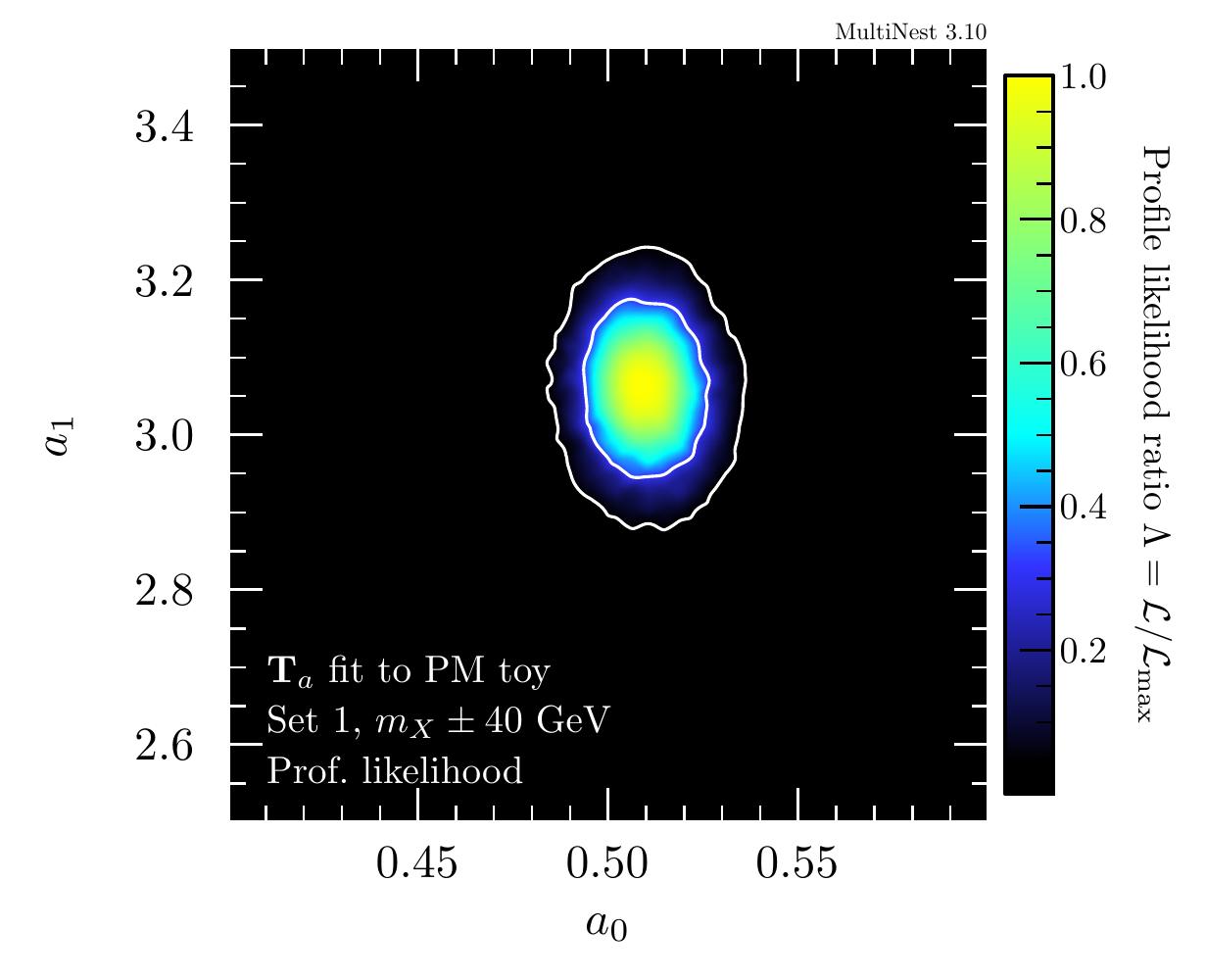}\\
	\includegraphics[width=0.49\textwidth]{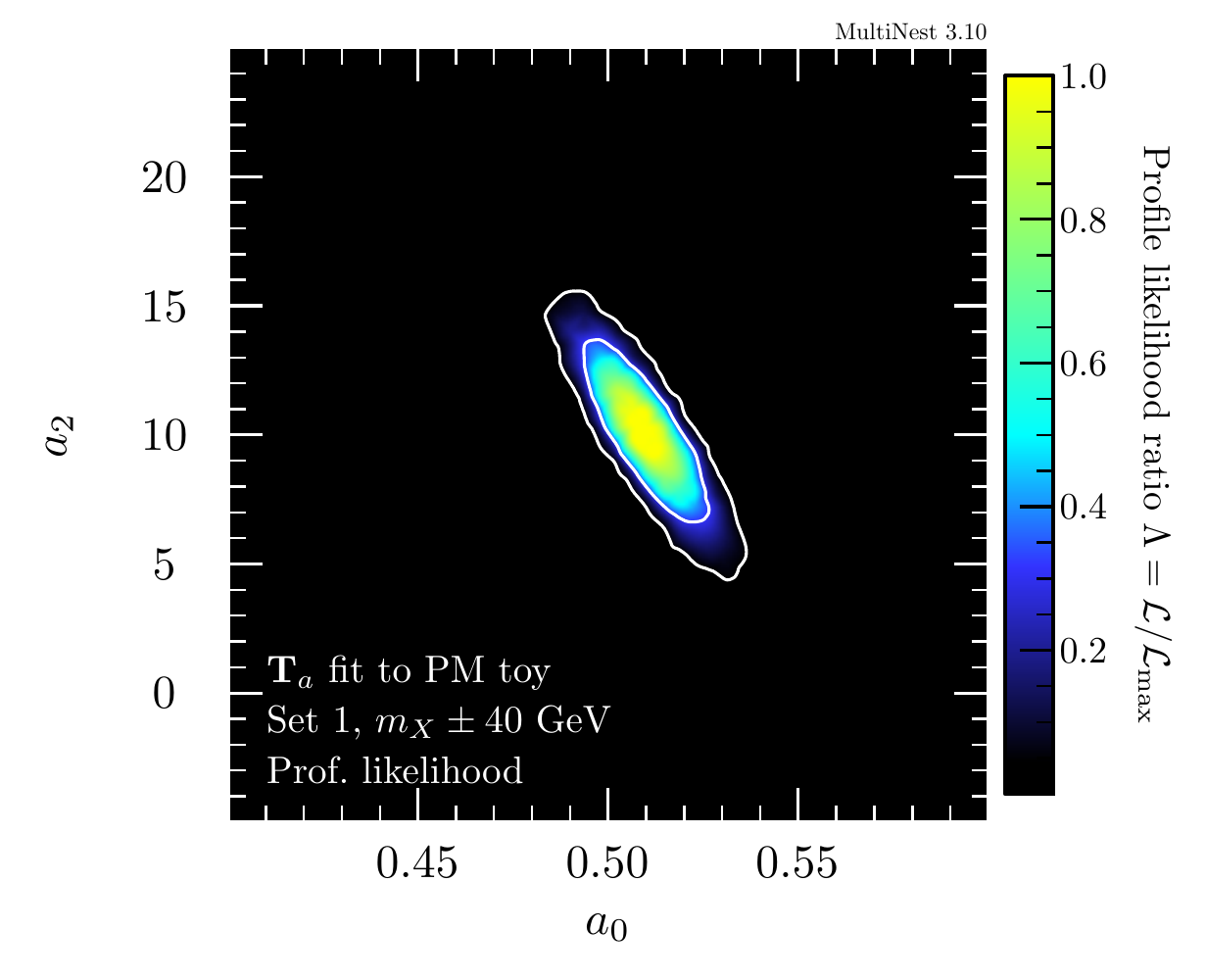}
	\includegraphics[width=0.49\textwidth]{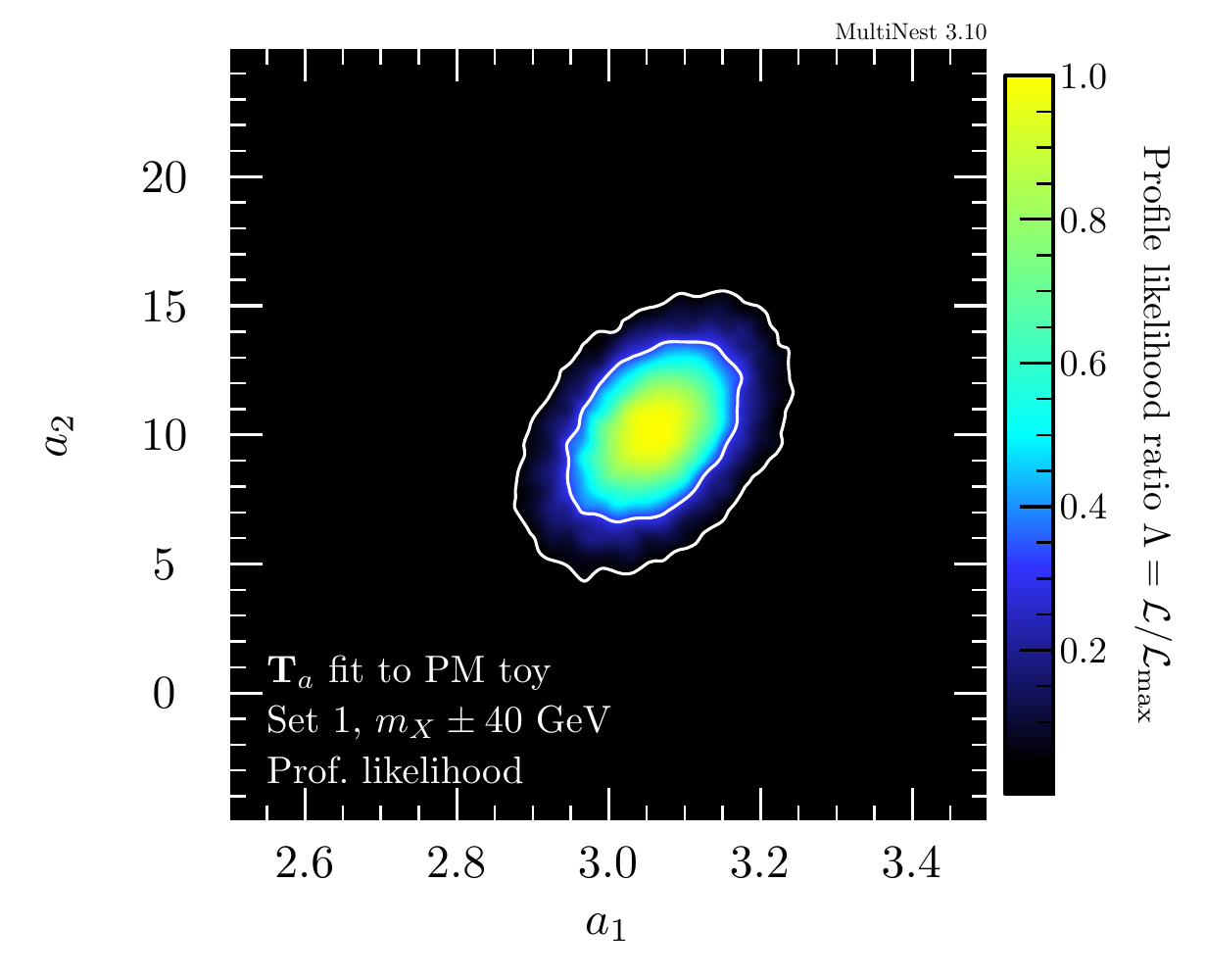}
	\caption{
		Fit result for the \textbf{T}$_a$ parameters to a PM Asimov toy over the $m_X\pm40$\,GeV window.
		The toy corresponds to the $m_X=700$\,GeV, $\Gamma_X/m_X=5\%$, and set~1 input parameters.
		Two-dimensional profile likelihood plots are presented, with 1$\sigma$ and 2$\sigma$ boundaries outlined in white.
		Where applicable, input PM parameter values are indicated with a red circle.
	}
	\label{ak2pm-700-pm40-set1}
\end{figure}

Fig.~\ref{ak2pm-700-pm40-set1} visualises the \textbf{T}$_a$ parameter space of this result as a collection of two-dimensional profile likelihood ratios.
Each point in a parameter plane corresponds to the ratio of local to global maximum likelihoods:
the former is found by profiling over the remaining \textbf{T}$_a$ parameters (i.e. by allowing them to adopt values that maximise the likelihood),
while the latter corresponds to the overall best-fit likelihood.
The white contours represent the 1$\sigma$ and 2$\sigma$ confidence boundaries,
and a red circle marks the PM input point $(m_X,\Gamma_X)$ in the $\langle m,\Gamma\rangle$ plane.
The expected input mass and width values are recovered in the fit to be well within the 1$\sigma$ confidence level,
and we find tightly-constrained contours with very mild correlations between the \textbf{T}$_a$ parameters.

\begin{figure}
	\centering
	\includegraphics[width=0.49\textwidth]{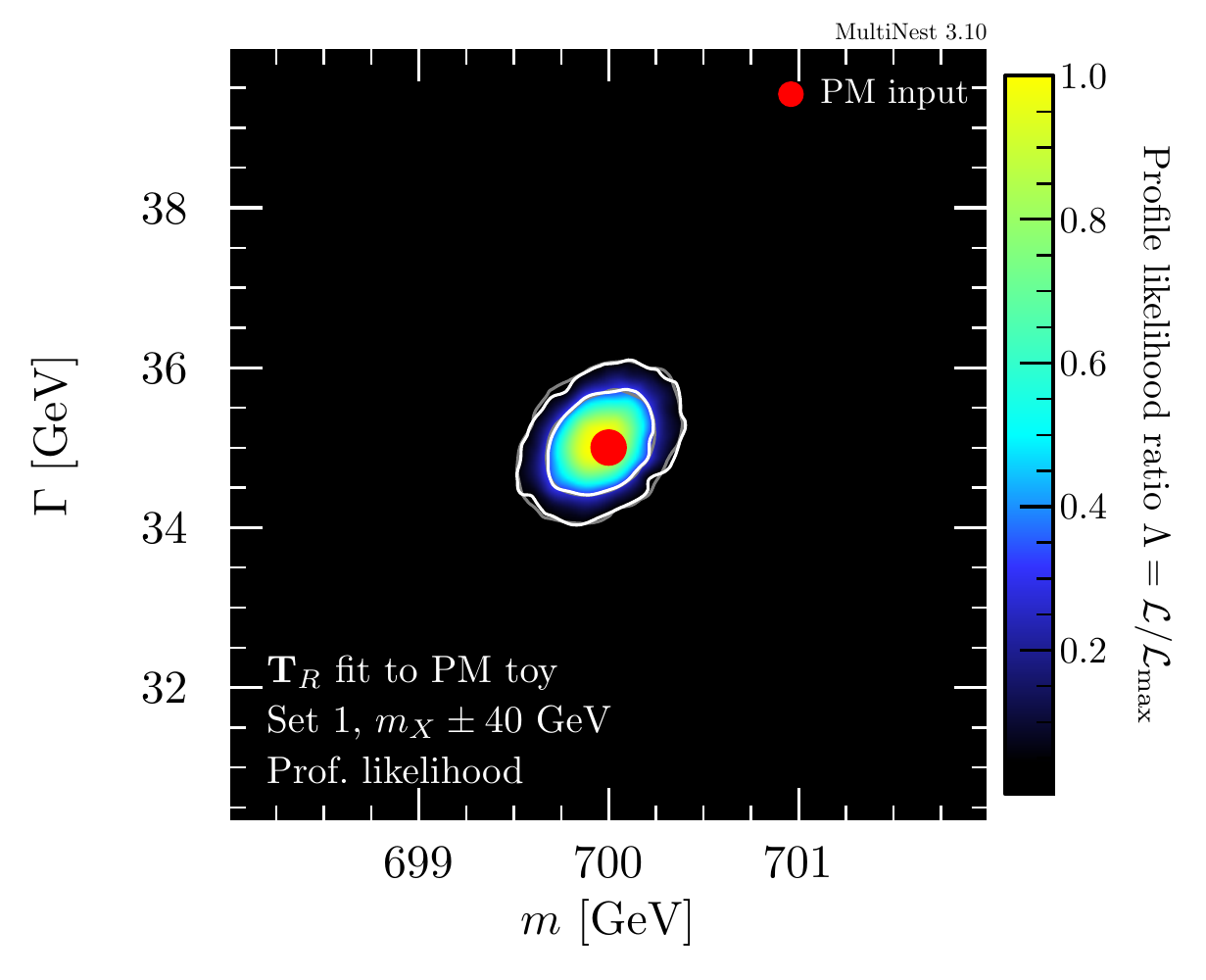}
	\includegraphics[width=0.49\textwidth]{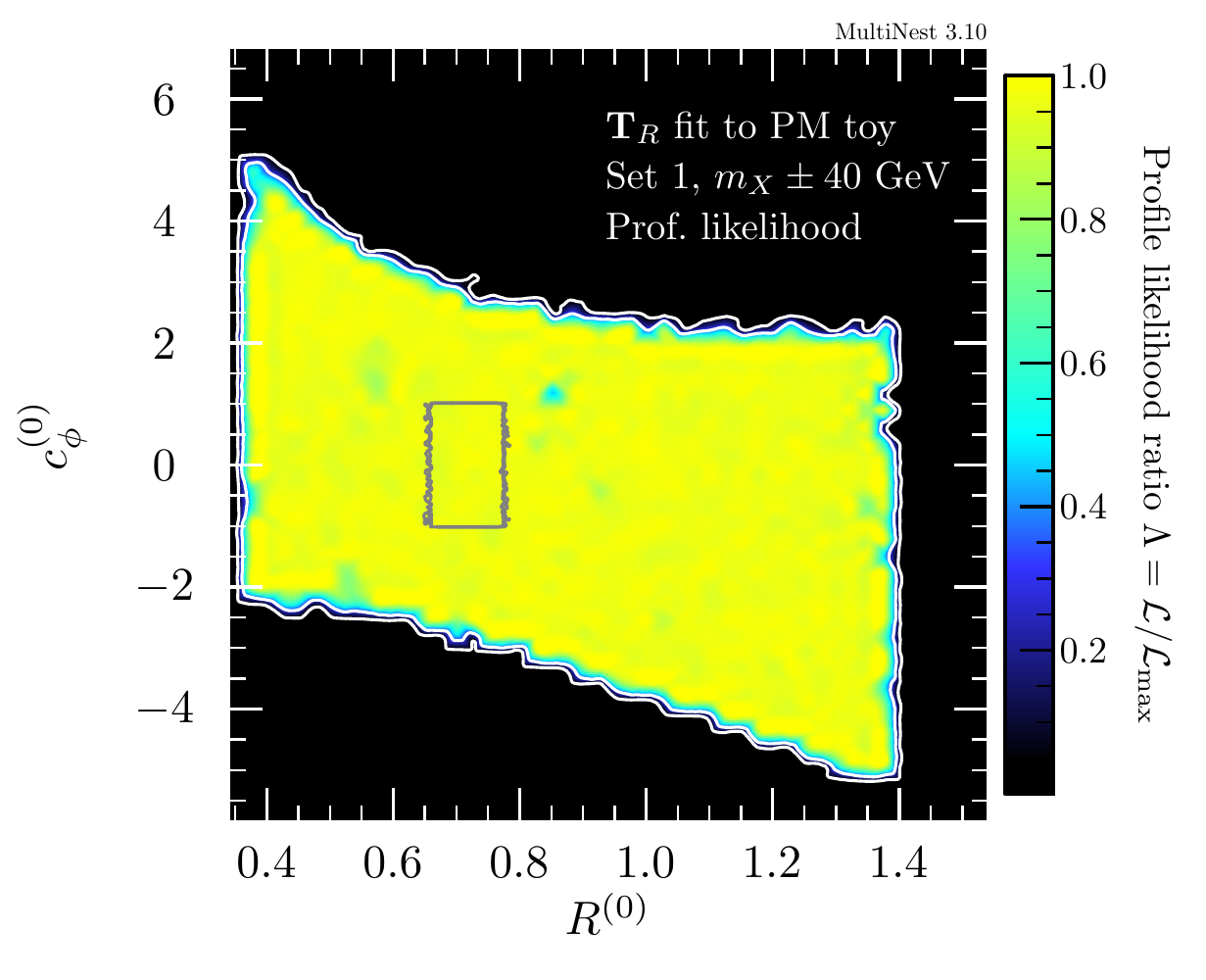}\\
	\includegraphics[width=0.49\textwidth]{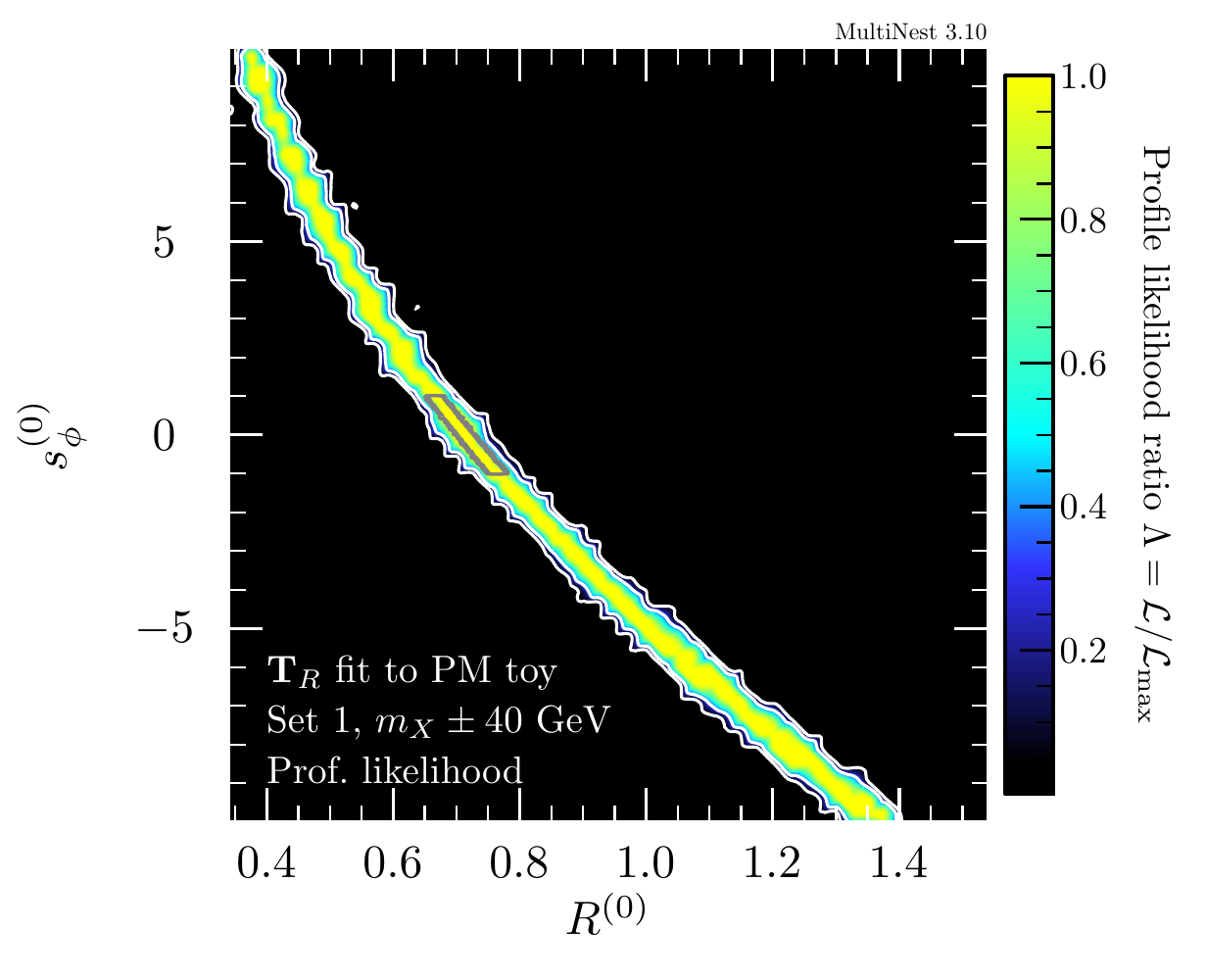}
	\includegraphics[width=0.49\textwidth]{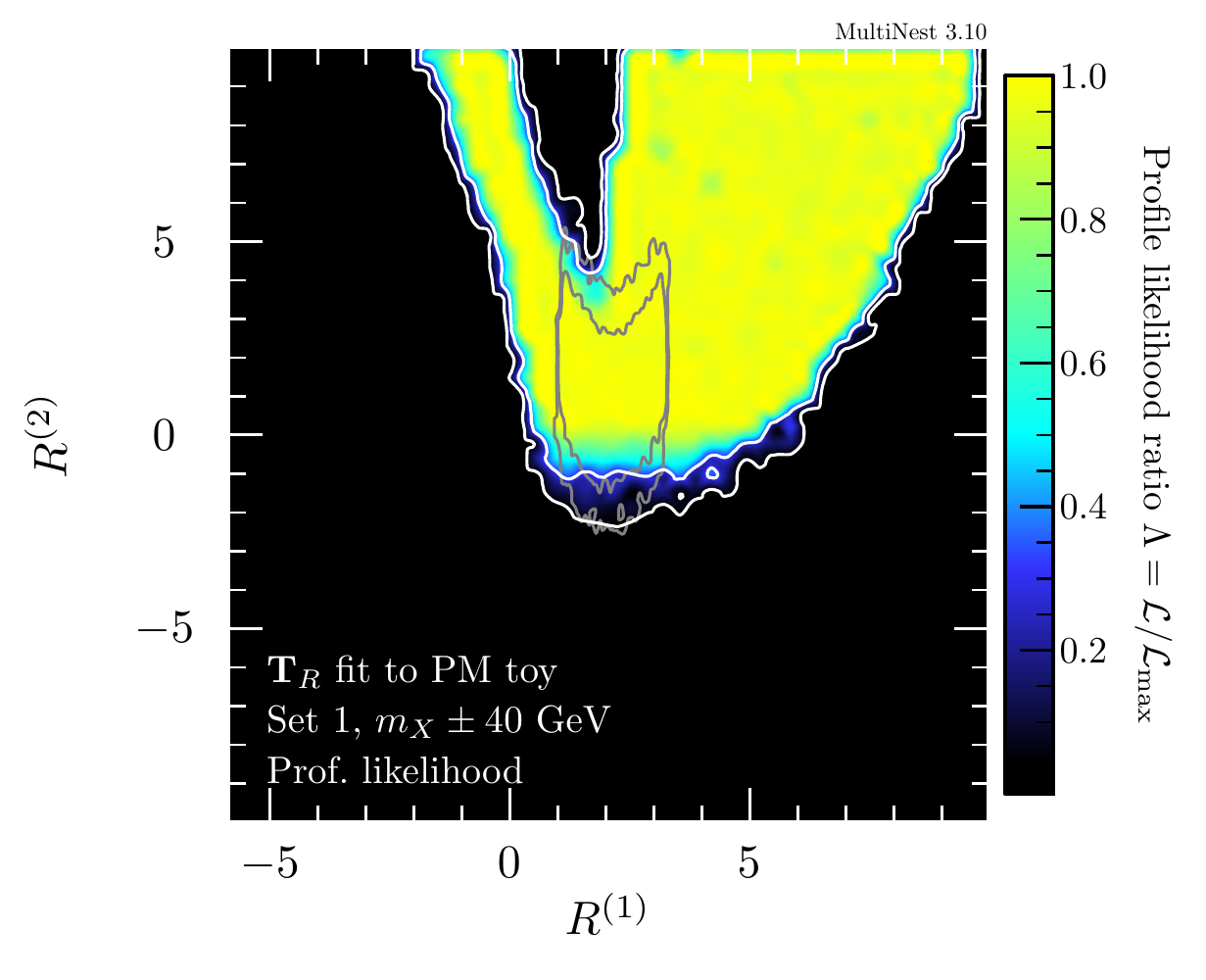}\\
	\includegraphics[width=0.49\textwidth]{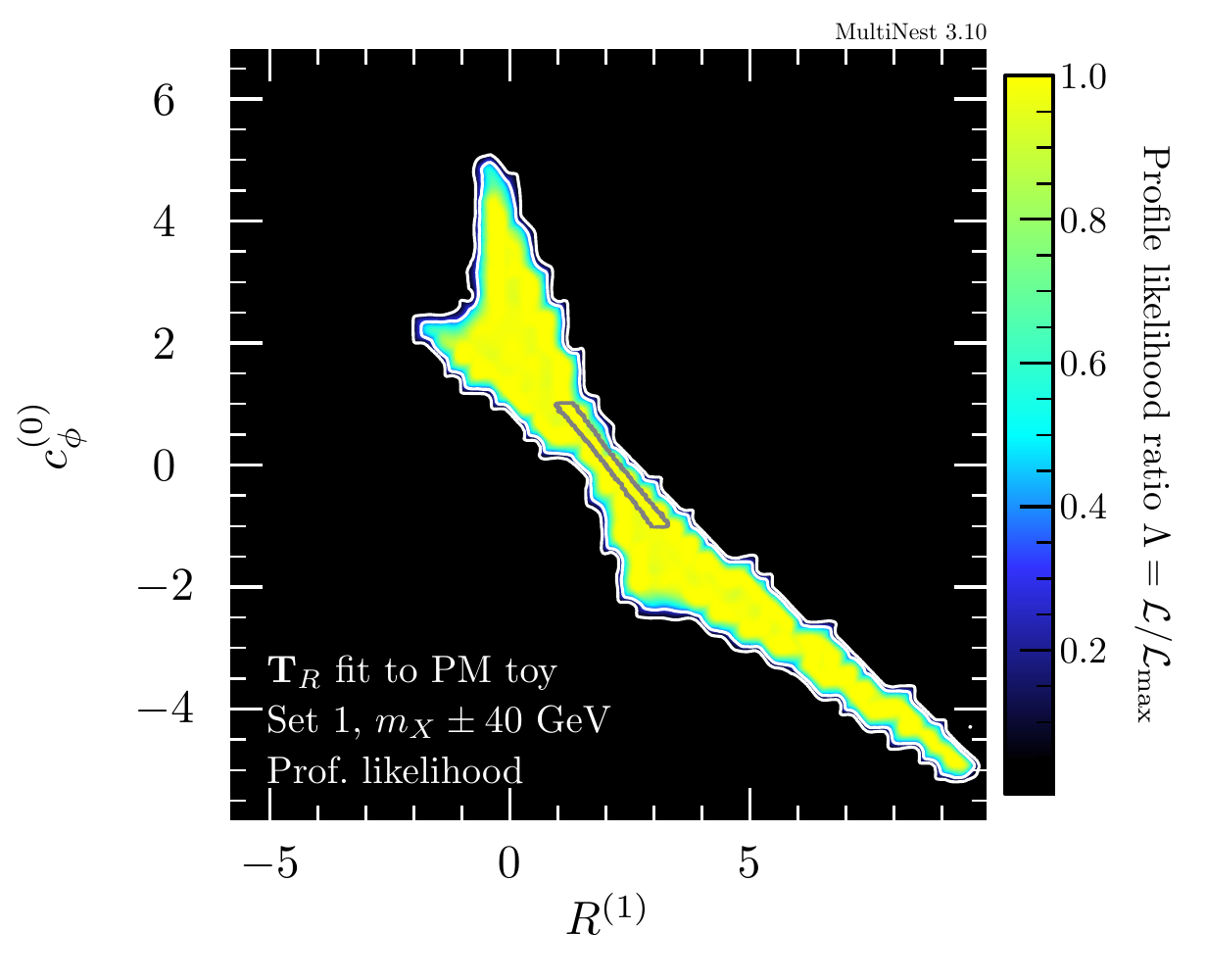}
	\includegraphics[width=0.49\textwidth]{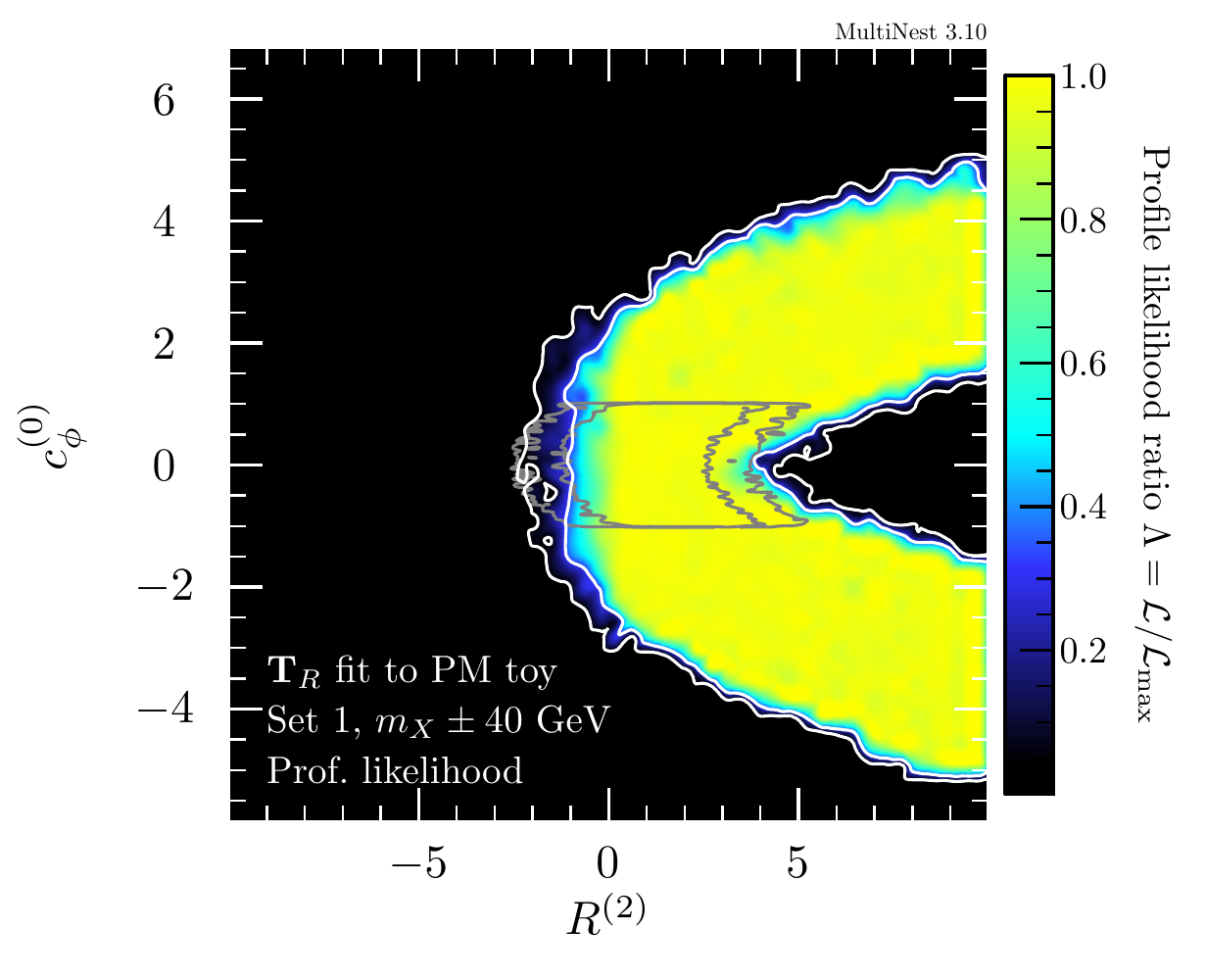}
	\caption{
		Fit result of the \textbf{T}$_R$ parameters to a PM Asimov toy over the $m_X\pm40$\,GeV window.
		The toy corresponds to the $m_X=700$\,GeV, $\Gamma_X/m_X=5\%$, and set~1 input parameters.
		Two-dimensional profile likelihood plots are presented, with 1$\sigma$ and 2$\sigma$ boundaries outlined in white.
		Where applicable, input PM parameter values are indicated with a red circle.
		Grey contours represent the 1$\sigma$ and 2$\sigma$ regions that correspond to physical solutions, $\cpz,\spz\in[-1,1]$.
	}
	\label{rc2pm-700-pm40-set1}
\end{figure}

Fig.~\ref{rc2pm-700-pm40-set1} shows the analogous result in terms of the \textbf{T}$_R$ parameter set, for a selection of possible two-parameter planes.
The $\langle m,\Gamma\rangle$ contour is essentially identical to that in fig.~\ref{ak2pm-700-pm40-set1},
which confirms the consistency of the two $m$ and $\Gamma$ determinations obtained by means of the \textbf{T}$_R$ and \textbf{T}$_a$ sets.
The $R^{(i)}$, $\cpz$ and $\spz$ parameter planes are of more interest:
we find large, relatively flat regions of high likelihood,
with highly non-trivial degeneracies between the parameters.
These are a sharp contrast to the neat solutions of \textbf{T}$_a$ space.
As we previously mentioned in section~\ref{sec:template},
such a behaviour is the expected consequence of the system being underconstrained by the \textbf{T}$_R$ parameters,
and affirms that the results obtained using this parametrisation cannot meaningfully be presented as a set of parameter values with associated uncertainties.
Furthermore, flat regions in the fit-parameter space may induce larger uncertainties in cases less ideal than those constituted by Asimov datasets.

From the perspective of a fit, $\cpz$ and $\spz$ are no different from the other parameters of the functional form;
as such, we have allowed them to vary beyond their physical ranges,
$\cpz,\spz\in[-1,1]$, in the fit.
The grey contours imposed over each plot represent the 1$\sigma$ and 2$\sigma$ boundaries of the physical region,
corresponding to points with $\cpz,\spz\in[-1,1]$.
In these results, the physical contours are found to be fully contained within the enlarged contours.
However, this is not necessarily the case in general,
since the underconstrained nature of the parametrisation implies that $\cpz$ and $\spz$ values can vary to compensate for the lack of higher order terms in the functional form.
This can lead to solutions that strongly prefer unphysical regions of the parameter space.

Given that the only reason one would favour the \textbf{T}$_R$ over the \textbf{T}$_a$ parameters lies in the more immediate physical interpretation of the former,
even at the cost of more severe degeneracies,
results presented in terms of \textbf{T}$_R$ parameters should preferably be consistent with physical constraints.
Nonphysicality of the trigonometric parameters can be used as a diagnostic tool:
results that do not admit any physical solutions must be discarded.
On the other hand, this implies that as long as physical regions are not rejected by the fit,
it is safe to present the result in which the physical considerations are imposed on $\cpz$ and $\spz$.
This has the benefit of greatly reducing the extent of degeneracy of the likelihood;
in particular, if the narrow correlations in the $\langle \ro,\spz\rangle$ and $\langle \ri,\cpz\rangle$ planes in fig.~\ref{rc2pm-700-pm40-set1} are symptomatic of the parametrisation,
then $\ro$ and $\ri$ will be greatly constrained upon bounding $\cpz$ and $\spz$.
We have verified that fit results yielding $\cpz$ and $\spz$ solutions outside of their physical ranges occur more readily with the set of eq.~\eqref{fitRcspar3K2} (which emerges from a strict first-order expansion of the function $R(q^2)$),
and for this reason, we suggest the baseline \textbf{T}$_R$ fitting procedure to involve the second-order set of eq.~\eqref{tempRcs}.

\begin{figure}
	\centering
	\includegraphics[width=0.49\textwidth]{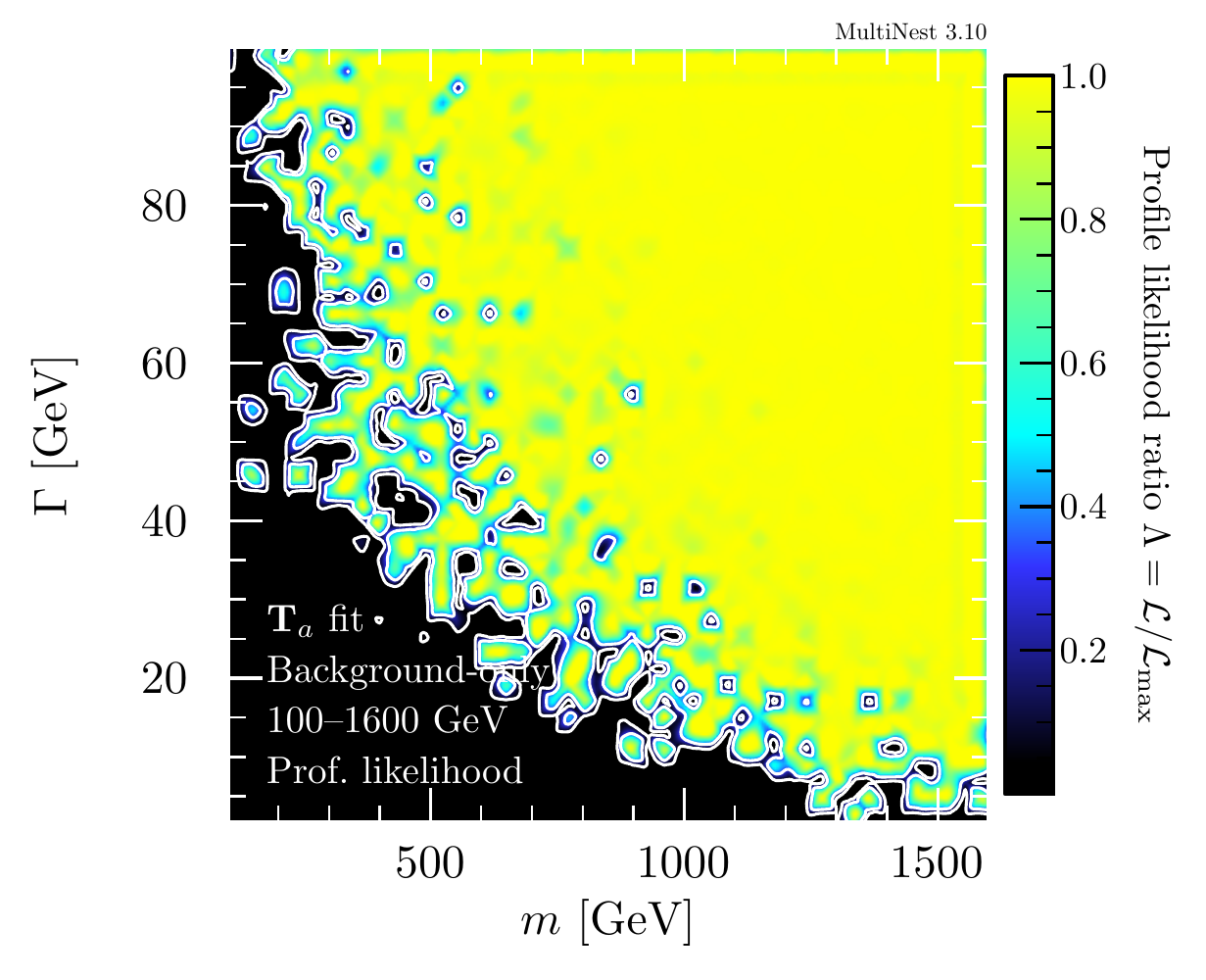}
	\includegraphics[width=0.49\textwidth]{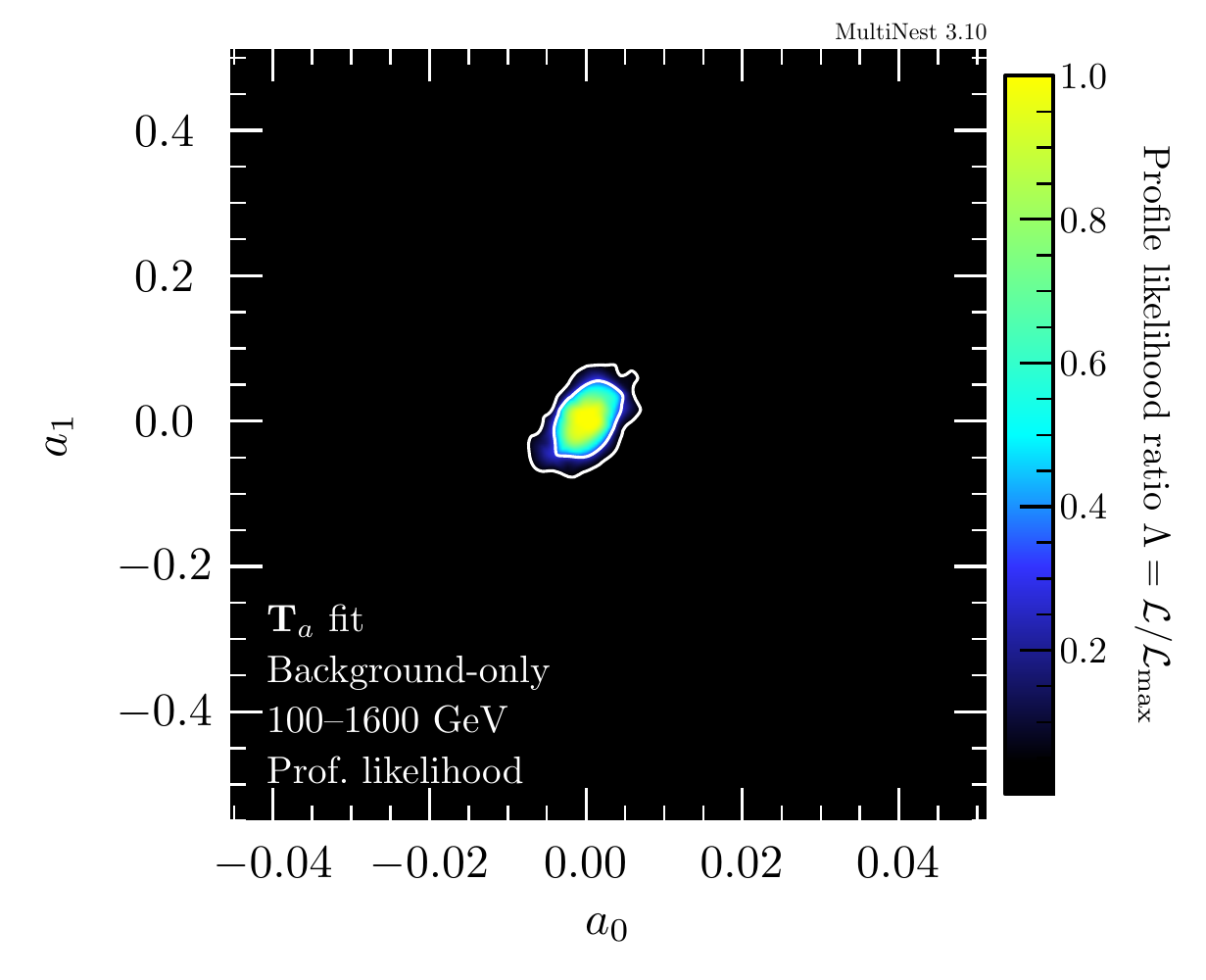}\\
	\includegraphics[width=0.49\textwidth]{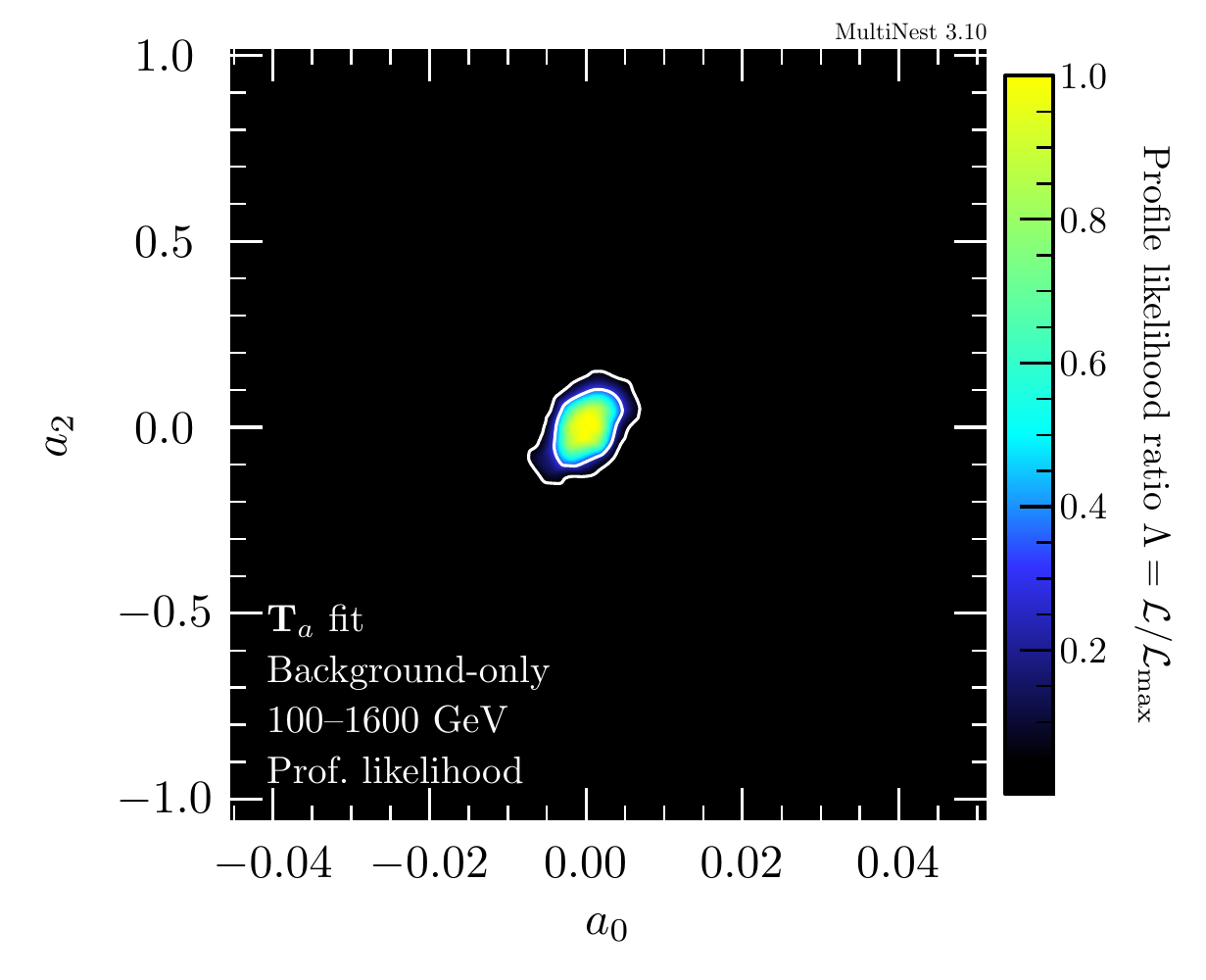}
	\includegraphics[width=0.49\textwidth]{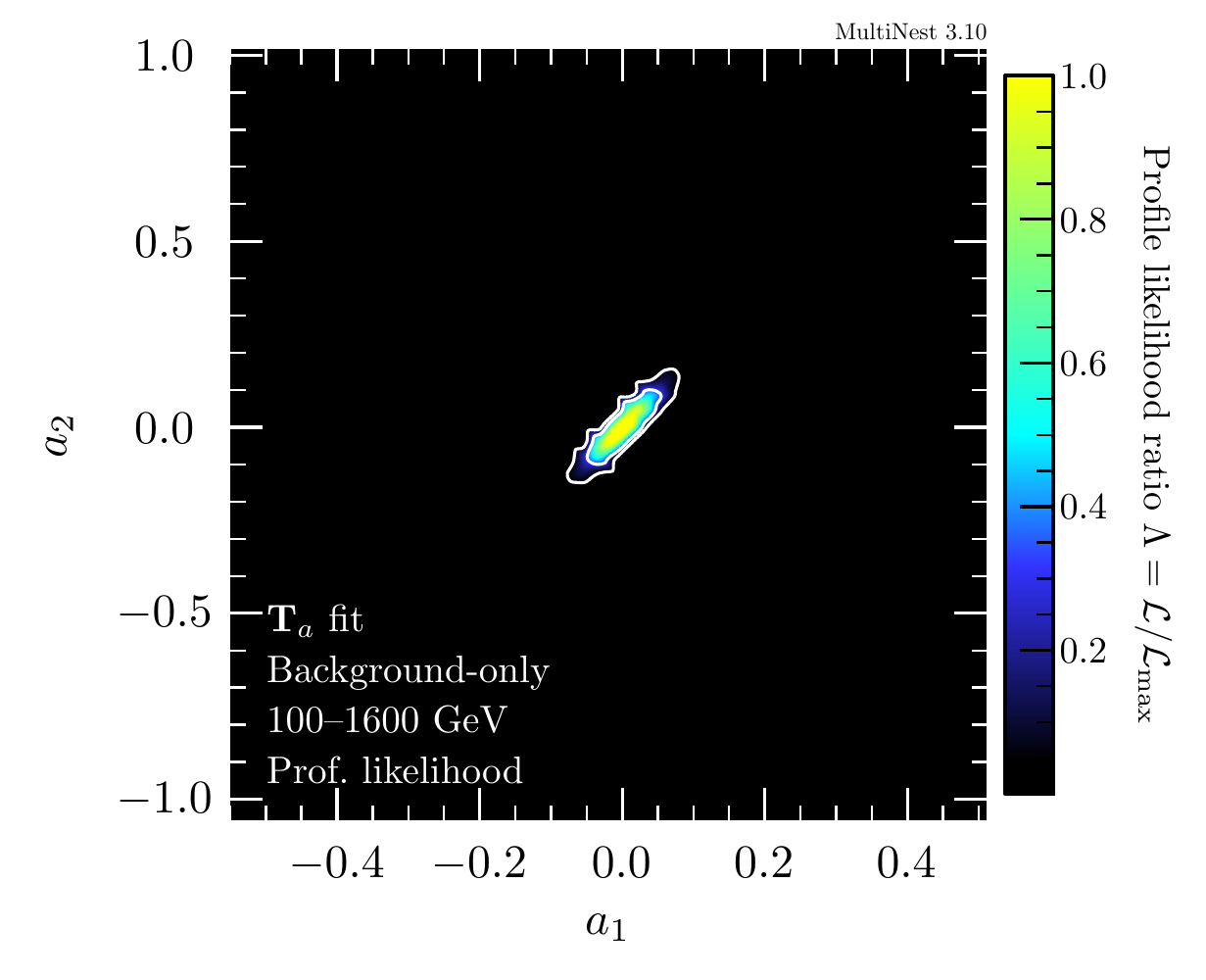}
	\caption{Fit of the \textbf{T}$_a$ parameters to a background-only Asimov over the 100--1600\,GeV mass range.}
	\label{ta-null}
\end{figure}

We conclude the section by remarking that both of the \textbf{T}$_R$ and \textbf{T}$_a$ parameter sets can accommodate background-only solutions,
and as such, can also be employed for the description of null signals.
The \textbf{T}$_a$ parameters provide the simplest demonstration of a null result:
the solution corresponding to $a_k=0$ for all $k$ is the background-only one.
The analogous solution in \textbf{T}$_R$ parameter space is less simple, owing to the complicated correlations between its parameters;
however, its advantage lies in that they admit the space of possible signal and interference contributions that sum to an apparent background-only distribution.
We present an example of a null result in fig.~\ref{ta-null},
corresponding to a fit of the \textbf{T}$_a$ parameters to a background-only Asimov histogram over the 100--1600\,GeV mass range.
Note that the result in the $\langle m,\Gamma\rangle$ plane is included for completeness,
but is largely meaningless (and can thus be neglected) in the case of a null result.

\subsection{Physics model closure test}
\label{sec:toytruth-subsec:pmclosure}

For a theorist to be able to test their specific physics model given a set of results presented in terms of either the \textbf{T}$_R$ or \textbf{T}$_a$ parameters,
there are two possible strategies.
Either one computes the \textbf{T}$_R$ or \textbf{T}$_a$ parameters directly from one's chosen theoretical model,
and compares the predictions to those measured by the experiments;
or one transforms the \textbf{T}$_R$ or \textbf{T}$_a$ experimental results into the parameters of one's physics theory.
While the former approach is more straightforward in that it follows the same procedure we have outlined so far,
we demonstrate in this section that the latter one is viable, too.
We adopt again the case corresponding to $m_X=700$\,GeV and $\Gamma_X/m_X=5\%$ as an example;
by using the results of figs.~\ref{ak2pm-700-pm40-set1} and~\ref{rc2pm-700-pm40-set1},
we show that the expected (i.e.~input) PM parameter values can be recovered.

We begin by using the \textsc{MultiNest} (see eq.~\eqref{log-like}) sampling outputs corresponding to the \textbf{T}$_a$ and \textbf{T}$_R$ fits to construct ``results'' histograms using their respective functional forms.
Such histograms are defined over an invariant mass range equal to the fitted invariant mass window (in this case, 660--740\,GeV),
with the same binning as that of the fitted dataset (2\,GeV).
The points in parameter space satisfying
\begin{equation} \label{rc-results-1sigma}
	\chi^2_\text{point} - \chi^2_\text{bf} \le \mathcal{Q}_{\chi^2_{\nu}}(p=0.68)\,,
\end{equation}
were extracted from the \textsc{MultiNest} files,
where $\nu=5$ for the \textbf{T}$_a$ parameters ($\mathcal{Q}_{\chi^2_5}(0.68)\approx5.86$)
and $\nu=7$ for the \textbf{T}$_R$ ones ($\mathcal{Q}_{\chi^2_7}(0.68)\approx8.14$).
These are the points that lie within the 1$\sigma$ confidence region of the respective $\nu$-dimensional parameter spaces.
For a given histogram bin, the range of functional form values prescribed by these points can thus be interpreted as the 1$\sigma$ confidence interval of the bin.
In this way, we construct histograms, corresponding to a diphoton invariant mass distribution,
that represent the band of \textbf{T}$_a$ and \textbf{T}$_R$ lineshapes agreeing with the original PM toy to within 1$\sigma$.

\begin{figure}
	\centering
	\includegraphics[width=0.49\textwidth]{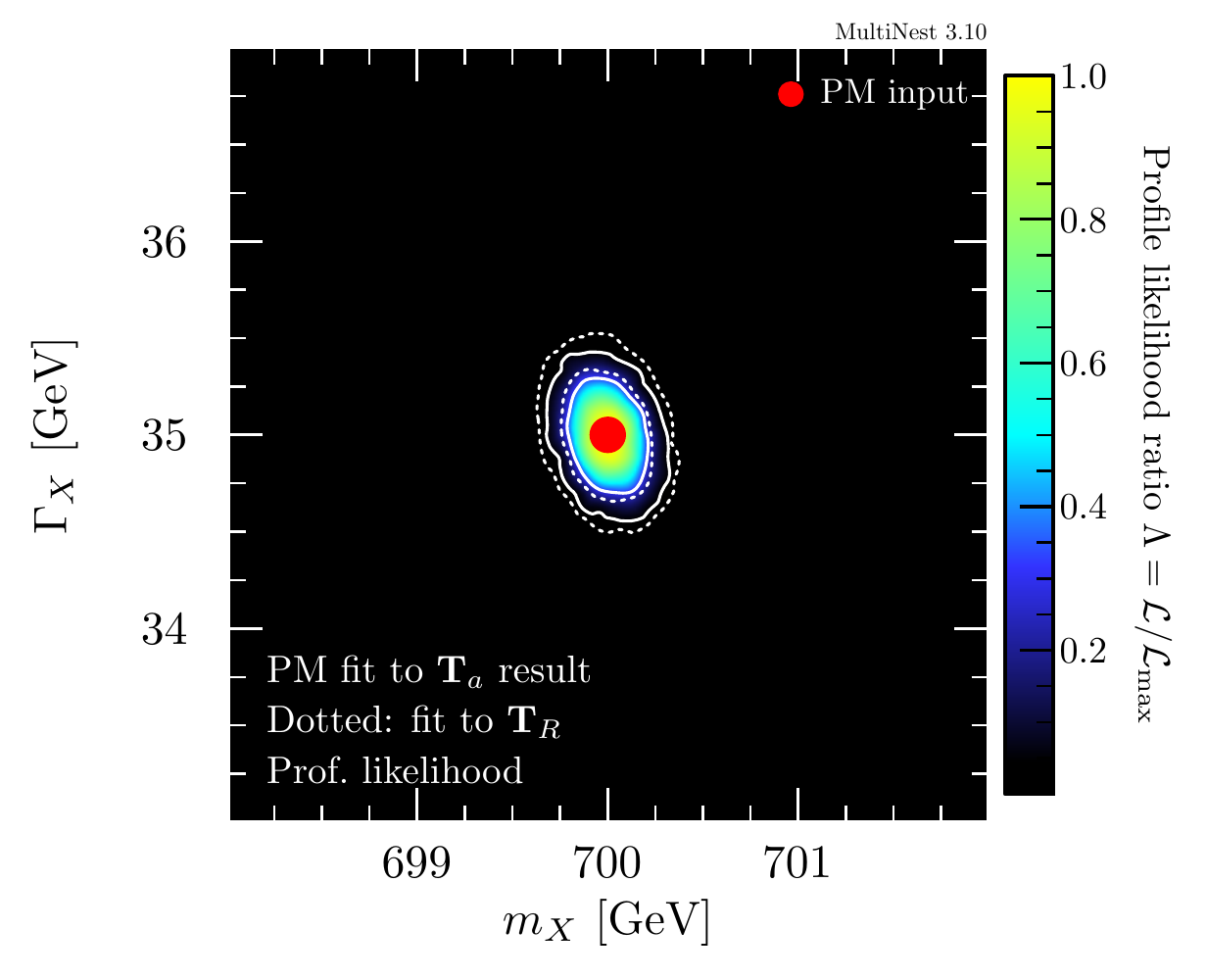}
	\includegraphics[width=0.49\textwidth]{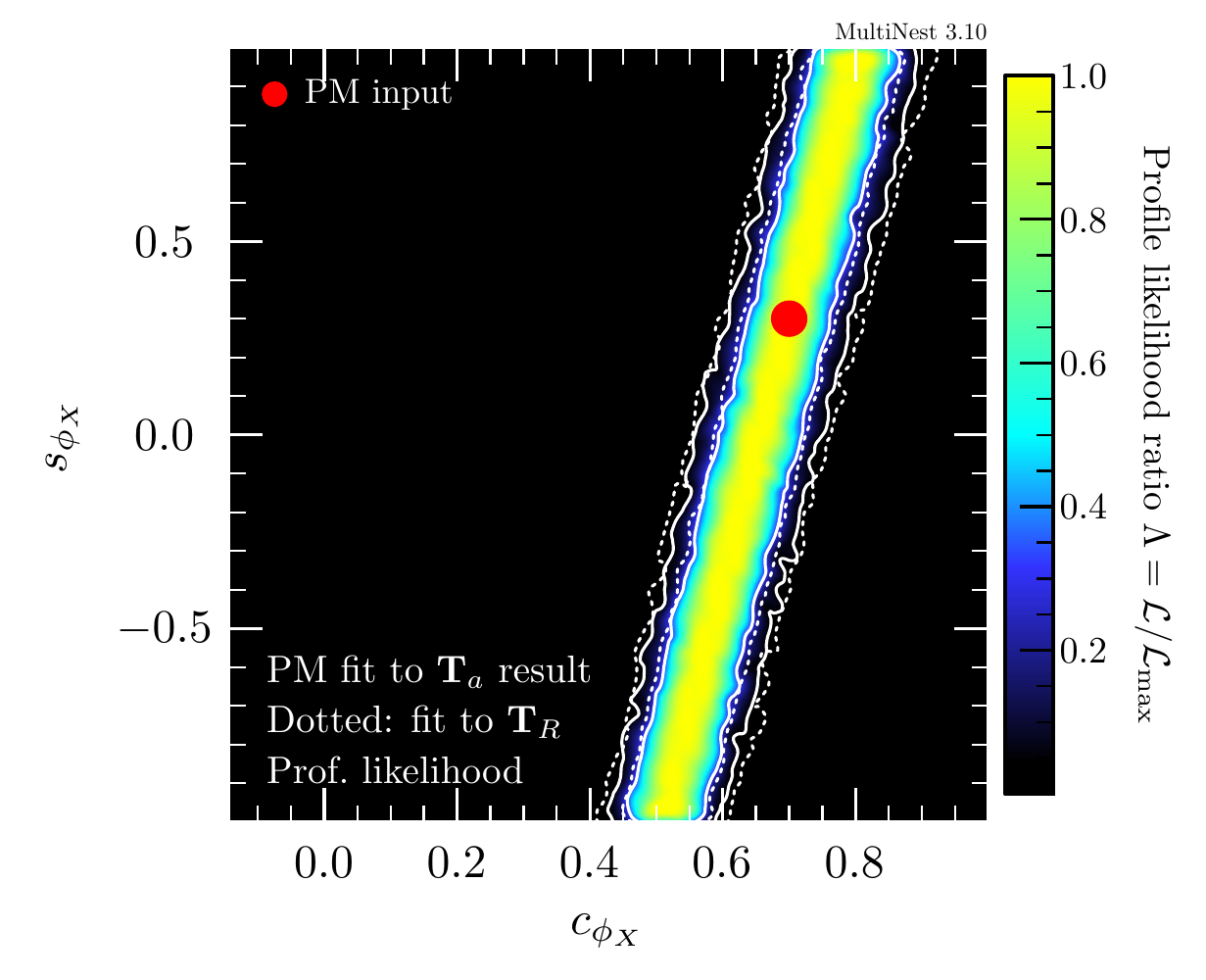}\\
	\includegraphics[width=0.49\textwidth]{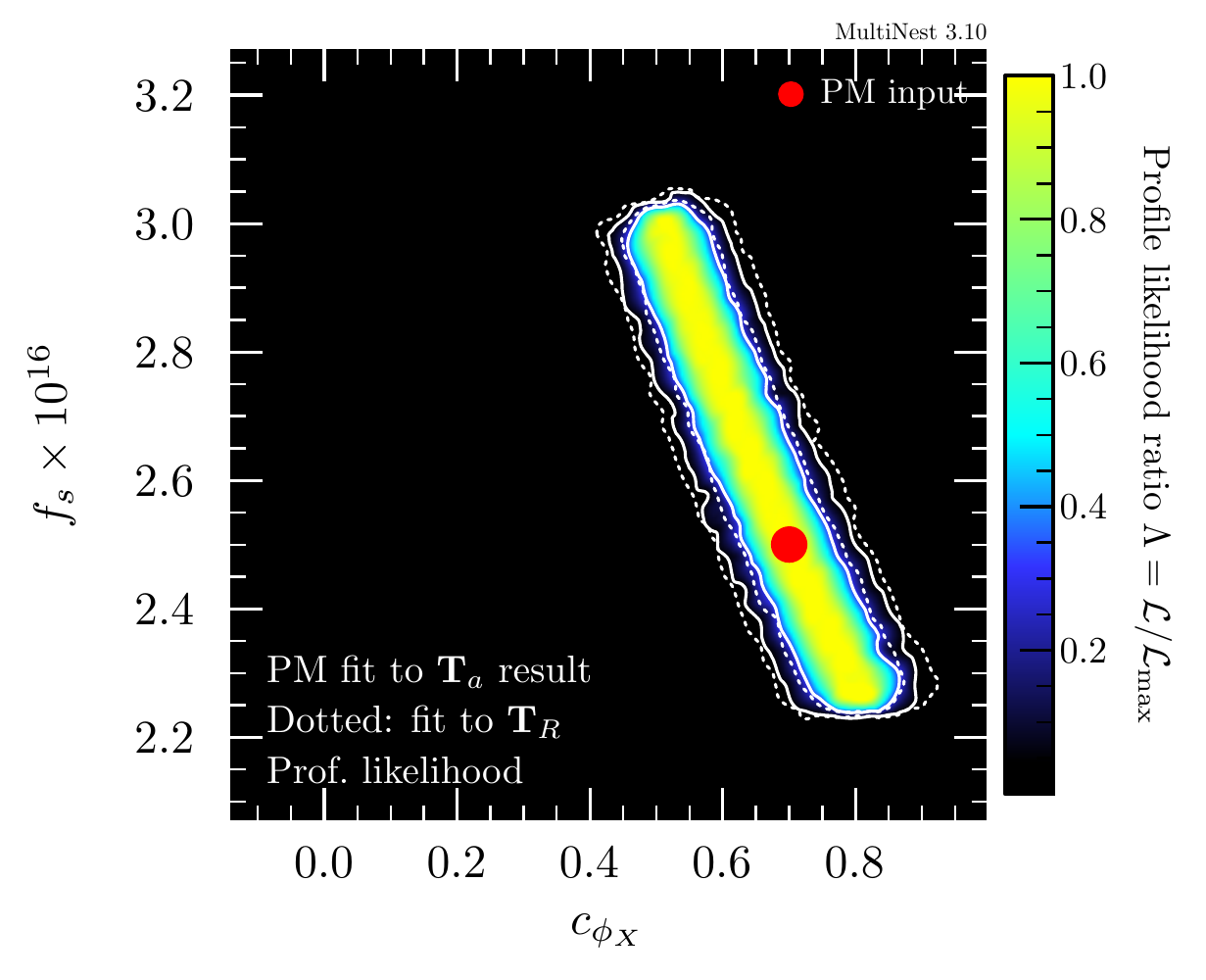}
	\includegraphics[width=0.49\textwidth]{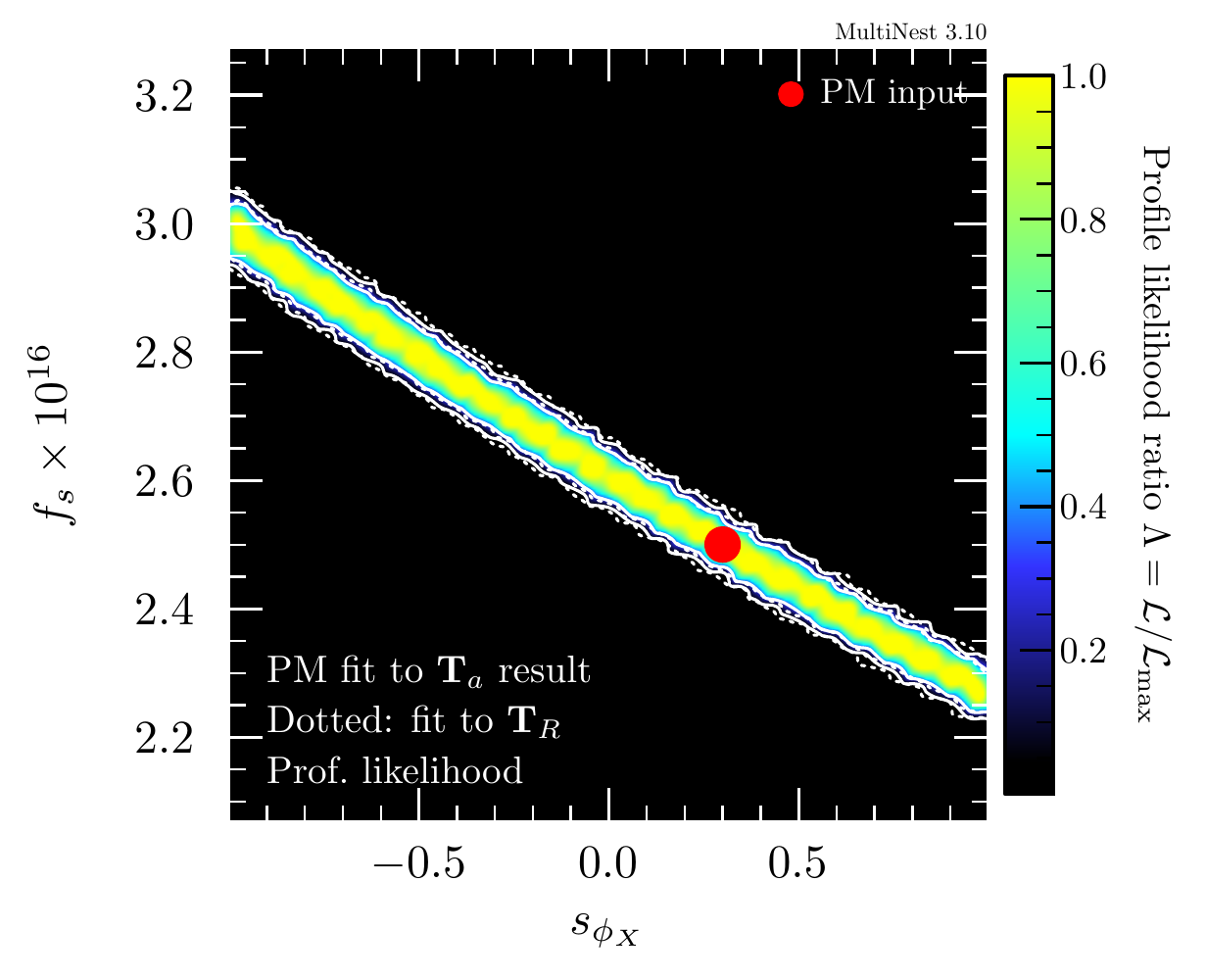}
	\caption{
		Profile likelihood plots of the PM parameters from a fit to \textbf{T}$_a$ and \textbf{T}$_R$ results histograms,
		corresponding to input $m_X=700$\,GeV, $\Gamma_X/m_X=5\%$ and set~1 parameter values.
		The fit is performed over the $700\pm40$\,GeV mass window.
		1$\sigma$ and 2$\sigma$ contours are outlined in solid white (\textbf{T}$_a$) and dotted white (\textbf{T}$_R$).
		Input PM points are marked with a red circle.
	}
	\label{pm2rc-700-pm40-set1}
\end{figure}

Fig.~\ref{pm2rc-700-pm40-set1} shows the profile likelihood ratios resulting from fits of the analytic PM description to the results histograms constructed as detailed above.
The main set of results correspond to the fit of the \textbf{T}$_a$ results histogram,
with dotted white contours showing the result in the \textbf{T}$_R$ case.
Since the \textbf{T}$_a$ and \textbf{T}$_R$ results both describe the same underlying physics,
the two sets of contours are very similar, as expected;
their slight difference arises due to the greater dimensionality of the \textbf{T}$_R$ parameter space,
which implies larger bin uncertainties being defined for its results histogram, in accordance with eq.~\eqref{rc-results-1sigma}.
For all of the parameter planes visualised, we find the corresponding PM inputs (eq.~\eqref{pm-input-sets}, set~1) to lie within the high-likelihood regions obtained.
Thus, the procedure described above can indeed be used to construct histograms representative of physical data given a set of results in terms of the general parameters,
and thence constrain the parameter space of the physics model of interest.

While we have utilised an analytic description of the physics model in our demonstration,
in the case that such a form is not readily available one can generate a Monte Carlo histogram for each point in the parameter space of the model one wishes to test, and compare that to the histograms derived from the \textbf{T}$_a$ or \textbf{T}$_R$ fit results. This comparison will define a likelihood for the fit of the new model parameters.

\subsection{Test of discovery significance}
\label{sec:toytruth-subsec:discovery}

The results obtained in the previous sections used knowledge of the true (input) resonance mass in selecting fit windows.
However, in a blind search, there is the lower-level question of whether or not a signal is even present.
Thus, some preliminary analysis is required to identify candidate signal regions.

For this purpose, a background-only hypothesis is commonly used to quantify the significance of a discovery.
The test statistic of interest is given by
\begin{equation} \label{q0-statistic}
	q_0=
	\begin{cases}
		-2\log\Lambda_0 & \text{ for } 0<\Lambda_0\le 1\,,\\
		0 & \text{ for } \Lambda_0 > 1\,,
	\end{cases}
\end{equation}
where
\begin{equation} \label{bg-test-like-ratio}
	\Lambda_0(\bsym\Theta) \equiv \frac{\lagr_0}{\lagr(\bsym\Theta)}\,,
\end{equation}
with $\mathcal{L}_0$ the likelihood of a background-only description against the data,
and $\mathcal{L}(\bsym{\Theta})$ the likelihood of a model that includes both signal and background, with model parameters $\bsym\Theta$ (in this case, the \textbf{T}$_a$ or \textbf{T}$_R$ parameter sets).
The $q_0$ statistic is thus a measure that compares the likelihoods of the null (background-only) and alternative (i.e. including a signal) hypotheses;
given a particular dataset and the observed value, $q_{0,\text{obs}}$, one can calculate the $p$-value,
\begin{equation} \label{probability-p0}
	p_0 = \int_{q_0,\text{obs}}^\infty f(q_0|0)\, dq_0\,,
\end{equation}
that subsequent data will exhibit an incompatibility with the background-only hypothesis to an equal or greater degree.
Here, $f(q_0|0)$ represents the probability distribution function for $q_0$ conditional on the null hypothesis being true, and as such is always non-negative;
a larger value of $q_{0,\text{obs}}$ thus yields a smaller $p_0$, and indicates a greater disagreement with the background-only hypothesis.

An alternative metric that quantifies the statistical incompatibility of the background-only assumption against the observed dataset is the local significance:
\begin{equation} \label{local-signif}
	\mathcal{Z}_0 = \mathcal{Q}_G(1-p_0)\,,
\end{equation}
where $\mathcal{Q}_G(p)$ is the quantile function for the standard Gaussian distribution evaluated at probability $p$.
For the test statistic of eq.~\eqref{q0-statistic}, the local significance is given by~\cite{Cowan:2010js}:
\begin{equation} \label{signif-simple}
	\mathcal{Z}_{0} = \sqrt{q_{0}}\,.
\end{equation}
The common threshold required for a claim of discovery in particle physics is $\mathcal{Z}_0 = 5$.

For the purposes of a background-only hypothesis test,
the entire invariant mass region of the search is typically fitted;
thus, the \textbf{T}$_R$ and \textbf{T}$_a$ parameter sets (eqs.~\eqref{tempRcs} and~\eqref{tempA}),
which are accurate only within a restricted mass range about the signal peak, are technically inadequate.
However, we note that one can still achieve the goal of identifying prospective signals using the second-order functional forms,
since the test only seeks to compare the background-only description with one that includes a signal,
and not to extract accurate estimations of the parameters of the latter.
To prevent an underestimation of the local significance,
one should define sufficiently large prior volumes for the parameters during a fit,
so as to give them the freedom needed to compensate for the lack of higher-order terms.

We demonstrate the procedure using an Asimov dataset generated from the physics model,
with the input parameters as we have adopted in the previous sections ($m_X=700$\,GeV, $\Gamma_X/m_X=5\%$, and set~1 parameter values).
The functional form of eq.~\eqref{rc-dxs} was used as the model alternative to the null hypothesis,
and was fitted over the full (100--1600\,GeV) invariant mass range of the toy.
We note that since parameter estimation is not of import,
one can choose to do this using either the \textbf{T}$_a$ or \textbf{T}$_R$ parameters;
in the interest of computational efficiency, we recommend the \textbf{T}$_a$ parameter set be employed,
as a fit of its parameters typically converges more quickly in comparison to the \textbf{T}$_R$ one.
The prior volumes for $m$ and $\Gamma$ were uniformly partitioned into 100 separate regions over the 100--1600\,GeV and 0--160\,GeV ranges, respectively.
A scan over all of the partitions was performed:
for each iteration,
the maximum likelihood was obtained by profiling over the remaining parameters,
and eqs.~\eqref{q0-statistic} and~\eqref{signif-simple} were then used to calculate the local significance,
in units of standard deviations ($\sigma$).

\begin{figure}
	\centering
	\includegraphics[width=0.5\textwidth]{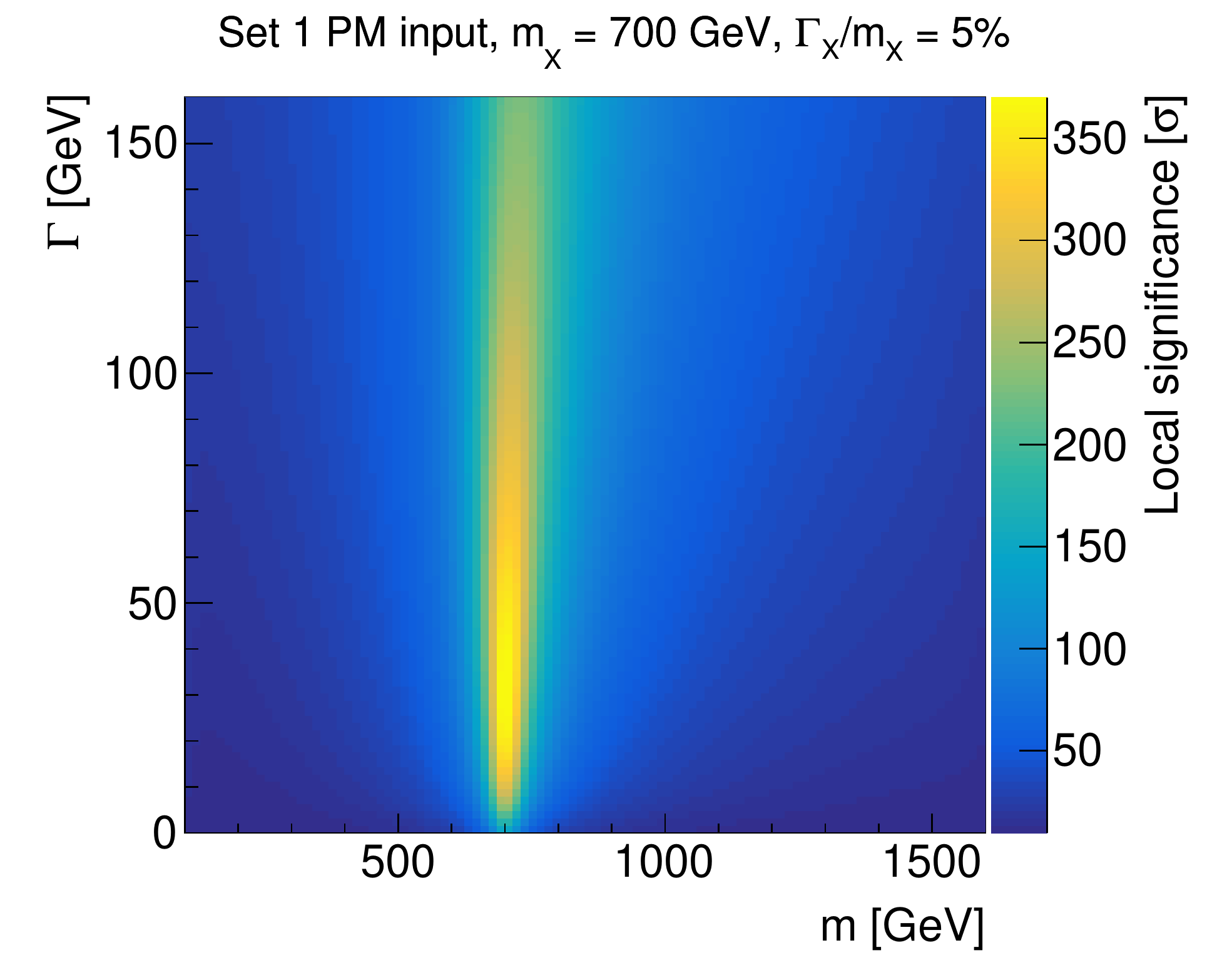}
	\caption{
		Local significance for rejecting the background-only hypothesis, for an Asimov dataset corresponding to input mass $m_X=700$\,GeV and width $\Gamma_X=35$\,GeV.
		The \textbf{T}$_a$ parametrisation of the template functional form is used as the alternative hypothesis.
	}
	\label{signif-plot}
\end{figure}

The results are visualised in fig.~\ref{signif-plot}.
A distinct region of high local significance can be identified at the mass value expected from the input value chosen.
The width is constrained less precisely by this result,
but in a blind search one would only require the former in order to presuppose the existence and approximate mass of a resonance.
The remaining general parameters (including the width) can then be more precisely obtained by following our procedure of performing a dedicated fit of the model-independent functional form within a suitable fit window (see in particular sections~\ref{sec:toytruth-subsec:fitwindows} and~\ref{sec:toytruth-subsec:proflike}).

\subsection{Procedure for a general resonance search}
\label{sec:toytruth-subsec:procedure}

Our tests of the \textbf{T}$_R$ and \textbf{T}$_a$ functional forms have demonstrated their flexibility in fitting a wide range of physical distributions,
provided one makes a suitable choice for the invariant mass window over which the fit is performed.

A summary of our results,
and an outline of the procedure that we propose for a general resonance search,
is as follows:
\begin{enumerate}
	\item Perform a background-only hypothesis test by scanning over the $\langle m,\Gamma\rangle$ plane of the \textbf{T}$_a$ parameters,
	and produce a plot of the local discovery significance.
	This can also be performed with the \textbf{T}$_R$ parameters,
	but we recommend the \textbf{T}$_a$ for simplicity and computational reasons.
	Apart from the mass and width,
	all of the other \textbf{T}$_a$ (or \textbf{T}$_R$) parameters are to be profiled over to yield the maximum significance for each iteration of the scan.
	
	\item \begin{enumerate}
	    \item If notably high significance regions exist: perform fits of the template functional form to the data in these regions over mass windows of increasing size,
	    using either of the \textbf{T}$_R$ or \textbf{T}$_a$ parameter sets.
	    A suitable mass window that yields a good fit quality should be identified from the results of these fits.
	    Then, extract the general parameters (\textbf{T}$_R$ or \textbf{T}$_a$) from a fit over the chosen mass window.
	    If this is performed using the \textbf{T}$_R$ parameters,
	    one should confirm that unphysical solutions are not strongly preferred for $\cpz$ or $\spz$.
	    Fit results are to be presented in the form of 2-dimensional profile likelihood contours in the general parameter space.
	
	    \item If notably high significance regions do \emph{not} exist:
	    perform a fit of the template functional form to the data over the entire mass region.
	    We recommend the \textbf{T}$_a$ parameter set for such a fit, due to its simpler interpretation in the case of null signals:
	    one should simply find $a_k\approx0$ for each $k$.
	    Fit results are to be presented in the form of 2-dimensional profile likelihood contours in the general parameter space.
	\end{enumerate}
\end{enumerate}
We will also reiterate our preliminary considerations:
a single resonance should contribute to the invariant mass range considered, and if multiple regions of high significance are found in step \#1,
they must contribute to sufficiently separated regions of the data, such that independent analyses can be conducted for each.
Furthermore, it is assumed that the data contains only backgrounds produced in the same partonic channel as the signal (and hence generating an interference).
Any additional backgrounds yielding the same final state but induced by a different partonic process from the signal should be subtracted from the data prior to the analysis.

\section{Incorporation of detector effects in template forms}
\label{sec:detector}

The results presented thus far have been obtained under the assumption of statistically perfect datasets,
whilst also neglecting event reconstruction effects stemming from the limitations of particle detectors,
which will affect the events observed in any physical experiment.
To account for these effects, one can choose one of two approaches.
The first option is to obtain an approximation of truth-level data using unfolding algorithms,
to which the procedure, as summarised in section~\ref{sec:toytruth-subsec:procedure}, can then be applied.
The alternative approach is to incorporate a parametrisation of the detector reconstruction effects within the template descriptions,
so that they can be fitted directly to reconstructed data.

In this section, we study the latter approach.
We work with MC event samples generated using \madgraph{} and the HC model,
as described in section~\ref{sec:bm},
coupled with a fast simulation of the CMS detector using the \delphes{} framework~\cite{deFavereau:2013fsa}.
In accordance with current CMS trigger requirements~\cite{Khachatryan:2016hje}, we designate photons passing a $p_T>60$\,GeV selection cut as candidates for the diphoton pair produced in hard scattering events.

We begin by describing a method of parametrising detector reconstruction effects,
before demonstrating that it holds for the reconstructed MC event samples.
In applying the procedure to the template functional form,
several assumptions are made to simplify the method.
It is important to note that these assumptions,
while expected to be reasonable,
may not be generalisable to all experimental scenarios.
As such, the content of this section should not be seen as a comprehensive guideline for an analysis,
but rather as a procedure that could be adopted, or adapted, if the assumptions are deemed reasonable.
Since our procedure for doing so is based on that used to mimic the full simulation of the ATLAS and CMS experiments,
we note that it should therefore be applicable within the LHC experimental collaborations.

\subsection{Description of detector effects through convolution}
\label{sec:detector-subsec:baseline}

The standard procedure of an analysis is to model the detector-smeared invariant mass lineshape, $\mathcal{R}(q)$,
as the convolution of the truth-level description, $\mathcal{T}(q)$,
with a detector resolution function, $\text{DR}(q)$~\cite{Aaboud:2016tru}:
\begin{equation} \label{convolution-dr}
	\mathcal{R}(q) = (\mathcal{T}*\text{DR})(q) = \int_0^\infty \mathcal{T}(Q)\, \text{DR}(q-Q)\, dQ\,.
\end{equation}
This convolution is assumed to hold for the separate modelling of smeared signal, interference, and background distributions,
although for the latter one would typically opt for the alternative of obtaining a direct parametrisation instead.
If $\mathcal{R}(q)$ is normalised to unity,
one can then write the reconstructed differential cross section as:
\begin{equation} \label{reco-lineshape}
    \frac{d\sigma}{dq} = \varepsilon(q)\,\sigma\mathcal{R}(q)\,,
\end{equation}
where $\varepsilon(q)$ is the (generally $q$-dependent) reconstruction efficiency,
and $\sigma$ the (truth-level) total integrated cross section.

We choose to parametrise $\text{DR}(q)$ as a Double-Sided Crystal Ball (DSCB) function, which comprises of a Gaussian core with power law tails:
\begin{equation} \label{dscb}
	\text{DSCB}(q) = N_\text{DSCB}\,
	\begin{cases} 
		\exp\left(-\frac{t^2}{2}\right) &  \text{ for } -\alpha_\text{low}\le t\le \alpha_\text{high}\,, \\
		\frac{\exp\left(-\frac{\alpha_\text{low}^2}{2}\right)} {\left[\frac{\alpha_\text{low}}{n_\text{low}} 
		\left(\frac{n_\text{low}}{\alpha_\text{low}} -\alpha_\text{low}-t\right)\right]^{n_\text{low}}} &  \text{ for } t<-\alpha_\text{low}\,, \\
		\frac{\exp\left(-\frac{\alpha_\text{high}^2}{2} \right)}{\left[\frac{\alpha_\text{high}}{n_\text{high}} 
		\left(\frac{n_\text{high}}{\alpha_\text{high}} -\alpha_\text{high}+t\right)\right]^{n_\text{high}}} &  \text{ for } t>\alpha_\text{high}\,,
	\end{cases}
\end{equation}
where
\begin{equation} \label{dscb-t-param}
	t = \frac{q-\mu_\text{DSCB}}{\sigma_\text{DSCB}}\,.
\end{equation}
The parameters $\mu_\text{DSCB}$ and $\sigma_\text{DSCB}$ are the mean and standard deviation of the Gaussian core.
$\alpha_\text{low}$ and $\alpha_\text{high}$ determine the mass point at which the Gaussian transitions into power law distributions,
with exponents $n_\text{low}$ and $n_\text{high}$ respectively.
$N_\text{DSCB}$ is a factor that normalises the DSCB to unity.

The parametrisation of the detector resolution function as a DSCB function
(whose parameters potentially depend on the invariant mass)
is largely inspired by the procedure adopted in ATLAS analyses~\cite{grevtsov}.
The exact description of the DSCB parameters is obtained using narrow-width signal samples generated at a range of mass points;
such samples assume sharply-peaked truth-level distributions, $\mathcal{T}(q)\sim\delta(q-m_X)$,
and it thus follows from eq.~\eqref{convolution-dr} that the reconstructed narrow-width signal distributions should be well described by the chosen detector resolution function.
Given the parametrised DSCB resolution function obtained in this way,
a reconstructed lineshape can then be modelled according to eqs.~\eqref{convolution-dr} and~\eqref{reco-lineshape},
for any appropriate truth-level lineshape $\mathcal{T}(q)$,
and for $\text{DR}(q)$ corresponding to the DSCB function with parameters evaluated at the resonance mass, $q=m_X$.

That being said, we point out that a more sophisticated, and fully correct,
approach would be that of modelling detector effects by means of a convolution analogous to that of eq.~\eqref{convolution-dr},
but where $\text{DR}(q)$ is replaced by a detector response function that depends on several variables
(e.g.~the photon transverse momenta in addition to the pair invariant
mass),
all of which would be integrated over.
Such a convolution would account for both the resolution and the efficiency contributions to the $q$-dependence manifest in the observed reconstructed lineshape.
Nevertheless, we will adopt the simpler parametrisation of eqs.~\eqref{convolution-dr} and~\eqref{reco-lineshape} for our current study.

\subsection{Verification of the convolution description}
\label{sec:detector-subsec:large-width-signal}

In this section, we show that eqs.~\eqref{convolution-dr} and~\eqref{reco-lineshape} can be used to describe reconstructed large-width signal,
background-only, and interference-only events.
We demonstrate that the description works separately for each of these components,
using events that are reconstructed from the corresponding truth-level samples of figs.~\ref{fit-pm-signal}, \ref{fit-background} and~\ref{fit-interference}.
In each case, the truth-level description that enters the convolution is taken to be that of eqs.~\eqref{pm-signal} (in conjunction with eq.~\eqref{glumin}), \eqref{bg-func}, and \eqref{dxs-interf}, respectively,
with all parameter values fixed to those reported from their corresponding fits of the truth-level data.
A parametrisation of the DSCB function was extracted by using narrow-width signal samples,
with its parameter values evaluated at the known resonance mass ($m_X=400$\,GeV),
and a numerical value for $N_\text{DSCB}$ computed and implemented accordingly to ensure normalisation of the DSCB to unity.

Requiring a minimum photon $p_T$ during the reconstruction of diphoton events
introduces a distortion (dependent on the physics model) in the invariant mass lineshape
that is most prominent at the lower end of its spectrum.
Thus, to model the lineshape of a reconstructed invariant mass distribution in its entirety,
one requires accurate knowledge of the reconstruction efficiency and how it varies with the invariant mass.
However, since this effect is expected to manifest mainly in the low mass region,
one can also posit that the approximation of a $q$-independent efficiency should suffice to describe reconstructed lineshapes above some invariant mass threshold.
In such a case, only a single free parameter remains in the parametrised description,
namely that of an overall normalisation factor corresponding to the reconstruction efficiency of the diphoton events, $\varepsilon$.
Note that in general this will differ for signal, interference and background-only events,
since the rejection of events below a hard $p_T$ threshold removes a different fraction of events from each component,
owing to their different photon $p_T$ distributions.

\begin{figure}
	\centering
	\includegraphics[width=0.32\textwidth]{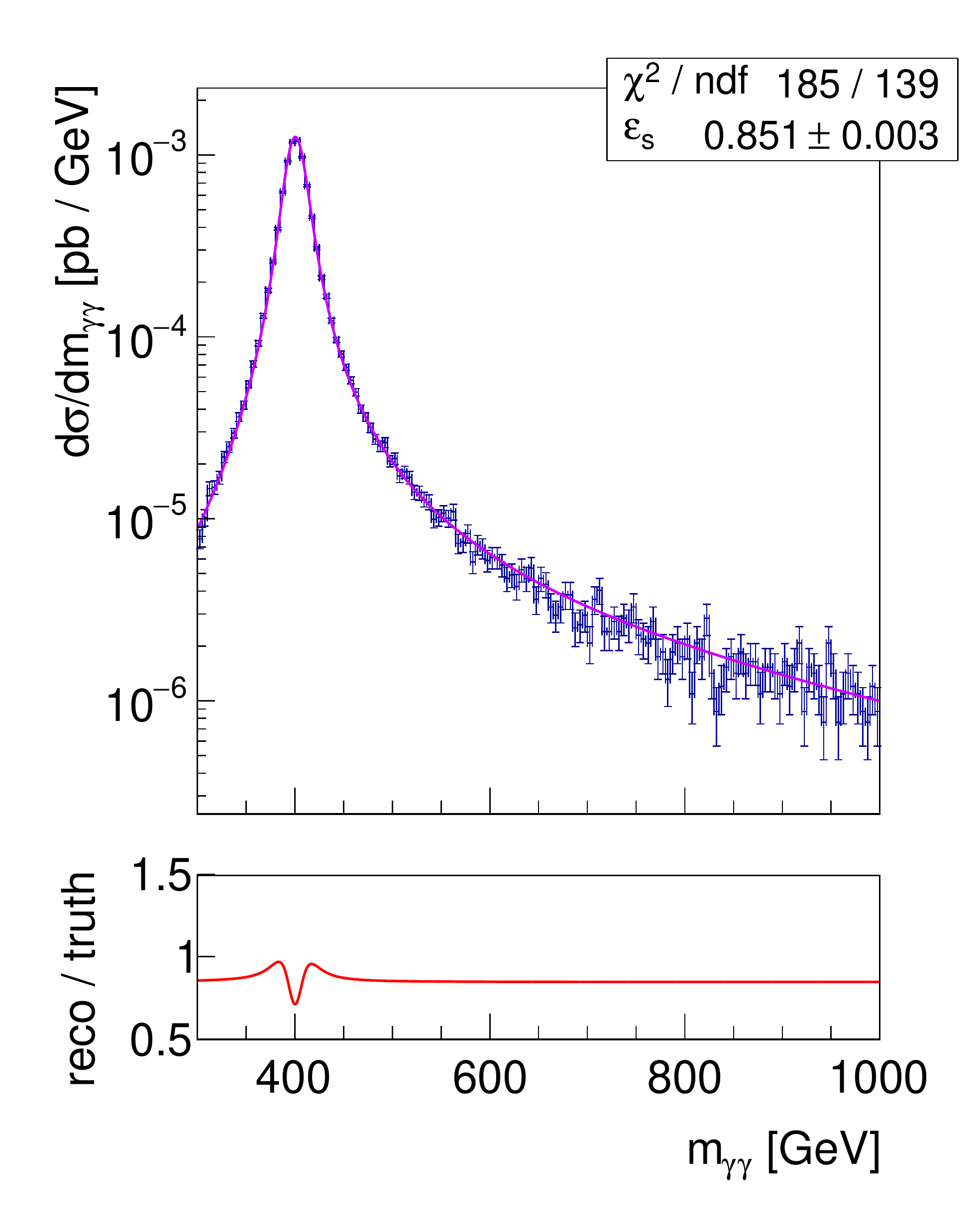}
	\includegraphics[width=0.32\textwidth]{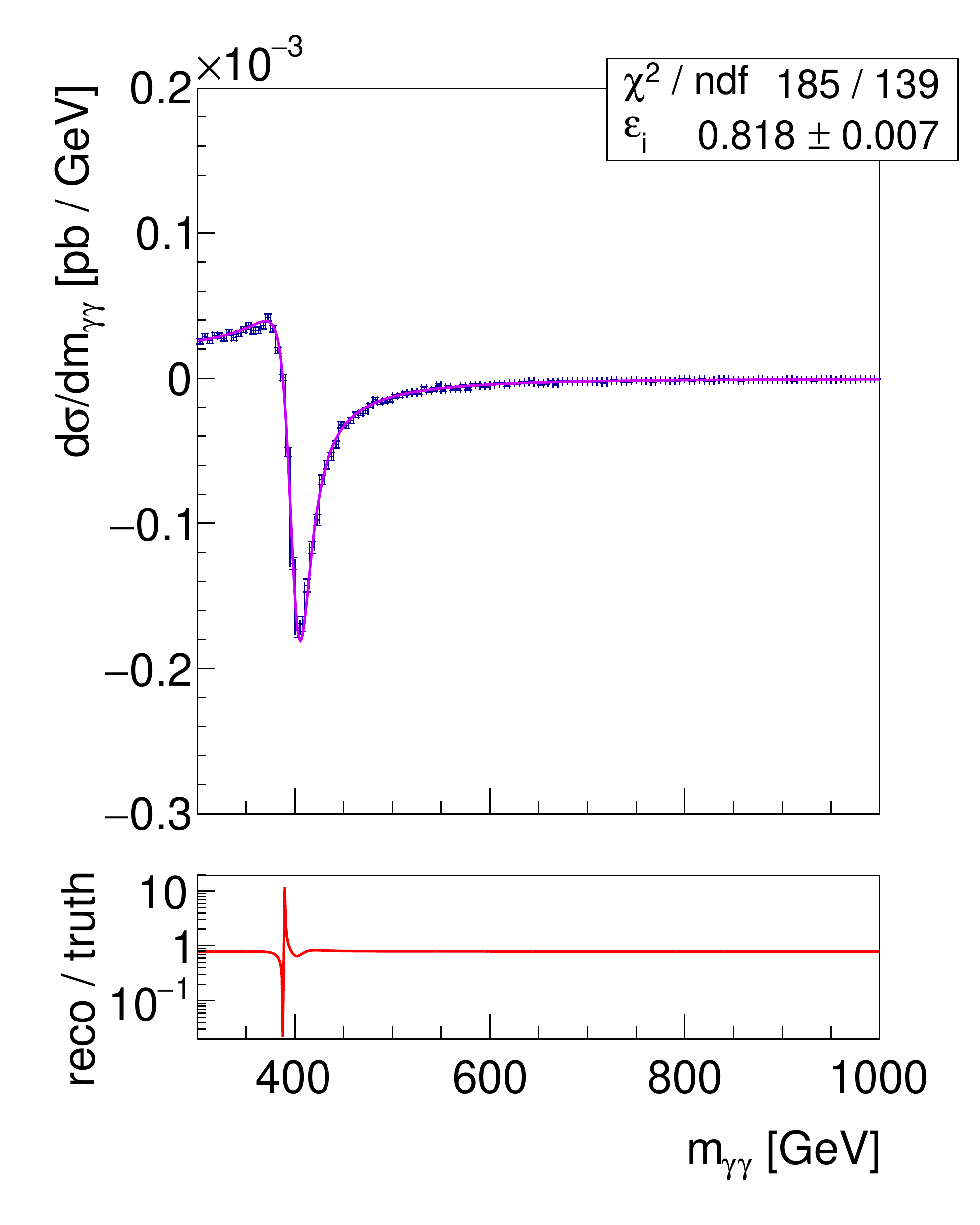}
	\includegraphics[width=0.32\textwidth]{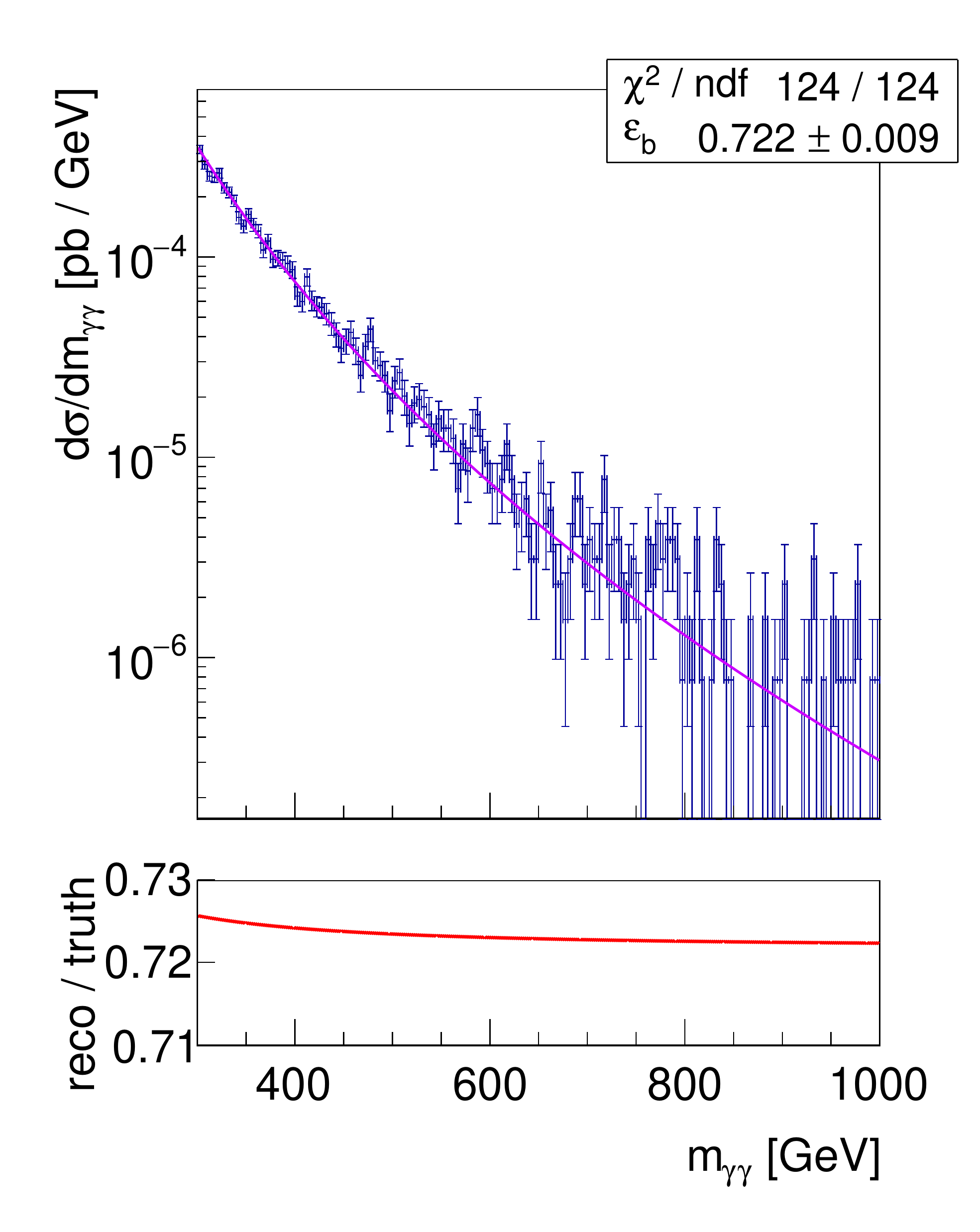}
	\caption{
		Fits to reconstructed signal, interference and background distributions,
		using a convolution of their respective truth-level analytic functions with a DSCB detector resolution parametrisation.
		The fits are performed over the 300--1000\,GeV invariant mass window.
		The free parameter in each of the fits is $\varepsilon$,
		the event reconstruction efficiency.
		The ratio of the best-fit function and its corresponding truth-level distribution is plotted beneath each result.
	}
	\label{reco-pm-convolution-fits}
\end{figure}

We test this assumption by performing one-parameter fits over the 300--1000\,GeV invariant mass range.
Results are presented in fig.~\ref{reco-pm-convolution-fits}.
The ratio of the reconstructed (i.e. fitted) function and the corresponding truth-level distribution is visualised beneath each plot;
in particular, we note that in the case of the background, this ratio varies slowly with the invariant mass.
We find that the simple convolution provides a good description of the reconstructed lineshape for all of the components,
and the approximation of $q$-independent efficiencies has not prevented a good fit being found in each result.
There is approximately a 15\% discrepancy, at most, between the efficiencies reported for each component,
which stems from the $p_T$ selection.
In a realistic analysis, and particularly as one begins to consider TeV-scale resonances,
this difference will diminish and most likely become sub-dominant in comparison to other sources of uncertainty.
For this reason, we expect that it will be valid firstly to adopt a $q$-independent reconstruction efficiency approximation,
and secondly, to neglect the difference between component efficiencies in realistic search scenarios.

\subsection{Convolving the template functional form}
\label{sec:detector-subsec:template-reco}

We now proceed to apply the convolution method of incorporating detector effects to the general functional form.
Note that the template description of eq.~\eqref{xsecrat1T} is formulated as a ratio of full to background-only differential cross sections.
However, we will again treat the (reconstructed) background as a quantity that is known exactly,
i.e. with zero associated uncertainty.
As was remarked in section~\ref{sec:toytruth}, this implies that it is equivalent to perform fits of the full differential cross sections,
and in this case, it allows us to work with the quantity that is consistent with eqs.~\eqref{convolution-dr} and~\eqref{reco-lineshape}.
	
To further simplify the procedure, we make the following additional assumptions,
based on the discussion of the previous section:
\begin{enumerate}		
	\item the reconstruction efficiencies of the various components are approximately equal,
	$\varepsilon_s\approx\varepsilon_b\approx\varepsilon_i\equiv\varepsilon$;
	
	\item the dependence of the efficiency on the invariant mass is negligible;
	
	\item the convolution of the background with the detector resolution function produces negligible change in its shape:
	\begin{equation} \label{reco-bg}
		F_B^\text{truth}(q) \approx \left(F_B^\text{truth}*\text{DR}\right)(q) = \frac{1}{\varepsilon}\,F_B^\text{reco}(q)\,,
	\end{equation}
	where $F_B^\text{truth}(q)$ and $F_B^\text{reco}(q)$ denote the truth and reconstructed backgrounds, respectively.
\end{enumerate}
Along with the assumption of a perfect reconstructed background description,\footnote
{Note, however, that this assumption differs from those enumerated in that it does not invalidate the equations we present in this section;
uncertainties on the background can always be incorporated by profiling over the associated nuisance parameters.}
these items constitute the assumptions underpinning the procedure we describe in this section.
It is important to point out that these assumptions need only hold within the invariant mass window over which a fit is to be performed.
	
Under the first assumption, one can exploit the linearity of the convolution operator
to convolve the full reconstructed differential cross section, and obtain what follows:
\begin{equation} \label{rc-simple-convolution}
	\frac{d\sigma_\text{full}^\text{reco}}{dq}(q) = \varepsilon \left(\frac{d\sigma_\text{full}^\text{truth}}{dq} * \text{DR}\right)(q)
	= \varepsilon \left(F_B^\text{truth}\, \frac{\left|\bar{A}\right|^2}{\left|\bar{B}\right|^2} * \text{DR}\right)(q) \,,
\end{equation}
where the rightmost equality is made using eq.~\eqref{rc-dxs},
with an according equality between $F_B^\text{truth}(q)$ and the $F_B(q)$ of eq.~\eqref{rc-dxs}.
The second and third assumptions can then be used to write eq.~\eqref{rc-simple-convolution}
in terms of only the reconstructed background:
\begin{equation} \label{reco-template}
	\frac{d\sigma_\text{full}^\text{reco}}{dq}(q)
	= \left(F_B^\text{reco}\, \frac{\left|\bar{A}\right|^2}{\left|\bar{B}\right|^2} * \text{DR}\right)(q) \,.
\end{equation}
In this way, the efficiency factor is absorbed into the description of the background,
and one obtains a parametrisation that can be expressed without pertaining to any additional free parameters:
a fit to reconstructed events can be performed using the same set of parameters as those in the truth scenario,
namely those of the \textbf{T}$_R$ or \textbf{T}$_a$ sets.
Note that returning to the ratio regime (cf. eq.~\eqref{xsecrat1T})
is a simple matter of dividing this equation through by the reconstructed background differential cross section.
	
A few comments are in order here.
Firstly, eq.~\eqref{reco-template} stems from eq.~\eqref{rc-simple-convolution}
if one can parametrise the background lineshape with the same functional form before and after a convolution with detector effects.
If the effect of the convolution is not entirely negligible,
the parameters relevant to the two scenarios will generally be different.
One might wonder, then, if detector effects are not double counted on the r.h.s. of eq.~\eqref{reco-template}.
We posit that this is not the case, since these must largely cancel in the ratio $F_B^\text{reco}(q)/\varepsilon$
(if that were not the case, the approximate equality between the leftmost and rightmost sides of eq.~\eqref{reco-bg} simply could not hold).
Secondly, if the convolution with detector effects is so significant for background lineshapes that eq.~\eqref{reco-bg} cannot be correct,
one needs to instead use eq.~\eqref{rc-simple-convolution};
this is acceptable, but it entails an increased dependence on theoretical predictions,
which would likely increase the overall systematics.
Finally, in the case of significant departure from all of the assumptions made,
it might be preferable to pursue the alternative procedure altogether,
namely that of fitting eq.~\eqref{xsecrat1T} directly to (the ratio of) unfolded datasets.

We perform a straightforward test of eq.~\eqref{reco-template} by fitting it to a sample of reconstructed events,
and comparing the extracted result to that of the corresponding truth-level fit.
An agreement between these two sets of results implies that eq.~\eqref{reco-template} has correctly characterised the underlying physics via a fit of the reconstructed distribution.
Following the discussion of the previous section,
we conduct this test using MC samples generated at a larger resonance mass of $m_X=800$\,GeV, with width $\Gamma_X/m_X=5\%$,
to diminish the impact of approximating equal selection efficiencies between the signal, interference, and background events.
The comparison is demonstrated using the \textbf{T}$_R$ parameter set.

\begin{figure}
	\centering
	\includegraphics[width=0.49\textwidth]{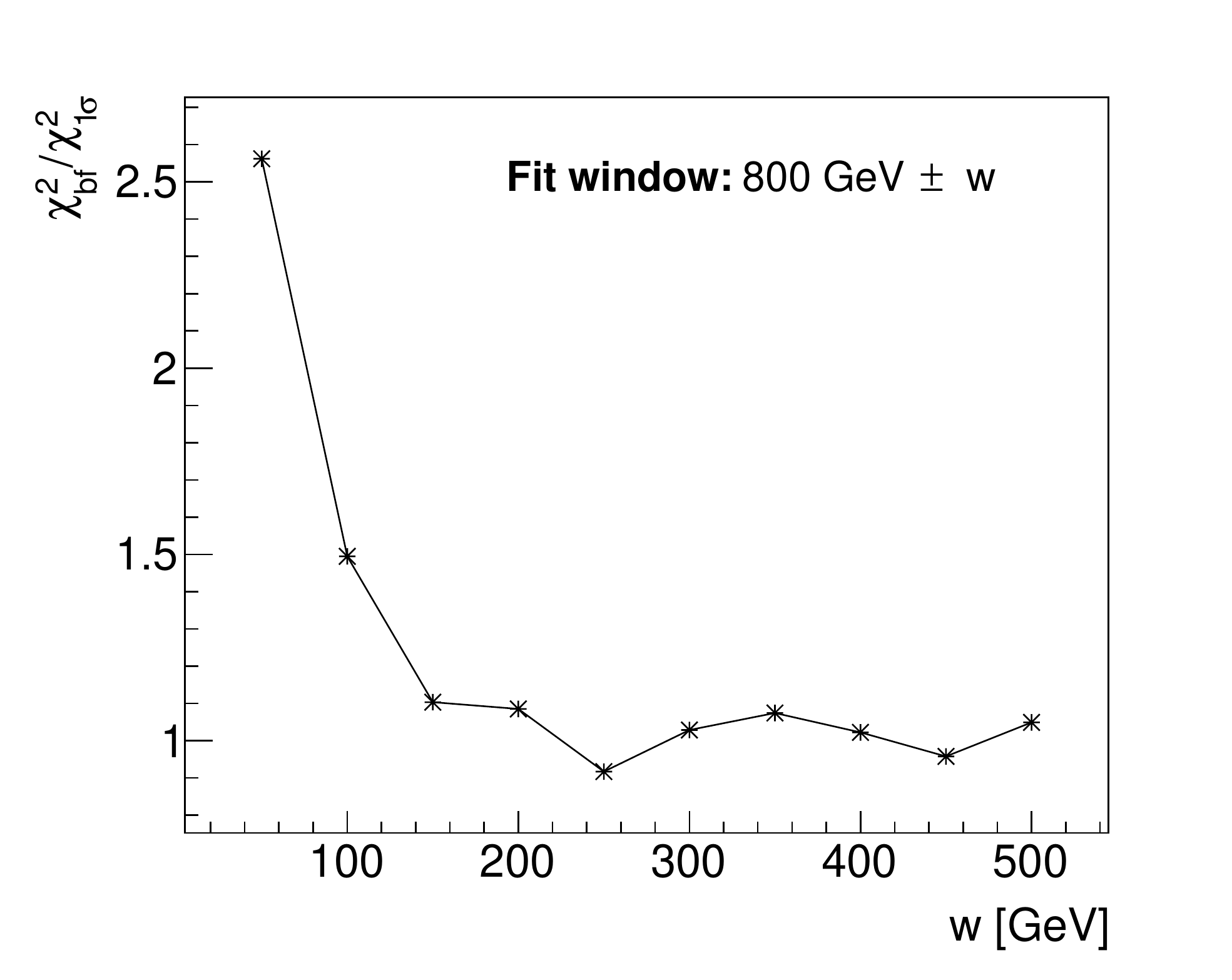}
	\includegraphics[width=0.49\textwidth]{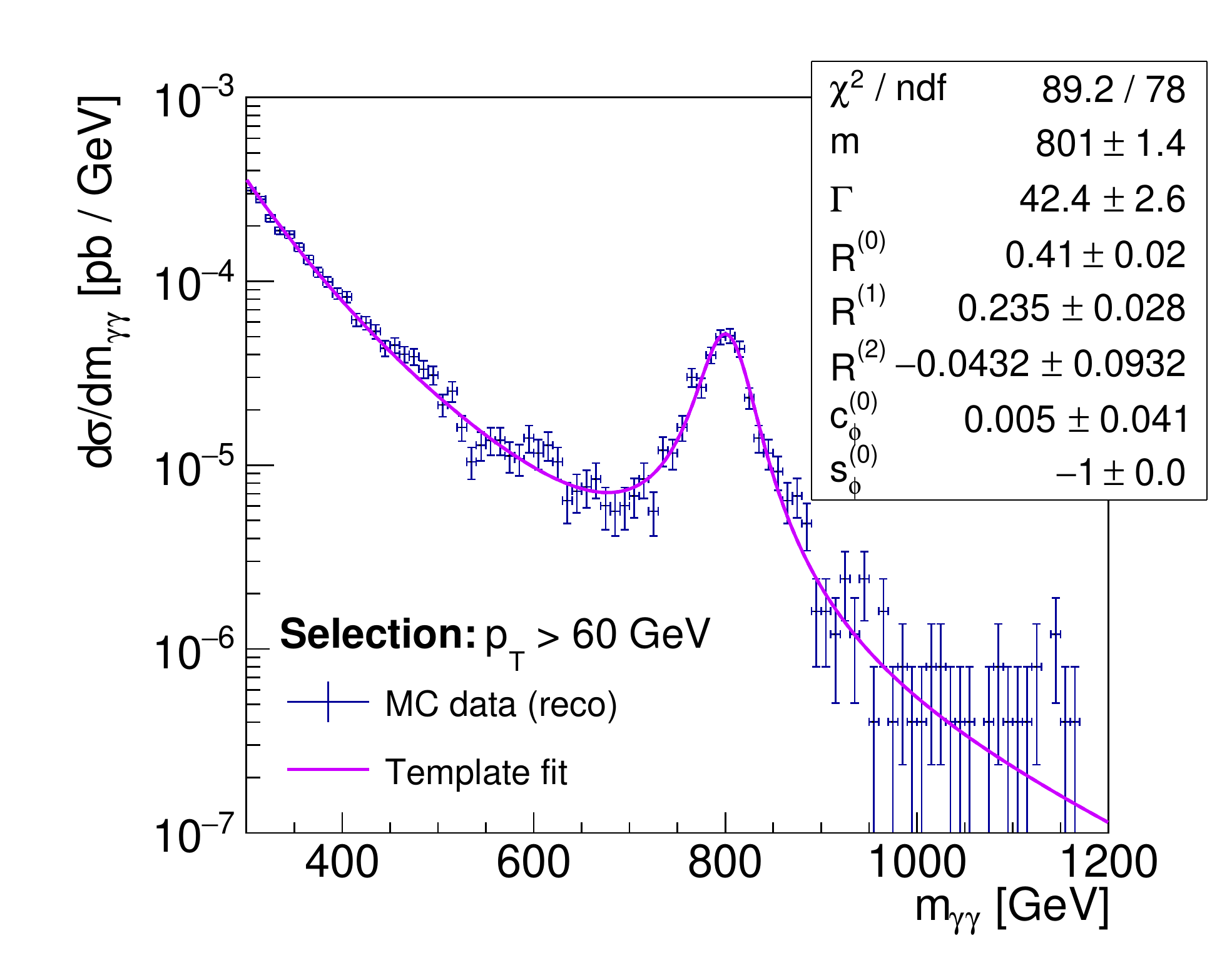}
	\caption{
		Left: the variation of $\chi^2_\text{bf}/\chi^2_{1\sigma}$ against the width of the fit window,
		for fits of the convolved second-order template functional form to reconstructed \madgraph{} events.
		Right: A fit of the convolved second-order template functional form to reconstructed \madgraph{} events over the 300--1200\,GeV mass range.
	}
	\label{rc-reco-fits-root}
\end{figure}

To determine a suitable mass window for estimating the parameters,
preliminary fits were performed using ROOT~\cite{Antcheva:2009zz} across a range of invariant mass windows centred on the true resonance mass.
The left panel of fig.~\ref{rc-reco-fits-root} shows the ratio of best-fit ($\chi^2_\text{bf}$) to 1$\sigma$ cut-off ($\chi^2_{1\sigma}$) chi-squared values against the fit windows tested.
Unlike the results of fig.~\ref{auto-5pc}, a monotonically increasing ratio with the fit window is not seen.
This is due to the stochastic nature of MC samples, in contrast to Asimovs that perfectly capture the underlying theory.\footnote
{In practice, the implication of this is simply that one loses the freedom to fit relatively small mass windows using non-Asimov datasets.}
To prevent random fluctuations from yielding volatile results, a sufficiently large fit window will thus be required;
for the current case, the ratio of chi-squares reaches a minimum at $w=250$\,GeV,
though the distribution effectively plateaus past $w\approx150$\,GeV and does not worsen significantly for larger windows.
This suggests a resonance small enough that a second-order approximation sufficiently characterises the entire range of its invariant mass lineshape.
For samples with larger signals, one would expect to instead see a relationship of the fit quality with a clear minimum in $w$.
The right panel of fig.~\ref{rc-reco-fits-root} shows that the \textbf{T}$_R$ parameter set is indeed able to fit the reconstructed distribution well over a large 300--1200\,GeV mass range.
However, for the purpose of parameter estimation,
it is still advisable to choose a restricted fit window
to avoid the possibility of parameter values drifting to compensate for the missing higher order terms of the functional form of eq.~\eqref{xsecrat1T}.

\begin{figure}
	\centering
	\includegraphics[width=0.49\textwidth]{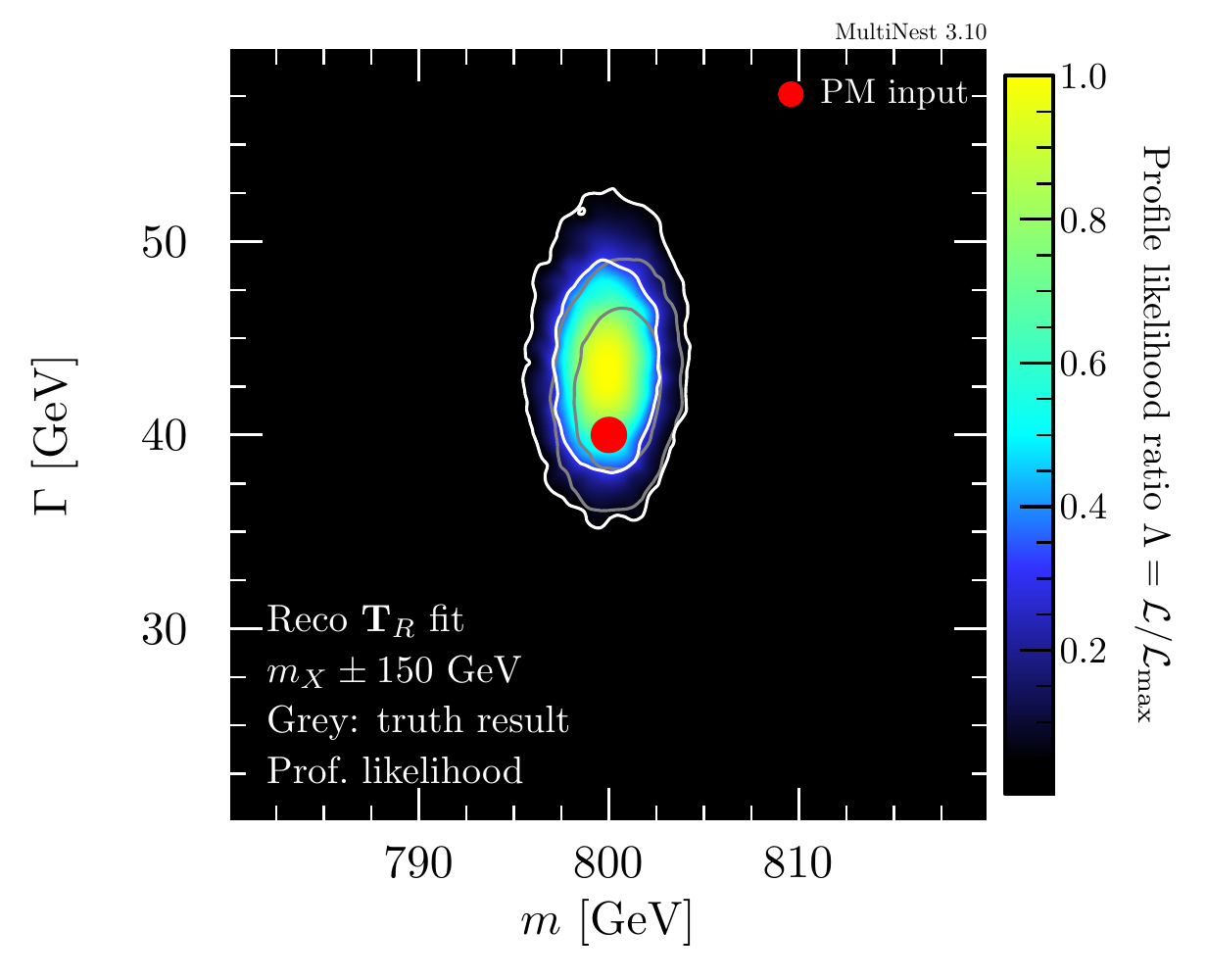}
	\includegraphics[width=0.49\textwidth]{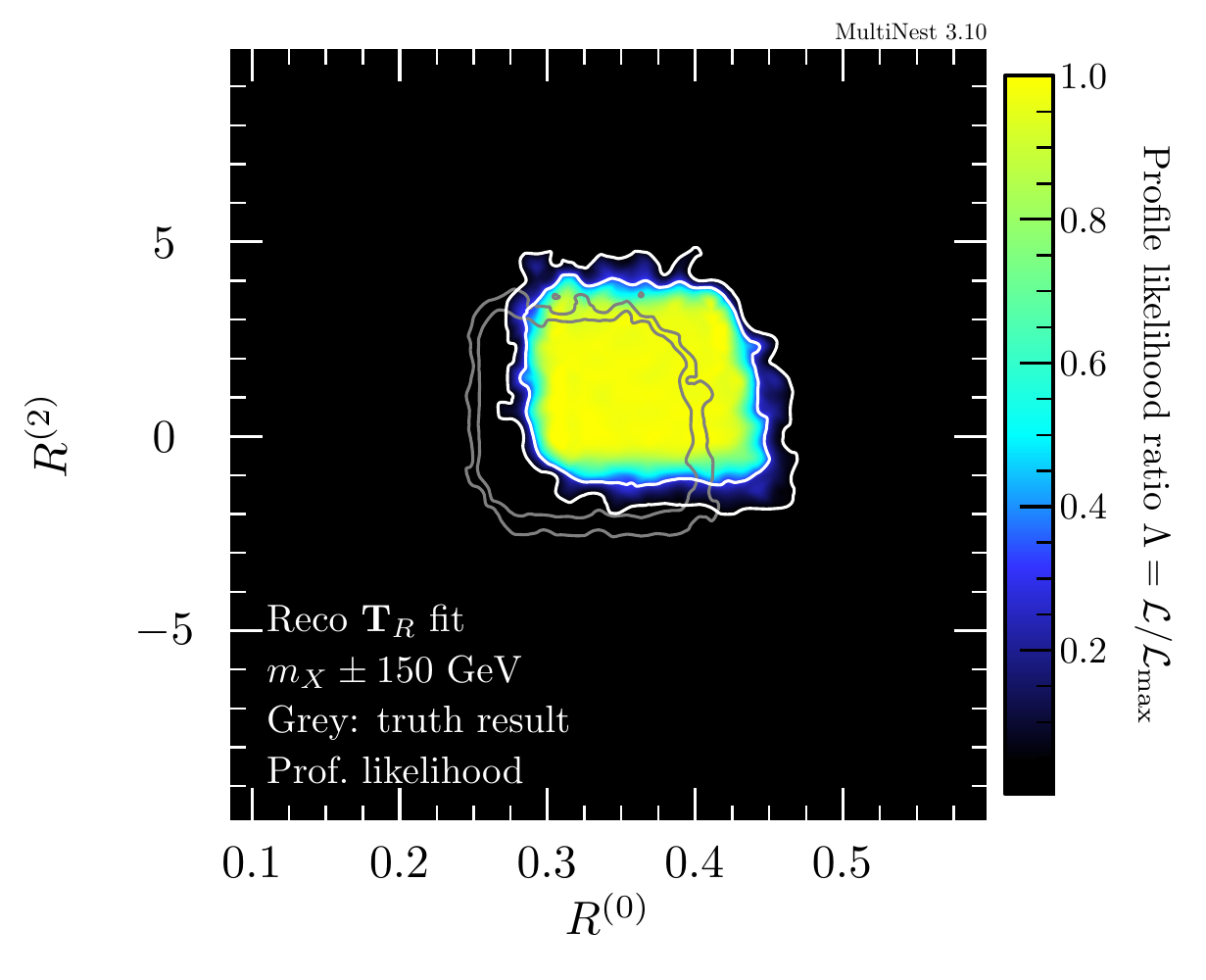}\\
	\includegraphics[width=0.49\textwidth]{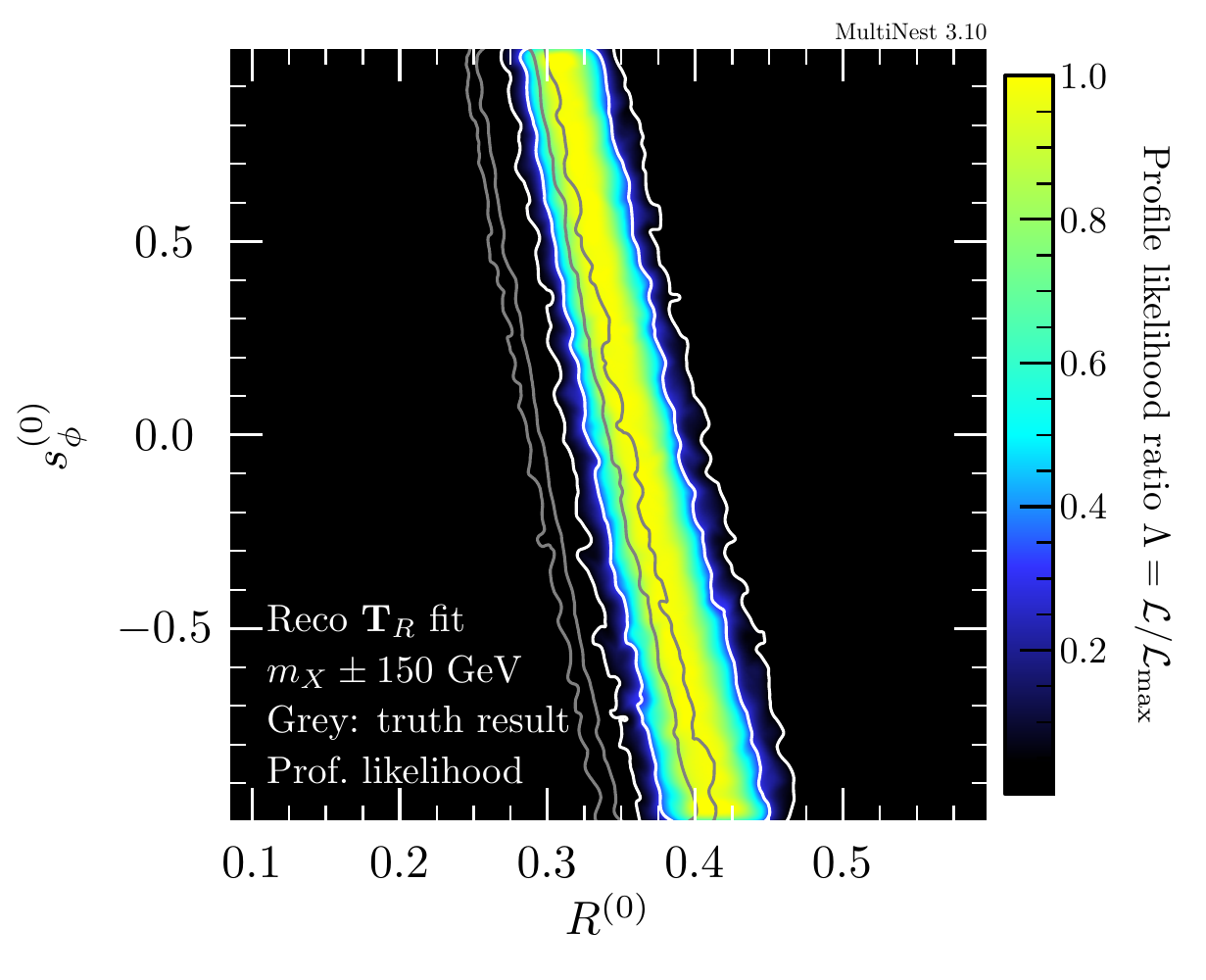}
	\includegraphics[width=0.49\textwidth]{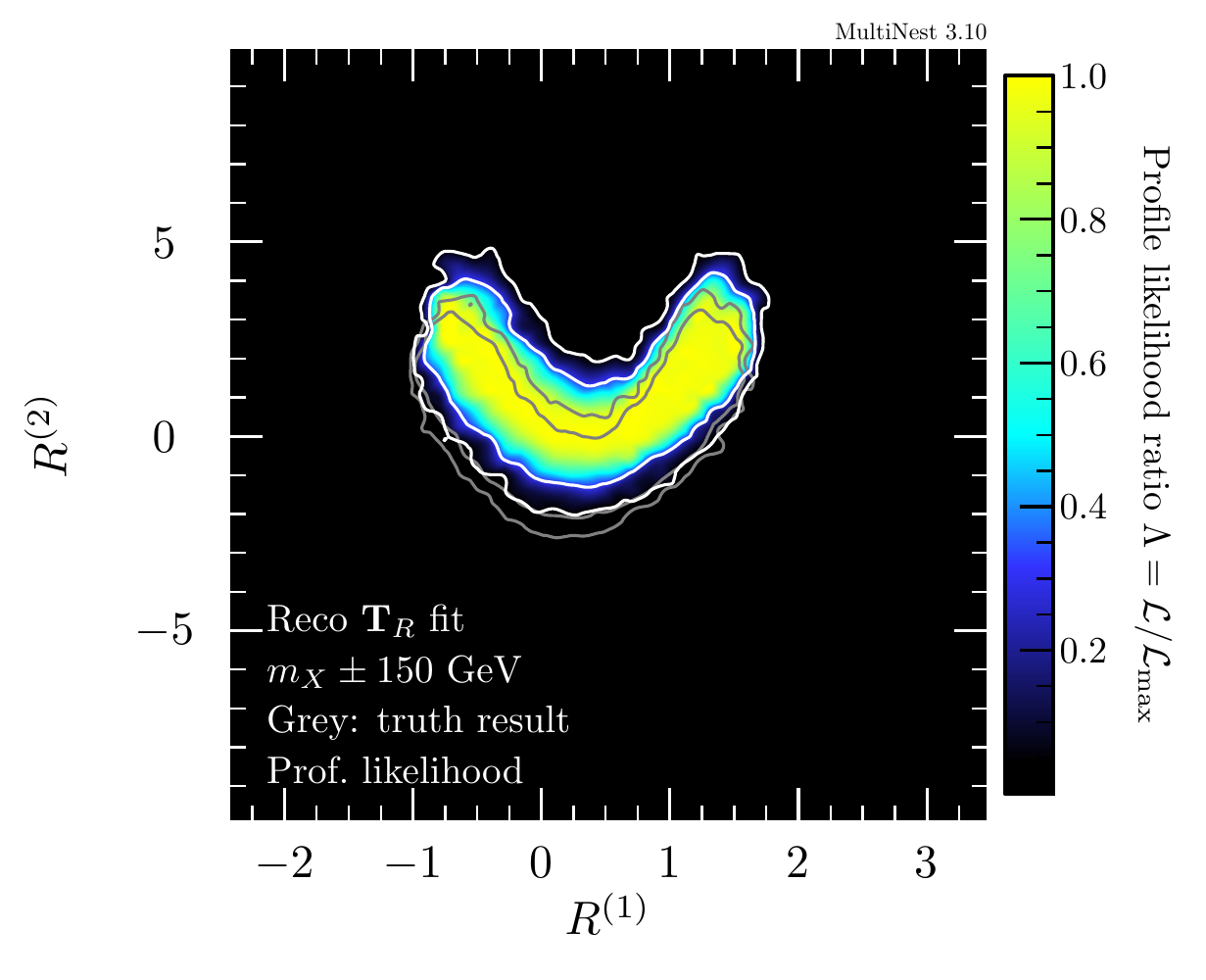}\\
	\includegraphics[width=0.49\textwidth]{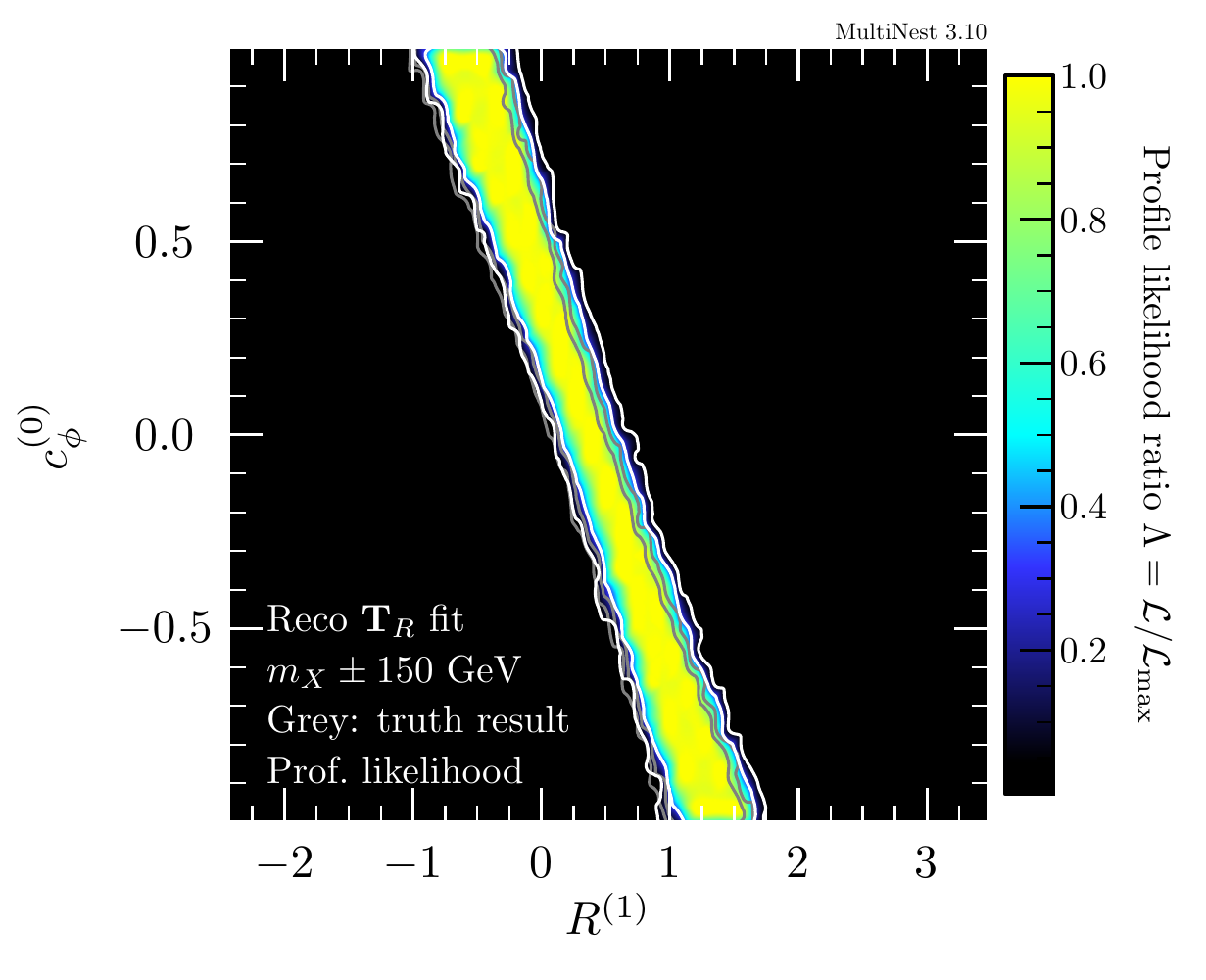}
	\includegraphics[width=0.49\textwidth]{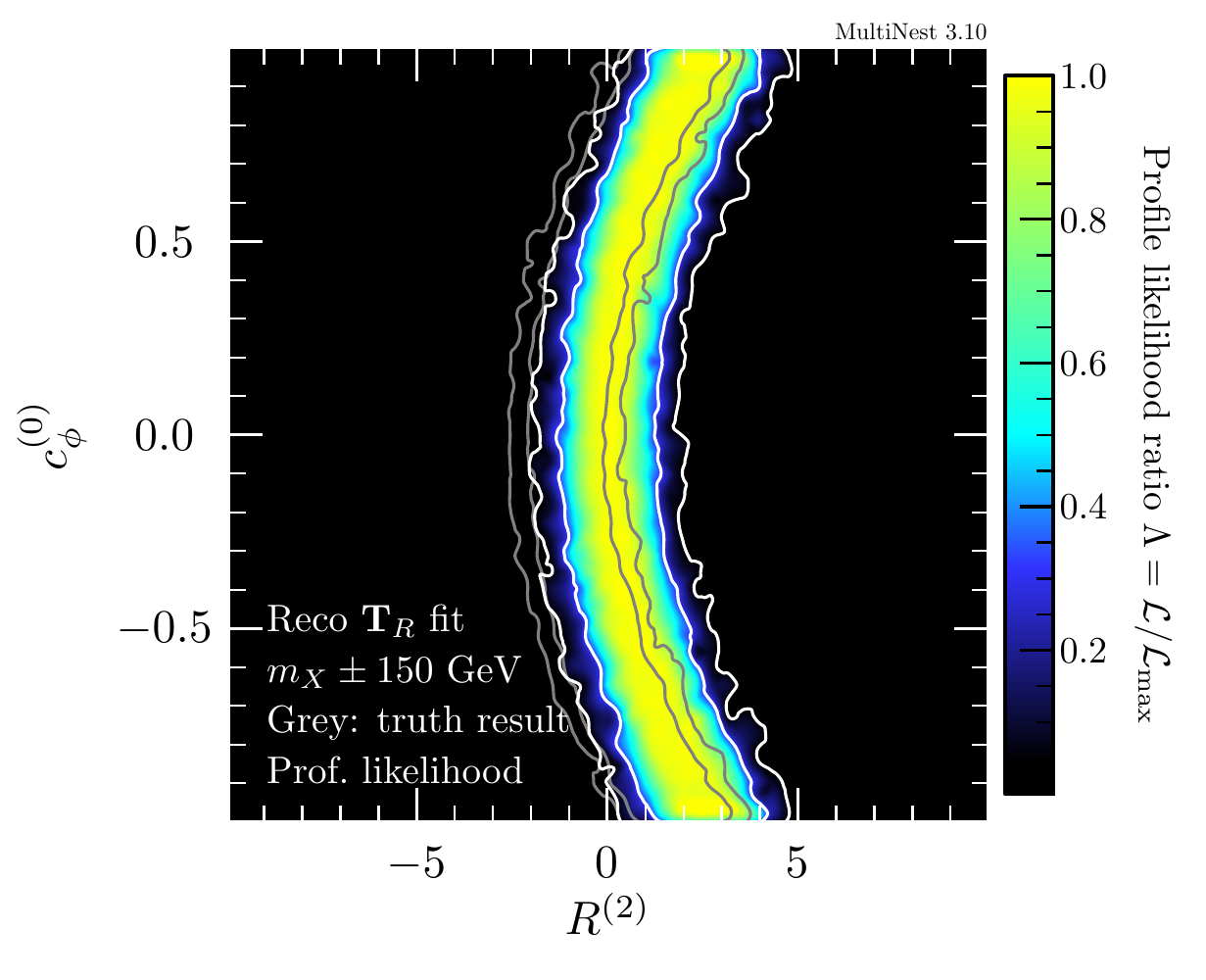}
	\caption{
	    Two dimensional profile likelihood plots of the \textbf{T}$_R$ parameters,
		for a fit of reconstructed diphoton events over the 650--950\,GeV invariant mass window.
		The grey boundaries correspond to 1$\sigma$ and 2$\sigma$ contours at the truth level.
		The physical constraint $-1\le \cpz,\, \spz \le1$ has been imposed.
	}
	\label{reco_rc2mc_pm150_physical}
\end{figure}

Choosing $w=150$\,GeV, we perform a fit of the \textbf{T}$_R$ parameters over the 650--950\,GeV mass window using \textsc{MultiNest}.
A fit of the truth-level template was also performed on the truth events.
A selection of resulting profile likelihood distributions in \textbf{T}$_R$ parameter space are presented and compared in fig.~\ref{reco_rc2mc_pm150_physical}.
The main set of results, represented by the colour gradient and white contours,
correspond to the fit to reconstructed events,
while truth-level contours are drawn in grey.
A red circle marks the expected mass and width in the corresponding parameter plane.

The reconstructed fit is able to recover the input mass and width values within 1$\sigma$ confidence.
We find the truth and reconstructed-level results to agree well,
with the white and grey contours largely overlapping in all of the parameter planes.
The largest discrepancies are seen in the $\ro$ and $\rz$ parameters;
the reconstructed fit yields slightly larger values for these,
with an approximately 10\% and 30\% increase for $\ro$ and $\rz$, respectively,
when comparing the truth-level contours to the reconstructed results.
This behaviour arises from approximating equal reconstruction efficiencies for the signal, interference and background components.
For the current benchmark point, where $\varepsilon_s\gtrsim \varepsilon_i\gtrsim \varepsilon_b$,
the approximation introduces a small enhancement to $R(q^2)$,
which manifests most noticeably in the zeroth and second order expansion coefficients in the current set of results.
However, as we have noted in section~\ref{sec:detector-subsec:large-width-signal},
such effects are expected to become negligible in realistic searches.
Our results thus show that the description of eq.~\eqref{reco-template} can be used to extract the general parameters by means of a fit directly to reconstructed events.

\section{Conclusion}
\label{sec:conclusions}

The top-down approach typically employed in the presentation of LHC resonance 
search results, while constituting a straightforward procedure in the scenario 
where there is strong motivation to believe {\em a priori} in a particular 
model, becomes less ideal for reporting the findings of a general search 
for BSM resonances. In the latter case, it is desirable to present results 
in a form that is sufficiently generalisable to any BSM models, such 
that a given theorist is able to relate them {\em a posteriori} to their 
particular model of interest.

In this paper, we have proposed a procedure that is apt to the presentation
of search results in the form advocated above. This procedure assumes
that data can be organised as the differential distribution in the 
invariant mass of the would-be resonance, e.g.~as is reconstructed by means 
of the four-momenta of its decay products, and is based on employing a
functional form for the lineshape of such a distribution that stems
from general quantum field theory considerations and is fully
model-independent. A definite advantage of the procedure is that it
allows the characterisation of the resonance and of its interference 
pattern with the Standard Model background, given in terms of the free 
parameters that enter the lineshape; the method is shown to work 
in the case of null search results as well.

The core of the procedure is described in section~\ref{sec:template}, and its results
in terms of the lineshape parameters are presented in section~\ref{sec:toytruth-subsec:proflike} using
as a test case the simple physics model that is summarised in section~\ref{sec:bm}.
While this is the essence of the method we have proposed, we have also
discussed various ancillary techniques that supplement it, and which can
either be adopted as they are, or replaced by alternative mechanisms if 
the latter will be deemed more convenient in certain working conditions.
More specifically, the lineshape-parameter determination works well if
one knows with a good approximation the value of the pole mass of the
resonance. In section~\ref{sec:toytruth-subsec:procedure} we have shown how a preliminary determination
of such a value can be achieved; however, nothing prevents experiments from 
pursuing a different determination strategy, whose result can then be used
in the context of our characterisation procedure. Likewise, when experimental
results are presented in terms of the lineshape parameters, any theorist
can check the compatibility of their models with the data by {\em computing}
the same parameters by following exactly the same procedure as is done by
experiments. However, in section~\ref{sec:toytruth-subsec:pmclosure} we have also discussed an alternative
approach, which employs the lineshape parameters extracted by the
experiments to determine allowed regions in the parameter space of
the theoretical model one wishes to test. Finally, while the procedure
we advocate would typically be applied to unfolded data, we have shown
in section~\ref{sec:detector} how, under certain conditions, it can be employed directly
on raw data, by means of a suitable and relatively simple description
of detector effects.

We have explicitly shown how the general lineshape functional form
can be parametrised by means of two different parameter sets that
are characterised by different features---one being more tightly
constrained but with a less direct physical interpretation, the other
being more closely connected with an underlying theoretical description
but liable to have flat directions in the parameter space. Ultimately,
the choice of which set to adopt depends on the emphasis that the analysers
will want to give to their searches and/or tests. In view of that, we 
conclude by pointing out that the two parameters sets we have discussed
constitute minimal options that can be systematically extended if necessary,
by following the methodology presented in section~\ref{sec:template}.

\section*{Acknowledgements}

The authors thank Fabio Maltoni and Yee Chinn Yap for their helpful comments and contributions during the early stages of the work.
SF thanks the CERN TH division for hospitality during the course of this work.
LR and MW are partly supported by the Australian Research Council (ARC) Discovery Project grant DP180102209.
ET is supported by an Australian government Research Training Program scholarship.
AGW is supported by the ARC Centre of Excellence for Particle Physics at the Terascale (CoEPP) (CE110001104) and the Centre for the Subatomic Structure of Matter (CSSM).
Ti$k$Z-Feynman~\cite{Ellis:2016jkw} and \texttt{pippi}~\cite{Scott:2012qh} were used to produce plots presented in this paper.

\bibliographystyle{JHEP}
\bibliography{bibliography.bib}

\end{document}

%% file: newCommands.tex
\def\abs#1{\left|#1\right|}

\newcommand{\qt}{q^2}

\newcommand{\bA}{\bar{A}}
\newcommand{\bB}{\bar{B}}
\newcommand{\hB}{B}
\newcommand{\hR}{R}
\newcommand{\hRz}{R^{(0)}}
\newcommand{\hRzt}{R^{{(0)}^2}}
\newcommand{\hRo}{R^{(1)}}
\newcommand{\hRot}{R^{{(1)}^2}}
\newcommand{\hRt}{R^{(2)}}
\newcommand{\ha}{a}

\newcommand{\cpz}{c_\phi^{(0)}}
\newcommand{\cpo}{c_\phi^{(1)}}

\newcommand{\spz}{s_\phi^{(0)}}
\newcommand{\spo}{s_\phi^{(1)}}
\newcommand{\spt}{s_\phi^{(2)}}

\newcommand{\madgraph}{\textsc{MadGraph5\_aMC@NLO}}
\newcommand{\pythia}{\textsc{Pythia8}}
\newcommand{\delphes}{\textsc{Delphes~3.4.1}}

\newcommand{\lagr}{\mathcal{L}}

\newcommand{\bsym}{\boldsymbol}

\newcommand{\ro}{R^{(0)}}
\newcommand{\ri}{R^{(1)}}
\newcommand{\rz}{R^{(2)}}




%% file: interfPaper.bbl
\providecommand{\href}[2]{#2}\begingroup\raggedright\begin{thebibliography}{10}

\bibitem{Aad:2012tfa}
{\scshape ATLAS} collaboration, \emph{{Observation of a new particle in the
  search for the Standard Model Higgs boson with the ATLAS detector at the
  LHC}}, \href{https://doi.org/10.1016/j.physletb.2012.08.020}{\emph{Phys.
  Lett. B} {\bfseries 716} (2012) 1}
  [\href{https://arxiv.org/abs/1207.7214}{{\ttfamily 1207.7214}}].

\bibitem{Chatrchyan:2012ufa}
{\scshape CMS} collaboration, \emph{{Observation of a New Boson at a Mass of
  125 GeV with the CMS Experiment at the LHC}},
  \href{https://doi.org/10.1016/j.physletb.2012.08.021}{\emph{Phys. Lett. B}
  {\bfseries 716} (2012) 30} [\href{https://arxiv.org/abs/1207.7235}{{\ttfamily
  1207.7235}}].

\bibitem{Djouadi:2016ack}
A.~Djouadi, J.~Ellis and J.~Quevillon, \emph{{Interference effects in the
  decays of spin-zero resonances into $\gamma \gamma$ and $ t\overline{t} $}},
  \href{https://doi.org/10.1007/JHEP07(2016)105}{\emph{JHEP} {\bfseries 07}
  (2016) 105} [\href{https://arxiv.org/abs/1605.00542}{{\ttfamily
  1605.00542}}].

\bibitem{Hespel:2016qaf}
B.~Hespel, F.~Maltoni and E.~Vryonidou, \emph{{Signal background interference
  effects in heavy scalar production and decay to a top-anti-top pair}},
  \href{https://doi.org/10.1007/JHEP10(2016)016}{\emph{JHEP} {\bfseries 10}
  (2016) 016} [\href{https://arxiv.org/abs/1606.04149}{{\ttfamily
  1606.04149}}].

\bibitem{BuarqueFranzosi:2017jrj}
D.~Buarque~Franzosi, E.~Vryonidou and C.~Zhang, \emph{{Scalar production and
  decay to top quarks including interference effects at NLO in QCD in an EFT
  approach}}, \href{https://doi.org/10.1007/JHEP10(2017)096}{\emph{JHEP}
  {\bfseries 10} (2017) 096}
  [\href{https://arxiv.org/abs/1707.06760}{{\ttfamily 1707.06760}}].

\bibitem{Djouadi:2019cbm}
A.~Djouadi, J.~Ellis, A.~Popov and J.~Quevillon, \emph{{Interference effects in
  $ t\overline{t} $ production at the LHC as a window on new physics}},
  \href{https://doi.org/10.1007/JHEP03(2019)119}{\emph{JHEP} {\bfseries 03}
  (2019) 119} [\href{https://arxiv.org/abs/1901.03417}{{\ttfamily
  1901.03417}}].

\bibitem{Xia:2019opu}
L.-G. Xia, \emph{{F2F, a model-independent method to determine the mass and
  width of a particle in the presence of interference}},
  \href{https://arxiv.org/abs/1908.04048}{{\ttfamily 1908.04048}}.

\bibitem{yee}
Y.~C. Yap, \emph{{Search for New Physics with Two Photons in the Final State
  with the ATLAS Detector}}, Ph.D. thesis, Paris Diderot University, 2017.

\bibitem{Bertone:2013vaa}
V.~Bertone, S.~Carrazza and J.~Rojo, \emph{{APFEL: A PDF Evolution Library with
  QED corrections}},
  \href{https://doi.org/10.1016/j.cpc.2014.03.007}{\emph{Comput. Phys. Commun.}
  {\bfseries 185} (2014) 1647}
  [\href{https://arxiv.org/abs/1310.1394}{{\ttfamily 1310.1394}}].

\bibitem{Ball:2013hta}
{\scshape NNPDF} collaboration, \emph{{Parton distributions with QED
  corrections}},
  \href{https://doi.org/10.1016/j.nuclphysb.2013.10.010}{\emph{Nucl. Phys. B}
  {\bfseries 877} (2013) 290}
  [\href{https://arxiv.org/abs/1308.0598}{{\ttfamily 1308.0598}}].

\bibitem{Alwall:2014hca}
J.~Alwall, R.~Frederix, S.~Frixione, V.~Hirschi, F.~Maltoni, O.~Mattelaer
  et~al., \emph{{The automated computation of tree-level and next-to-leading
  order differential cross sections, and their matching to parton shower
  simulations}}, \href{https://doi.org/10.1007/JHEP07(2014)079}{\emph{JHEP}
  {\bfseries 07} (2014) 079} [\href{https://arxiv.org/abs/1405.0301}{{\ttfamily
  1405.0301}}].

\bibitem{Frixione:2002ik}
S.~Frixione and B.~R. Webber, \emph{{Matching NLO QCD computations and parton
  shower simulations}},
  \href{https://doi.org/10.1088/1126-6708/2002/06/029}{\emph{JHEP} {\bfseries
  06} (2002) 029} [\href{https://arxiv.org/abs/hep-ph/0204244}{{\ttfamily
  hep-ph/0204244}}].

\bibitem{Sjostrand:2014zea}
T.~Sj{\"o}strand, S.~Ask, J.~R. Christiansen, R.~Corke, N.~Desai, P.~Ilten
  et~al., \emph{{An Introduction to PYTHIA 8.2}},
  \href{https://doi.org/10.1016/j.cpc.2015.01.024}{\emph{Comput. Phys. Commun.}
  {\bfseries 191} (2015) 159}
  [\href{https://arxiv.org/abs/1410.3012}{{\ttfamily 1410.3012}}].

\bibitem{Artoisenet:2013puc}
P.~Artoisenet et~al., \emph{{A framework for Higgs characterisation}},
  \href{https://doi.org/10.1007/JHEP11(2013)043}{\emph{JHEP} {\bfseries 11}
  (2013) 043} [\href{https://arxiv.org/abs/1306.6464}{{\ttfamily 1306.6464}}].

\bibitem{Antcheva:2009zz}
I.~Antcheva et~al., \emph{{ROOT: A C++ framework for petabyte data storage,
  statistical analysis and visualization}},
  \href{https://doi.org/10.1016/j.cpc.2009.08.005}{\emph{Comput. Phys. Commun.}
  {\bfseries 180} (2009) 2499}
  [\href{https://arxiv.org/abs/1508.07749}{{\ttfamily 1508.07749}}].

\bibitem{Aaboud:2016tru}
{\scshape ATLAS} collaboration, \emph{{Search for resonances in diphoton events
  at $\sqrt{s}$=13 TeV with the ATLAS detector}},
  \href{https://doi.org/10.1007/JHEP09(2016)001}{\emph{JHEP} {\bfseries 09}
  (2016) 001} [\href{https://arxiv.org/abs/1606.03833}{{\ttfamily
  1606.03833}}].

\bibitem{Feroz:2008xx}
F.~Feroz, M.~Hobson and M.~Bridges, \emph{{MultiNest: an efficient and robust
  Bayesian inference tool for cosmology and particle physics}},
  \href{https://doi.org/10.1111/j.1365-2966.2009.14548.x}{\emph{Mon. Not. Roy.
  Astron. Soc.} {\bfseries 398} (2009) 1601}
  [\href{https://arxiv.org/abs/0809.3437}{{\ttfamily 0809.3437}}].

\bibitem{Cowan:2010js}
G.~Cowan, K.~Cranmer, E.~Gross and O.~Vitells, \emph{{Asymptotic formulae for
  likelihood-based tests of new physics}},
  \href{https://doi.org/10.1140/epjc/s10052-011-1554-0}{\emph{Eur. Phys. J. C}
  {\bfseries 71} (2011) 1554}
  [\href{https://arxiv.org/abs/1007.1727}{{\ttfamily 1007.1727}}].

\bibitem{deFavereau:2013fsa}
{\scshape DELPHES 3} collaboration, \emph{{DELPHES 3, A modular framework for
  fast simulation of a generic collider experiment}},
  \href{https://doi.org/10.1007/JHEP02(2014)057}{\emph{JHEP} {\bfseries 02}
  (2014) 057} [\href{https://arxiv.org/abs/1307.6346}{{\ttfamily 1307.6346}}].

\bibitem{Khachatryan:2016hje}
{\scshape CMS} collaboration, \emph{{Search for Resonant Production of
  High-Mass Photon Pairs in Proton-Proton Collisions at $\sqrt s$ =8 and 13
  TeV}}, \href{https://doi.org/10.1103/PhysRevLett.117.051802}{\emph{Phys. Rev.
  Lett.} {\bfseries 117} (2016) 051802}
  [\href{https://arxiv.org/abs/1606.04093}{{\ttfamily 1606.04093}}].

\bibitem{grevtsov}
K.~Grevtsov, \emph{{Exploring the diphoton final state at the LHC at 13 TeV:
  searches for new particles, and the Higgs boson mass measurement with the
  ATLAS detector}}, Ph.D. thesis, Grenoble Alpes University, 2017.

\bibitem{Ellis:2016jkw}
J.~Ellis, \emph{{TikZ-Feynman: Feynman diagrams with TikZ}},
  \href{https://doi.org/10.1016/j.cpc.2016.08.019}{\emph{Comput. Phys. Commun.}
  {\bfseries 210} (2017) 103}
  [\href{https://arxiv.org/abs/1601.05437}{{\ttfamily 1601.05437}}].

\bibitem{Scott:2012qh}
P.~Scott, \emph{{Pippi - painless parsing, post-processing and plotting of
  posterior and likelihood samples}},
  \href{https://doi.org/10.1140/epjp/i2012-12138-3}{\emph{Eur. Phys. J. Plus}
  {\bfseries 127} (2012) 138}
  [\href{https://arxiv.org/abs/1206.2245}{{\ttfamily 1206.2245}}].

\end{thebibliography}\endgroup
